\documentclass[twoside,twocolumn,english,aps,showpacs,pra,superscriptaddress,10pt,floatfix]{revtex4-1}
\usepackage[T1]{fontenc}
\usepackage{lmodern}
\usepackage[utf8]{inputenc}
\usepackage{color}
\usepackage{bm}
\usepackage{amsmath}
\usepackage{amsthm}
\usepackage{amssymb}
\usepackage{mathtools}
\usepackage{graphicx}
\usepackage{esint}
\usepackage{mathrsfs}
\usepackage{booktabs}
\usepackage{multirow}
\usepackage{makecell}
\usepackage{array}
\usepackage{placeins}
\usepackage{needspace}
\usepackage{microtype}
\usepackage[plainpages=false,pdfpagelabels,bookmarks,bookmarksopen=true,
 bookmarksnumbered=true,breaklinks=true,linktocpage,colorlinks=true,
 linkcolor=blue,urlcolor=blue,citecolor=blue,anchorcolor=green,
 hyperindex=true,hyperfigures]{hyperref}
\usepackage[nameinlink,noabbrev]{cleveref}

\makeatletter
\newcommand{\startsupplementarycontents}{%
  \section*{Contents}%
  \let\SI@addcontentsline\addcontentsline
  \renewcommand{\addcontentsline}[3]{%
    \def\SI@extension{##1}\def\SI@toc{toc}%
    \ifx\SI@extension\SI@toc
      \SI@addcontentsline{sitoc}{##2}{##3}%
    \else
      \SI@addcontentsline{##1}{##2}{##3}%
    \fi}%
  \@starttoc{sitoc}%
}
\makeatother

\newtheorem{theorem}{Theorem}
\newtheorem{corollary}{Corollary}
\newtheorem{lemma}{Lemma}[section]
\newtheorem{supplementarylemma}{Supplementary Lemma}[section]
\theoremstyle{definition}
\newtheorem{definition}{Definition}
\newtheorem{supplementarydefinition}{Supplementary Definition}
\theoremstyle{remark}
\newtheorem{remark}{Remark}
\newtheorem{supplementaryremark}{Supplementary Remark}

\newcommand{\cC}{\mathcal C}
\newcommand{\cN}{\mathcal N}
\newcommand{\cP}{\mathcal P}
\newcommand{\cR}{\mathcal R}
\newcommand{\cV}{\mathcal V}
\newcommand{\Tr}{\operatorname{Tr}}
\newcommand{\supp}{\operatorname{supp}}
\newcommand{\dist}{\operatorname{dist}}
\newcommand{\sech}{\operatorname{sech}}
\newcommand{\AQEC}{D_{\mathrm{AQEC}}}
\newcommand{\eps}{\varepsilon}
\DeclareMathOperator{\diag}{diag}
\crefname{theorem}{Theorem}{Theorems}
\crefname{lemma}{Lemma}{Lemmas}
\crefname{corollary}{Corollary}{Corollaries}

\crefname{supplementarylemma}{Supplementary Lemma}{Supplementary Lemmas}
\Crefname{supplementarylemma}{Supplementary Lemma}{Supplementary Lemmas}
\crefname{supplementarydefinition}{Supplementary Definition}{Supplementary Definitions}
\Crefname{supplementarydefinition}{Supplementary Definition}{Supplementary Definitions}

\begin{document}

\title{Local enhancement of finite-strength code preparation:\texorpdfstring{\\}{ }recoverability and Wilson coherence}

\author{Tong Xu}
\affiliation{State Key Laboratory of Low Dimensional Quantum Physics, Department of Physics, Tsinghua University, Beijing, 100084, China}
\affiliation{Frontier Science Center for Quantum Information, Beijing 100184, China}

\author{Yuanchen Zhao}
\affiliation{State Key Laboratory of Low Dimensional Quantum Physics, Department of Physics, Tsinghua University, Beijing, 100084, China}
\affiliation{Frontier Science Center for Quantum Information, Beijing 100184, China}

\author{Dong E. Liu}
\affiliation{State Key Laboratory of Low Dimensional Quantum Physics, Department of Physics, Tsinghua University, Beijing, 100084, China}
\affiliation{Frontier Science Center for Quantum Information, Beijing 100184, China}
\affiliation{Beijing Academy of Quantum Information Sciences, Beijing 100193, China}
\affiliation{Hefei National Laboratory, Hefei 230088, China}

\date{September 2026}

\begin{abstract}
Stabilizer measurements provide a direct route to preparing quantum error-correcting codes and long-range entangled states. In realistic implementations, however, finite measurement strength can coherently deform the encoding map away from the projective limit. Certain nonlocal signatures associated with long-range order can survive such a deformation, but this alone does not guarantee that the resulting state protects encoded quantum information. We show that, at fixed strength, local logical leakage obstructs asymptotically faithful recovery; along logical operators' supports, its accumulation forces the relevant measurement strength to grow logarithmically with code distance. We then prove that this obstruction can be removed locally: for fixed toric registers under sufficiently weak local-stochastic errors, enhancing only a vanishing fraction of checks near selected logical supports restores recovery while the bulk remains at finite  measurement strength. With recoverability restored, we prove a further dimensional distinction in logical operations. In the strict three-dimensional deconfined phase, suitable enhanced tubes restore the equivalence of macroscopically separated representatives, so either representative asymptotically implements the same controlled logical action on a fixed encoding. In two dimensions, recoverability alone does not restore this equivalence: the representatives can remain inequivalent even when both associated encodings are recoverable. 
\end{abstract}

\maketitle

\section{Introduction}\label{sec:introduction}
Stabilizer measurements provide a direct way to prepare quantum error-correcting codes by imposing commuting constraints on a many-body state~\cite{Kitaev2003,Dennis2002}. Preparing topological order from a product state by bounded local unitary dynamics requires a time that grows with system size~\cite{BravyiHastingsVerstraete2006}. Measurements and classical feed-forward can instead prepare such states with constant quantum depth~\cite{Piroli2021,Lu2022}, with constructions extending to a broad range of topological orders~\cite{Tantivasadakarn2024,Tantivasadakarn2023}. A toric-code implementation using mid-circuit measurements and feed-forward has also been demonstrated experimentally~\cite{Iqbal2024}. In the ideal code space, the measured constraints both protect information against local errors and make different physical representatives of a logical operator act identically.

In practice, imperfect control of the system--ancilla interaction can produce deviations from the intended projective measurement~\cite{Brun2002,Zhu2023}. We describe such deviations through finite-strength measurement operators that retain coherent amplitudes across different check sectors and thereby deform the encoding map itself~\cite{Yang2022,PNAS2025}. In many-body states, such measurements can drive phase transitions~\cite{Zhu2023,Chen2025Nishimori,Eckstein2024Teleportation,LeeJianXu2023,Su2024HigherForm} yet preserve long-range entanglement, including in a three-dimensional toric code~\cite{Zhu2023,Lee2022}.
Finite-strength protocols can nevertheless retain long-range entanglement, including in a three-dimensional toric code~\cite{Zhu2023,Lee2022}. For the all-plus branch, the corresponding ferromagnetic $\mathbb Z_2$ gauge theory has an area law at every finite strength in two dimensions but a finite-strength deconfined perimeter-law regime in three dimensions~\cite{Wegner1971,Kogut1979,Forsstrom2025,DuncanSharp2025}. Yet the survival of these nonlocal signatures or of bulk order does not by itself show that the prepared state is a useful quantum code, which must protect an unknown input and keep its logical representatives equivalent. We ask whether both operational requirements are governed by the same bulk order when stabilizer measurements are only approximate.

Approximate quantum error correction (AQEC) relates recoverability to logical information leaked to the noise environment~\cite{Schumacher2002,Kretschmann2008,Beny2010}. Local correctability and subsystem variance connect protection to code geometry and complexity~\cite{Flammia2017,Yi2024}, while studies of noisy topological phases relate quantum-memory stability to mixed-state order and reversibility~\cite{Fan2024MixedStateDiagnostics,SangHsieh2025Markov,Bao2026ErrorfieldDouble}. Finite-strength preparation can obstruct recovery~\cite{Zhao2024} and make the encoding depend on the chosen logical representative~\cite{PNAS2025}. Inspired by earlier work on quasi-locally dressed and locally error-corrected Wilson observables~\cite{HastingsWen2005,Cong2024LocalErrorCorrection}, we ask whether AQEC recoverability can similarly be restored through spatially selective control: can measurement accuracy be concentrated near selected logical supports while the bulk remains at fixed finite strength? If such selective control is sufficient, how must its strength and spatial extent scale with code size? Even then, a second question remains: does the same local control make macroscopically separated representatives act identically on a fixed encoded state, or does their equivalence require an additional condition on the intervening bulk?

\begin{figure}[!tbp]
 \centering\includegraphics[width=\columnwidth]{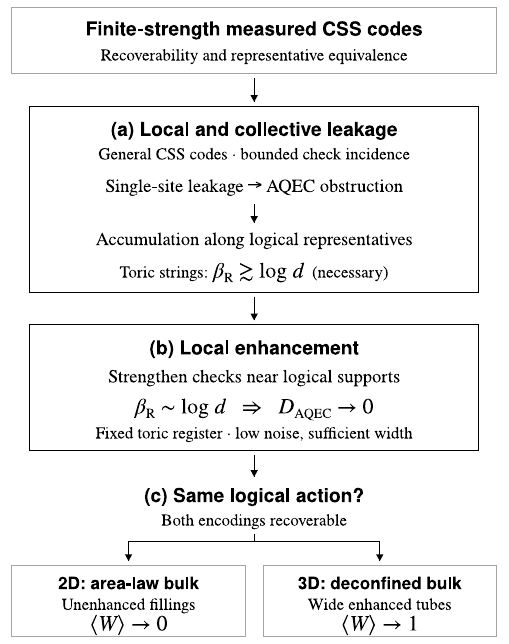}
 \caption[Logical structure of the results]{\textbf{Logical structure of the results.} \textbf{(a)} Local errors reveal selected logical information in general CSS encodings. Along toric representatives, accumulated leakage makes logarithmic measurement strength necessary. \textbf{(b)} Strengthening nearby checks restores AQEC at low noise and sufficient width. \textbf{(c)} Once both encodings are recoverable, their representative equivalence still depends on the bulk: the Wilson overlap vanishes in 2D with macroscopic unenhanced fillings, whereas sufficiently enhanced tubes restore it to one in the strict 3D deconfined phase.}
 \label{fig:logic-overview}
\end{figure}

We identify a local source of logical leakage in the conditioned Calderbank--Shor--Steane (CSS) encodings [Fig.~\ref{fig:logic-overview}(a)]. A single-qubit $X$ error on a selected toric support reverses only $O(1)$ measured checks. At finite strength, each logical-basis codeword has a nonzero overlap with its error image, with a sign set by the logical input. This makes logical inputs locally distinguishable and allows independent noise to transfer information to the environment. At a fixed nonzero error rate, the leakage persists as the code grows if the incident strengths remain bounded, even in the three-dimensional deconfined phase. The same mechanism leaves nonzero leakage per selected logical qubit in CSS quantum low-density parity-check (qLDPC) encodings at uniformly bounded strength. Along a straight toric representative, $d$ local signals can retain finite collective leakage even when each vanishes, requiring the common incident-check strength to grow at least logarithmically with ideal code distance $d$ (Theorem~\ref{thm:local} and Corollary~\ref{cor:obstruction}).

We prove that enhancing neighborhoods of selected logical representatives restores recovery while the bulk remains at fixed finite strength [Fig.~\ref{fig:logic-overview}(b)]. Increasing the measurement strength of the enhanced checks suppresses error overlaps whenever their syndrome enters the protected region, while increasing its width forces any logical crossing that leaves no protected syndrome to involve a larger support. Under sufficiently weak local-stochastic noise, without assuming independent error locations, these two mechanisms give entanglement fidelity approaching one for a fixed logical register. For straight toric representatives, enhancement strength $O(\log d)$ suffices, with widths $O(\log d)$ in two dimensions and $O(\sqrt{\log d})$ in three; both constructions use $O(d\log d)$ enhanced checks, a vanishing fraction of the total. For CSS codes with good linear confinement, the sufficient width is reduced to $O(\log\log U)$, where $U$ is the union size of the selected supports. The toric logarithmic strength matches the independent-noise lower bound in order, while the recovery construction requires no bulk phase condition (Theorem~\ref{thm:enhancement}; Corollary~\ref{cor:resources}).


With recoverability established, we compare two straight, parallel, macroscopically separated representatives of the same toric logical class in the same enhanced preparation [Fig.~\ref{fig:logic-overview}(c)]. Their equivalence is measured by the Wilson overlap $w=\langle Z^{\ell_1}Z^{\ell_2}\rangle$, which tests whether the two physical strings act the same on a fixed encoding. In two dimensions, both fillings relating the strings retain macroscopic unenhanced area, so finite-strength factors accumulate and force $w\to0$, even with projective strips. In the strict three-dimensional deconfined phase, full enhanced tubes suppress the dual obstructions that distinguish the representatives; logarithmic radius and strength suffice to give $w\to1$. The same overlap directly quantifies logical action: substituting one representative for the other in an ideal controlled operation incurs the exact operator-norm error $\sqrt{2(1-w)}$. Thus recoverability and coherent representative equivalence are distinct requirements (Theorem~\ref{thm:wilson}; Corollary~\ref{cor:cnot}).

Two implications follow for code preparation and logical control:
\begin{itemize}
    \setlength{\itemsep}{0.5em}
    \setlength{\parsep}{0pt}

    \item \textbf{Preparation accuracy can be concentrated near selected logical supports.}For the stated toric protocols, strengthening a vanishing fraction of checks suffices for recovery at fixed logical-register size. The required accuracy scales with accumulated local leakage, rather than the location of a bulk phase boundary.
    \item \textbf{Local enhancement and three-dimensional deconfinement jointly restore representative equivalence.} Once local enhancement establishes recoverability, sufficiently strong full tubes in the three-dimensional deconfined phase also restore representative equivalence. The two-dimensional comparison shows that recoverability alone does not ensure this agreement.
\end{itemize}


\section{Results}\label{sec:results}
We first identify the leakage that the enhanced checks must suppress, then bound the strength and width needed for recovery. With recoverability established, we compare the coherent actions of distant representatives.

\begin{figure*}[!t]
 \centering\includegraphics[width=\textwidth]{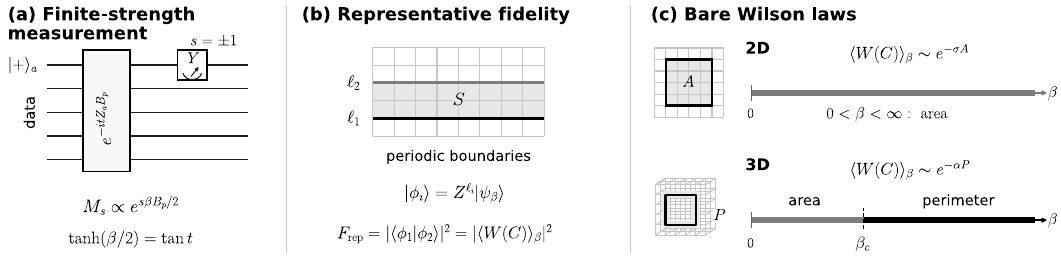}
 \caption[Finite-strength measurement, logical representatives, and Wilson laws]{\textbf{Finite-strength measurement, logical representatives, and Wilson laws.} \textbf{(a)} Ancilla-assisted measurement of a plaquette check $B_p$. All-plus postselection over the commuting checks prepares $\lvert\psi_\beta\rangle$ from $\lvert+\rangle^{\otimes n}$~\cite{Zhu2023,Zhao2024}. \textbf{(b)} Homologous logical strings bound a filling $S$, with $C=\partial S$ and $W(C)=Z^{\ell_1}Z^{\ell_2}$. Their squared state fidelity $F_{\mathrm{rep}}$ tests representative equivalence. \textbf{(c)} At uniform strength, 2D has an area law for every fixed $0<\beta<\infty$, whereas 3D undergoes an area-to-perimeter transition at finite $\beta_c$~\cite{Forsstrom2025,DuncanSharp2025,DuncanTopological2025}. Here $A$ and $P$ denote filling area and boundary length. The displayed laws give leading large-loop scaling away from criticality. A bare perimeter law still allows $F_{\mathrm{rep}}\to0$ for growing, macroscopically separated strings and does not by itself restore their equivalence.}
\label{fig:background}
\label{fig:weak-circuit}
\label{fig:preparation}
\end{figure*}

\subsection*{Background: finite-strength preparation and logical representatives}
\phantomsection\label{sec:model}
\addcontentsline{toc}{subsection}{Background: finite-strength preparation and logical representatives}
Consider a Calderbank--Shor--Steane (CSS) code on $n$ physical qubits, with $X$ checks $A_\mu=X^{a_\mu}$ and measured $Z$ checks $B_j=Z^{h_j}$ satisfying $[A_\mu,B_j]=0$~\cite{CalderbankShor1996,Steane1996}. The vectors $h_j$ are the rows of the binary matrix $H_Z$; all checks are unsigned, and redundant checks are allowed. For the edge-qubit toric codes, $A_v$ is a vertex star and $B_p=\prod_{e\in\partial p}Z_e$ is a plaquette check~\cite{Kitaev2003,Dennis2002}. Starting from $|+\rangle^{\otimes n}$ fixes every $A_\mu=+1$. Measuring the commuting $B_j$ then imposes the remaining constraints. Finite-strength measurements replace these projective constraints by conditional filters~\cite{Zhu2023,Zhao2024}; coherent deviations in syndrome extraction motivate studying this preparation error separately from subsequent noise~\cite{Yang2022,PNAS2025}.

Figure~\ref{fig:background}(a) gives a circuit for one check. Prepare an ancilla in $|+\rangle_a$, apply $U_j(t_j)=\exp[-it_jZ_a\otimes B_j]$, and measure $Y_a$ with result $s_j=\pm1$. Here $t_j\in[0,\pi/4]$ is a dimensionless interaction angle. Taking the ancilla matrix element gives, up to a branch-global phase~\cite{Brun2002,Zhu2023,Zhao2024},
\begin{equation}
 \begin{aligned}
 M_{j,s_j}
 &=\frac{\cos t_j\,I+s_j\sin t_j\,B_j}{\sqrt2}
 =\frac{e^{s_j\beta_j B_j/2}}{\sqrt{2\cosh\beta_j}},\\
 \tanh(\beta_j/2)&=\tan t_j,\qquad
 \tanh\beta_j=\sin(2t_j).
 \end{aligned}
 \label{eq:weak-kraus-circuit}
\end{equation}
Thus $\beta_j=2\operatorname{artanh}(\tan t_j)$ is the dimensionless filter strength. At $t_j=0$ ($\beta_j=0$), $M_{j,s_j}=I/\sqrt2$ and the readout carries no check information; at $t_j=\pi/4$ ($\beta_j\to\infty$), it becomes the projector $(I+s_jB_j)/2$. At intermediate strength, the conditional operation preserves coherence between check sectors while changing their relative amplitudes. The exact ancilla contraction and its outcome probabilities are derived in Supplementary Note~1.

We condition throughout on the all-plus record, $s_j=+1$ for every measured check. Multiplying the commuting Kraus operators and normalizing gives
\begin{equation}
 \begin{aligned}
 |\psi_\beta\rangle
 &=\frac{\exp(\frac12\sum_j\beta_jB_j)|+\rangle^{\otimes n}}
 {\sqrt{Z_\beta}},\\
 Z_\beta&=\langle+|^{\otimes n}
 \exp\!\left(\sum_j\beta_jB_j\right)|+\rangle^{\otimes n},\\
 p_+&=\frac{Z_\beta}{\prod_j2\cosh\beta_j}.
 \end{aligned}
 \label{eq:weakstate}
\end{equation}
Here $p_+$ is the probability of obtaining that record, and $\beta$ in subscripts abbreviates the full strength profile. This is a conditional pure-state preparation. Averaging over, or discarding, the readout instead defines a different quantum channel~\cite{Zhu2023,Lee2022}. We do not assume that a generic outcome record can be converted to this branch at no cost; Supplementary Note~1 gives the general-record state, the postselection probability including redundant checks, and the distinction from outcome averaging.

Choose $k$ ideal $Z$-logical representatives that are independent modulo the $Z$ stabilizers, and set $K=2^k$. They define an encoding $V_\beta:\mathcal H_L\to\mathcal H_Q$, where $L$ is the logical input register and $Q$ is the $n$-qubit physical system,
\begin{equation}
 \begin{aligned}
 V_\beta|q\rangle_L&=|q_\beta\rangle=Z^{\ell(q)}|\psi_\beta\rangle,\\
 \ell(q)&=\bigoplus_{a=1}^kq_a\ell_a,\qquad q\in\mathbb F_2^k.
 \end{aligned}
 \label{eq:encoding}
\end{equation}
Here $|q\rangle_L$ is an abstract logical basis state and $|q_\beta\rangle$ is its physical codeword. Complementary physical logical $X$ operators distinguish these codewords, so $V_\beta^\dagger V_\beta=I_K$. Postselection has input-independent success probability, including for reference-entangled inputs (\hyperref[sec:m-encoding]{Appendix} and Supplementary Note~1). Write $P_\beta=V_\beta V_\beta^\dagger$. The register can contain some or all ideal logical qubits; $d$ denotes the ideal code distance, not an exact distance of the imperfect subspace. In the toric examples, qubits lie on edges of a periodic square or cubic lattice and the measured checks are plaquette boundaries. Here $d=L$, the torus side length.

The focus on the $B_j$ sector follows from the asymmetry of this preparation. Every $A_\mu$ remains an exact stabilizer of the entire encoded subspace: it stabilizes the initial state and commutes with both the filter and the chosen logical $Z$ operators. In the toric language, the gauge charges remain absent while plaquette fluxes fluctuate, as in the finite-strength measurement construction of Ref.~\cite{Zhu2023}. More strongly, the filtered encoding retains exact correction of any set of pure $Z$ errors whose pairwise differences have weight below the ideal $Z$ distance $d_Z$. Supplementary Note~1, ``Why the $B_j$ sector is the relevant recovery problem,'' proves this statement from the Knill--Laflamme condition~\cite{Knill1997}; it does not require the $B_j$ to be sharp. Physical $X$ errors, by contrast, anticommute with the weakened $B_j$ checks and acquire the residual logical overlaps studied below. We therefore analyze the quantum recovery problem for subsequent $X$ noise. This sector-specific recovery problem concerns arbitrary encoded quantum inputs, including their entanglement with a reference.

The Born weights of Eq.~\eqref{eq:weakstate} define a classical multispin model for any CSS code in this construction, with the generalized Ising form of Ref.~\cite{Wegner1971}. For a computational-basis configuration $s\in\mathbb F_2^n$, its action and probability measure are
\begin{equation}
 \begin{aligned}
 \mathcal S_\beta(s)&=-\sum_j\beta_j(-1)^{(H_Zs)_j},\\
 \pi_\beta(s)&=\frac{e^{-\mathcal S_\beta(s)}}
 {\sum_{s'}e^{-\mathcal S_\beta(s')}}.
 \end{aligned}
 \label{eq:general-wilson-action}
\end{equation}
Diagonal quantum expectations equal averages in this measure. For toric codes, each check eigenvalue is the plaquette product $\prod_{e\in\partial p}(-1)^{s_e}$, so $\mathcal S_\beta$ is the ferromagnetic $\mathbb Z_2$ Wilson action on the square or cubic lattice. The fixed all-plus branch introduces no measurement-record disorder.

Two representatives $\ell,\ell'$ of one logical class differ by a check product $W=Z^c$, with $c=\ell\oplus\ell'\in\operatorname{im}H_Z^T$. Their difference on the prepared state satisfies
\begin{equation}
 \|(Z^{\ell}-Z^{\ell'})|\psi_\beta\rangle\|^2
 =2(1-\langle W\rangle_\beta).
 \label{eq:representative}
\end{equation}
The states $|\phi_1\rangle=Z^\ell|\psi_\beta\rangle$ and $|\phi_2\rangle=Z^{\ell'}|\psi_\beta\rangle$ therefore have squared fidelity $F_{\rm rep}=|\langle W\rangle_\beta|^2$ [Fig.~\ref{fig:background}(b)]. Expanding each check weight gives the exact general-code expression
\begin{equation}
 \langle W\rangle_\beta
 =\frac{\displaystyle\sum_{\substack{a\\H_Z^Ta=c}}\quad
              \prod_{j:a_j=1}\tanh\beta_j}
 {\displaystyle\sum_{\substack{a\\H_Z^Ta=0}}\quad
              \prod_{j:a_j=1}\tanh\beta_j}.
 \label{eq:general-wilson}
\end{equation}
Here $a$ selects a binary subset of measured checks. The numerator weights every check product implementing the change of representative; the denominator accounts for relations among the checks.

For the two-dimensional toric code, the support of $c$ forms a boundary $C$. The only closed plaquette sets on the torus are the empty set and the full lattice. A filling of $A$ plaquettes has a unique complementary filling of $L^2-A$ plaquettes. At uniform strength, Eq.~\eqref{eq:general-wilson} reduces to the finite-torus expression~\cite[Eq.~(65)]{Wegner2014}
\begin{equation}
 \langle W(C)\rangle_\beta
 =\frac{(\tanh\beta)^A+(\tanh\beta)^{L^2-A}}
 {1+(\tanh\beta)^{L^2}}.
 \label{eq:2dwilson}
\end{equation}
As both fillings grow, $-\log\langle W(C)\rangle_\beta$ equals $-\log(\tanh\beta)$ times the smaller area, up to a bounded correction. This area coefficient is positive at every fixed $0<\beta<\infty$; there is no finite-strength deconfined phase.

In three dimensions, closed plaquette surfaces include cube boundaries and noncontractible surfaces. At uniform strength, summing over all plaquette sets $S$ on the cubic torus gives
\begin{equation}
 \langle W(C)\rangle_\beta
 =\frac{\displaystyle\sum_{S:\,\partial S=C}(\tanh\beta)^{|S|}}
 {\displaystyle\sum_{S:\,\partial S=0}(\tanh\beta)^{|S|}},
 \label{eq:3dwilson}
\end{equation}
where $|S|$ counts the plaquettes in $S$. The local cube constraints force plaquette defects into closed loops on the dual lattice. At sufficiently large strength these loops are dilute, and their linking with $C$ produces a perimeter law. For large rectangular loops of fixed aspect ratio in the thermodynamic limit, the two phases satisfy $-\log\langle W(C)\rangle_\beta\sim\sigma_\beta A(C)$ for $0<\beta<\beta_c$ and $-\log\langle W(C)\rangle_\beta\sim\tau_\beta|C|$ for $\beta>\beta_c$, with positive coefficients and a finite $\beta_c$ [Fig.~\ref{fig:background}(c)]~\cite{Forsstrom2025,DuncanSharp2025,DuncanTopological2025}. Here $A(C)$ is the enclosed area and $|C|$ the perimeter. A perimeter law slows the decay of $F_{\rm rep}$ but still allows it to vanish as the boundary grows. Deconfinement therefore does not by itself restore representative equivalence, or establish recovery of a chosen encoding after noise. \hyperref[sec:m-wilson]{Appendix} and Supplementary Notes~1 and 4 give the gauge mapping, the loop conditions for these asymptotic laws, and the finite-torus estimates used below.

\begin{figure*}[!t]
 \centering\includegraphics[width=\textwidth]{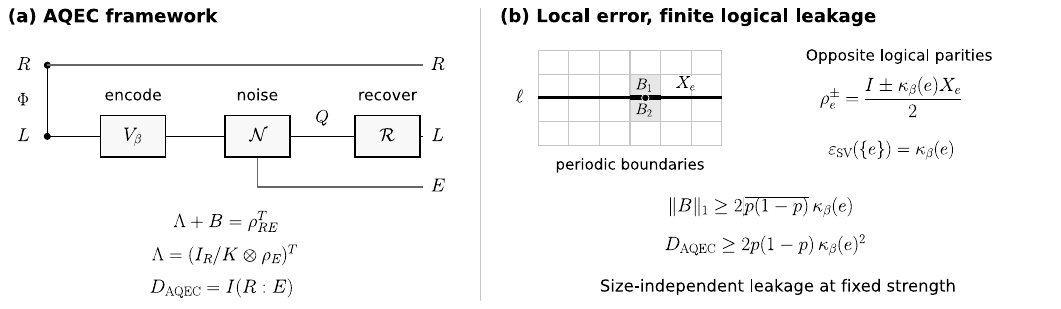}
 \caption[AQEC framework and local logical leakage]{\textbf{AQEC framework and local logical leakage.} \textbf{(a)} A logical input $L$ is maximally entangled with a reference $R$, encoded by $V_\beta$, and subjected to noise $\mathcal N$. Recovery $\mathcal R$ acts on the physical output $Q$. The normalized error-overlap matrix $\Lambda+B$ describes the joint reference--environment state, whereas $\Lambda$ retains its logical average; $T$ denotes the full transpose. The AQEC relative entropy measures correlations between $R$ and the noise environment $E$. \textbf{(b)} A single-edge error $X_e$ on a selected toric logical support reverses its incident plaquettes. Opposite logical parities give the reduced states $\rho_e^\pm$ and the single-edge subsystem variance shown. The leakage bounds assume independent $X$ flips with fixed $0<p<1$. At uniform finite strength, $\kappa_\beta(e)\ge e^{-s\beta}>0$, with $s=2$ in 2D (shown) and $s=4$ in 3D, so this local obstruction persists as the code grows.}
 \label{fig:leakage}
\end{figure*}

\subsection{Local logical leakage at finite measurement strength}
\label{sec:obstruction}\label{sec:tools}
We characterize recovery through the logical information transferred to the noise environment~\cite{Zhao2024}. For the encoding $V_\beta$ in Eq.~\eqref{eq:encoding}, consider the physical noise channel
\begin{equation}
 \cN(\rho)=\sum_{x\subseteq[n]}P(x)X_x\rho X_x,
 \qquad X_x=\prod_{e\in x}X_e.
 \label{eq:noise}
\end{equation}
Throughout this subsection, the flips are independent, with $P(x)=p^{|x|}(1-p)^{n-|x|}$ and fixed $0<p<1$. Writing its Kraus operators as $E_x=\sqrt{P(x)}X_x$, the overlaps between corrupted code states form the normalized Knill--Laflamme matrices~\cite{Knill1997,Zhao2024},
\begin{align}
 (\Lambda+B)_{xy,qq'}&=\frac1K\langle q_\beta|E_x^\dagger E_y|q'_\beta\rangle,\nonumber\\
 \Lambda_{xy,qq'}&=\frac{\delta_{qq'}}{K^2}\sum_u
 \langle u_\beta|E_x^\dagger E_y|u_\beta\rangle.
 \label{eq:matrixelements}
\end{align}
Here $x,y$ label error branches and $q,q'$ label logical inputs, with rows indexed by $(q,x)$ and columns by $(q',y)$. The matrix $\Lambda$ retains the logical average, while $B$ records the dependence on the encoded state. Exact correction is possible precisely when $B=0$.

For a logical input maximally entangled with a reference $R$, the same matrices describe correlations with the noise environment $E$ [Fig.~\ref{fig:leakage}(a)]: $\Lambda+B=\rho_{RE}^T$ and $\Lambda=(I_R/K\otimes\rho_E)^T$, where $T$ denotes the full transpose in the chosen basis. The AQEC relative entropy and the trace-norm deviation therefore measure the remaining correlations between $R$ and $E$:
\begin{equation}
 \begin{aligned}
 \AQEC&=D(\Lambda+B\Vert\Lambda)=I(R:E),\\
 \eps&=\frac12\|B\|_1
 =\frac12\left\|\rho_{RE}-\frac{I_R}{K}\otimes\rho_E\right\|_1.
 \end{aligned}
 \label{eq:metrics}
\end{equation}
At fixed register size $k$, vanishing $\AQEC$ or $\|B\|_1$ is equivalent to $F_e^\star\to1$, where $F_e^\star$ is the optimal squared entanglement fidelity for recovery of the whole register~\cite{Schumacher2002,Beny2010,Holevo1973,Winter2016}. Recovery acts on the physical output and has no access to $E$. Its definition and quantitative bounds are given in \hyperref[sec:m-recovery]{Appendix}.

The prepared code can also reveal logical information through a local measurement before noise acts. For a physical region $A$, this dependence is measured by the subsystem variance~\cite{Yi2024},
\begin{equation}
 \varepsilon_{\mathrm{SV}}(A)
 =\max_{\rho_L}\left\|\operatorname{Tr}_{A^c}\!\left(V_\beta\rho_LV_\beta^\dagger-\frac{P_\beta}{K}\right)\right\|_1.
 \label{eq:subsystem-variance}
\end{equation}
The maximization includes all logical states, including coherent superpositions, and $P_\beta/K$ is the maximally mixed code state. Here $\|\cdot\|_1$ is the full trace norm. Thus $\varepsilon_{\mathrm{SV}}(A)=0$ means that every encoded state has the same marginal on $A$; no measurement confined to that region can distinguish logical inputs. This is a property of the prepared subspace, while $\AQEC$ also depends on the noise channel.

For the encoding prepared at finite strength, both quantities are controlled by the residual overlap of physical errors. Let $\sigma(x)=\supp(H_Zx)$ be the syndrome of $x$ and $g(x)=(|x\cap\ell_a|\bmod2)_{a=1}^k$ its parity on the selected representatives. The exact logical sectors give
\begin{equation}
 \begin{aligned}
 \langle q_\beta|X_x|q'_\beta\rangle
 &=\delta_{qq'}\chi_q(x)\kappa_\beta(x),\\
 \kappa_\beta(x)&=\langle\psi_\beta|X_x|\psi_\beta\rangle,
 \qquad \chi_q(x)=(-1)^{q\cdot g(x)}.
 \end{aligned}
 \label{eq:sharedkernel}
\end{equation}
For $g(x)\ne0$, the sign distinguishes logical inputs. A single edge on a toric logical string already has this property [Fig.~\ref{fig:leakage}(b)]: opposite logical parities give reduced states $\rho_e^\pm=(I\pm\kappa_\beta(e)X_e)/2$. In a dilation of the independent noise channel, the environment qubit associated with this edge has the same diagonal probabilities $1-p,p$ for both inputs, but opposite off-diagonal elements $\pm\sqrt{p(1-p)}\,\kappa_\beta(e)$. Logical information is therefore carried by environment coherences even though the error probabilities are input-independent. Since $X_e$ reverses only finitely many checks, this dependence survives at fixed finite strength as the code grows, including in the three-dimensional Wilson perimeter-law regime. A longer representative also exposes more sites carrying the same logical parity. The following bounds quantify both the information revealed at one site and its accumulation along the selected support.

\begin{theorem}[Local leakage and its accumulation]\label{thm:local}\label{thm:collective}
For the encoding in Eq.~\eqref{eq:encoding} with nonnegative measurement strengths, any site with $g(e)\ne0$ obeys
\begin{equation}
 \varepsilon_{\mathrm{SV}}(\{e\})\ge\kappa_\beta(e)
 \ge\exp\!\left[-\sum_{j\in\sigma(e)}\beta_j\right].
 \label{eq:localobstruction}
\end{equation}
Under independent $X$ noise with fixed $0<p<1$,
\begin{equation}
 \begin{aligned}
 \|B\|_1&\ge2\sqrt{p(1-p)}\,\kappa_\beta(e),\\
 \AQEC&\ge2p(1-p)\kappa_\beta(e)^2.
 \end{aligned}
 \label{eq:localentropy}
\end{equation}
If $M$ sites share the same nonzero logical parity, have pairwise disjoint syndrome supports of size $s$, and every check in these supports has strength $\beta$, then
\begin{equation}
 \AQEC\ge
 \frac{2p(1-p)M e^{-2s\beta}}{1+2^{2s+2}p(1-p)M e^{-2s\beta}}.
 \label{eq:collective}
\end{equation}
The remaining strengths may be arbitrary and nonnegative. For straight toric representatives, $M=d$ and $s=2$ or $4$; at fixed $k$ and $p$, asymptotically faithful recovery requires $\beta-(\log d)/(2s)\to+\infty$.
\end{theorem}

The obstruction depends only on the checks reversed by the local error. Increasing measurements elsewhere cannot remove it while these incident strengths remain bounded [Fig.~\ref{fig:leakage}(b)]. In edge-qubit toric codes, a single edge reverses two plaquettes in two dimensions and four in three dimensions. For $g(e)\ne0$, its subsystem variance equals $\kappa_\beta(e)$ exactly. At uniform strength and $L\ge3$,
\begin{equation}
 \begin{aligned}
 \kappa_\beta(e)&=\frac{\sech^2\beta}{1+(\tanh\beta)^{L^2}}&& (2D),\\
 e^{-4\beta}\le\kappa_\beta(e)&\le\sech^4\beta&& (3D).
 \end{aligned}
 \label{eq:2dkernel}
\end{equation}
Both remain bounded away from zero at every fixed finite strength. The three-dimensional deconfined phase therefore retains a local recovery obstruction despite its perimeter-law Wilson correlations.

The same mechanism applies to CSS quantum low-density parity-check (qLDPC) encodings, where bounded check incidence takes the place of lattice geometry. Physical sites can be chosen that probe independent logical parities of the selected register. Their combined contribution leaves a finite leakage per logical qubit, even when the register grows.
\begin{corollary}\label{cor:obstruction}
Suppose each physical qubit belongs to at most $\Delta_Z$ measured checks and all strengths satisfy $\beta_j\le\beta_0<\infty$, uniformly in size. For any nonempty selected register and fixed independent noise rate $0<p<1$,
\begin{equation}
 \begin{aligned}
 \frac{\AQEC}{k}&\ge2p(1-p)e^{-2\Delta_Z\beta_0},\\
 1-F_e^\star&\ge\frac{p(1-p)}4e^{-2\Delta_Z\beta_0}.
 \end{aligned}
 \label{eq:densitylower}
\end{equation}
\end{corollary}

This result shows the fixed-strength obstruction for genral CSS qLDPC encodings. Increasing the ideal distance therefore cannot compensate for uniformly bounded preparation strength under fixed nonzero noise. The infidelity bound concerns recovery of the entire selected register and remains uniform in $k$. The entropy-density bound further shows that the obstruction cannot be diluted by storing more logical qubits.

Even when the preparation strength grows, suppressing each local signal separately need not suppress their combined information. For example, in two dimensions $\beta=\frac14\log L$ gives $\max_e\varepsilon_{\mathrm{SV}}(\{e\})\sim4/\sqrt L\to0$, while Eq.~\eqref{eq:collective} leaves $\AQEC\ge2p(1-p)/[1+64p(1-p)]>0$. Thus every single-qubit marginal can become independent of the logical input in the limit, while the noise environment retains a finite amount of logical information.

For straight toric strings, the number of such signals is $M=d$. At fixed $k$ and $p$, their total leakage can vanish only if
\begin{equation}
 \beta-\tfrac14\log d\to+\infty\quad(2D),\qquad
 \beta-\tfrac18\log d\to+\infty\quad(3D).
 \label{eq:lognecessary}
\end{equation}
The necessary strength concerns only checks incident on the selected string, suggesting that stronger preparation measurements can be concentrated near the chosen supports. \hyperref[sec:m-leakage]{Appendix} and Supplementary Note~2 give the local and collective leakage derivations.

\subsection{Restoring recoverability by local measurement enhancement}
\label{sec:enhancement}\label{sec:geometry}

\begin{figure*}[t]
 \centering\includegraphics[width=\textwidth]{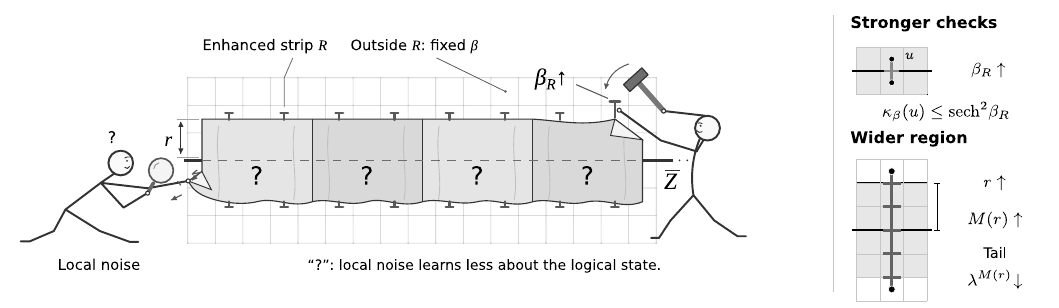}
 \caption[Local measurement enhancement]{\textbf{Local enhancement suppresses logical leakage.} Checks $\mathcal P_r$ in the shaded strip $R$ are prepared at strength $\beta_R$; the bulk remains at fixed $\beta$. Question marks indicate reduced logical information available to local noise. In the upper inset, stronger checks suppress the overlap of a short crossing $u$ with syndrome on two enhanced checks. In the lower inset, a longer crossing has syndrome only outside the enhanced region. Widening the region raises the minimum connected support size $M(r)$ of crossings with nonzero selected logical parity that remain silent inside, suppressing their contribution through $\lambda^{M(r)}$. Here $\lambda\le\lambda_0<1$ accounts for noise strength and connected-support counting; colored dots denote syndrome checks.}
 \label{fig:enhancement}
\end{figure*}

The preceding result requires logarithmic measurement strength near a straight toric representative. {\bfseries\boldmath Is the scaling $\beta\sim\log d$ also sufficient to make the AQEC relative entropy vanish? How large a neighborhood of the selected logical representatives must be strengthened?} For a fixed logical register, logarithmic strengthening does suffice under sufficiently weak noise. Only a vanishing fraction of the toric checks need be enhanced; the required width is determined by the geometry of logical crossings.

\begin{samepage}
We increase the preparation strength on a set of checks $\mathcal P_r$ surrounding the selected logical supports, with $r$ denoting the width of this neighborhood:
\begin{equation}
 \beta_j=\begin{cases}
 \beta_R,&j\in\mathcal P_r,\\
 \beta,&j\notin\mathcal P_r,
 \end{cases}
 \qquad \beta_R\ge\beta\ge0.
 \label{eq:enhancement-profile}
\end{equation}
\end{samepage}
The bulk strength $\beta$ remains fixed and finite. Figure~\ref{fig:enhancement} illustrates the resulting preparation around a chosen representative. Stronger checks reduce the residual overlap of short error crossings with syndrome in $\mathcal P_r$. Increasing the width catches syndrome that would lie outside a narrower region. A crossing can still escape these checks, but in the geometries considered below it must then extend across a larger support. At low noise, these extended crossings make a small contribution to logical leakage.

The relevant size is $M(r)$, the smallest connected support of an error difference $u$ with nonzero selected logical parity, $g(u)\ne0$, and no protected syndrome, $\sigma(u)\cap\mathcal P_r=\varnothing$. For a toric string, the parity condition is an odd intersection with the representative. The error may have syndrome outside $\mathcal P_r$. Connectedness is defined in the bounded-degree graph $G_Z$ joining qubits that share a measured check; we set $M(r)=\infty$ when no such support exists.

For correlated $X$ noise, we assume the local-stochastic bound~\cite{Fawzi2018}
\begin{equation}
 \Pr[A\subseteq x]\le\eta^{|A|}\quad\text{for every }A\subseteq[n].
 \label{eq:localstochastic}
\end{equation}
There are at most $\mu^m$ connected supports of size $m$ through a given qubit, with $\mu$ fixed by the local check structure. Their contributions can be summed at sufficiently small $\eta$, with $\lambda=4\mu\sqrt\eta\le\lambda_0$ for a fixed $0<\lambda_0<1$. The number of possible starting sites is controlled by the selected-support union $U=|\bigcup_a\ell_a|$. We can therefore bound leakage under the full physical noise channel by the sizes of crossings near this union, even though errors may occur throughout the code.

\begin{theorem}[Recovery under finite-strength local enhancement]\label{thm:enhancement}\label{thm:soft}\label{thm:hard}
For the preparation in Eq.~\eqref{eq:enhancement-profile} and noise satisfying Eq.~\eqref{eq:localstochastic} with $\lambda\le\lambda_0<1$,
\begin{equation}
 \begin{aligned}
 \|B\|_1\le{}&\exp\!\left[\frac{CU\lambda^{M(r)}}{1-\lambda}\right]-1\\
 &+4\sqrt{|\mathcal P_r|}\,e^{-\beta_R},
 \end{aligned}
 \label{eq:finite-recovery}
\end{equation}
where $C$ depends only on the bounded local check structure and $\lambda^\infty=0$. The bound also holds for nonuniform protected strengths $\beta_j\ge\beta_R$ and finite, nonnegative exterior strengths. Projective enhancement removes the last term. For fixed $k$,
\begin{equation}
 U\lambda^{M(r)}\longrightarrow0,\qquad
 \sqrt{|\mathcal P_r|}\,e^{-\beta_R}\longrightarrow0
 \label{eq:enhancement-criterion}
\end{equation}
are sufficient for $\AQEC\to0$ and $F_e^\star\to1$.
\end{theorem}

The two terms distinguish errors that evade the enhanced checks from the residual effect of checks measured at finite strength. The first includes crossings whose syndrome lies outside the protected region. The second follows by comparing the prepared state with its projection onto the enhanced-check sector; this projection is used only in the proof. No phase condition is imposed on the bulk. What matters for recovery is that the protected neighborhood intercepts short logical crossings and that its checks are sufficiently sharp. The bound guarantees the existence of recovery on the physical output; it does not prescribe a decoder confined to that neighborhood. The estimates are derived in \hyperref[sec:m-projective]{Appendix} and Supplementary Note~3.

\begin{table*}[t]
 \centering\small
 \begin{tabular*}{\textwidth}{@{\extracolsep{\fill}}lccc@{}}
 \toprule
 Encoding & Sufficient width $r$ & Sufficient strength $\beta_R$ & \makecell{Fraction of enhanced checks\\$|\mathcal P_r|/N_{\rm chk}$}\\
 \midrule
 2D toric strip & $b_2\log d$ & $(\tfrac12+\delta)\log d$ & $O(\log d/d)$\\[4pt]
 3D toric full tube & $b_3\sqrt{\log d}$ & $(\tfrac12+\delta)\log d$ & $O(\log d/d^2)$\\[4pt]
 Good linear confinement & $O(\log\log U)$ & $(\tfrac12+\delta)\log U$ & $\le\min\!\left\{1,\dfrac{CU(\log U)^\nu}{N_{\rm chk}}\right\}$\\[2pt]
 \bottomrule
 \end{tabular*}
 \caption{Sufficient resources for vanishing AQEC relative entropy at fixed logical-register size and uniformly low local-stochastic $X$ noise. All rows use finite enhancement strength with $\delta>0$. The toric width coefficients $b_2,b_3$ are sufficiently large, and $d=L$ is the ideal code distance. The confinement row uses the actual support union $U$, bounded local degree, and the cutoff condition in Corollary~\ref{cor:resources}; $C,\nu$ are independent of code size. Its enhanced fraction vanishes if $U(\log U)^\nu=o(N_{\rm chk})$. The numerator and denominator count the same measured $Z$ checks. Widths are rounded up and are sufficient, with no claim of optimality.}
 \label{tab:resources}
\end{table*}

In toric codes, the protected plaquette ribbons relate the crossing parities of neighboring parallel representatives. An error silent on these ribbons must cross every such string with the same parity. A full strip in two dimensions contains a number of disjoint strings proportional to its width, whereas a full transverse tube in three dimensions contains a number proportional to its cross-sectional area. With the neighborhoods specified in \hyperref[sec:m-geometry]{Appendix}, this gives
\begin{equation}
 M(r)\ge r\quad(2D),\qquad M(r)\ge r^2\quad(3D).
 \label{eq:toricM}
\end{equation}
Thus, for a fixed set of straight representatives with $U=O(d)$, widths $r=O(\log d)$ and $r=O(\sqrt{\log d})$ suffice in two and three dimensions, respectively. The three-dimensional gain comes from protecting the full tube; a single sheet would not enforce the same number of crossings.

For general CSS codes, linear confinement provides a different mechanism for suppressing silent crossings. Let $S_X=\operatorname{span}_{\mathbb F_2}\{a_\mu\}$ be the support space of the $X$ stabilizers. Linear confinement in the $X$-error sector means that
\begin{equation}
 \begin{aligned}
 |x|_{\mathrm{red}}&=\min_{s\in S_X}|x\oplus s|,\\
 |\sigma(x)|&\ge\kappa|x|_{\mathrm{red}}
 \quad\text{for }|x|_{\mathrm{red}}<T.
 \end{aligned}
 \label{eq:confinement}
\end{equation}
Here $|\sigma(x)|$ counts violated measured $Z$ checks, and $|x|_{\mathrm{red}}$ is the smallest error weight after multiplication by $X$ stabilizers. The constant $\kappa>0$ is independent of code size, and $T$ is the cutoff below which this linear syndrome growth holds. A family has \emph{good linear confinement} if $T\ge c_Tn^\alpha$ for constants $c_T,\alpha>0$ independent of $n$~\cite{Quintavalle2021,Hong2025Preparation}. Only the $X$-error sector is required here; Supplementary Note~3 states the definition and the reduction convention explicitly.

The linear bound is the robustness property established for quantum expander codes~\cite{Leverrier2015,Fawzi2018}. Good linear confinement was subsequently shown to suffice for single-shot correction under local stochastic noise~\cite{Quintavalle2021}, extending earlier confinement arguments for subsystem codes~\cite{Bombin2015}. It also enters the final decoding step of single-shot preparation via a dimension jump~\cite{Hong2025Preparation} and has been used to construct passive quantum memories at nonzero temperature~\cite{Hong2025Memory}. Its classical analogue has been used to establish eigenstate localization in quantum-perturbed code Hamiltonians~\cite{Yin2024Localization}.

We enhance all checks touching the radius-$(r-1)$ neighborhood of the selected supports in $G_Z$. To remain silent, an error must cancel the syndrome produced in each layer by extending into the next. If a qubit meets at most $\Delta_Z$ measured checks, this requires
\begin{equation}
 M(r)\ge\min\!\left\{T,(1+\kappa/\Delta_Z)^r\right\}.
 \label{eq:confM}
\end{equation}
Our recovery bound needs only $T\ge c\log U$ with $c>1/|\log\lambda_0|$, a weaker requirement than good linear confinement because $U\le n$. A width $O(\log\log U)$ then suffices. Bounded degree limits the corresponding enhanced set to $CU(\log U)^\nu$ checks, or the total check count if smaller, with $C,\nu$ depending only on the local degree and confinement constants. This also makes logarithmic strength in $U$ sufficient at fixed $k$. Combining these neighborhood counts with the finite-strength correction gives the resources in Table~\ref{tab:resources}.

\begin{corollary}[Geometric resources for local enhancement]\label{cor:resources}
For a fixed nonempty logical register, uniform local constants, and $\lambda\le\lambda_0<1$, the three constructions in Table~\ref{tab:resources} give $\|B\|_1\to0$, $\AQEC\to0$, and $F_e^\star\to1$ at fixed finite bulk strength. The width coefficients are sufficiently large and $\delta>0$ is fixed. The confinement construction assumes $U\to\infty$ and $T\ge c\log U$, with $c>1/|\log\lambda_0|$.
\end{corollary}

Both toric constructions use $O(d\log d)$ enhanced checks. This is a vanishing fraction of the $\Theta(d^2)$ checks in two dimensions and the $\Theta(d^3)$ checks in three dimensions, as shown in Table~\ref{tab:resources}. Their sufficient strength, $\beta_R=(\tfrac12+\delta)\log d$, has the same logarithmic order as the necessary bounds from the preceding section. Those lower bounds apply to the common protected strength $\beta_R$ under independent flips at fixed $0<p<1$. Hence logarithmic strengthening is necessary and sufficient in order for these toric protocols; equality of the coefficients is not established.

For confinement, the smaller radius does not by itself imply a vanishing enhanced fraction. The enhanced fraction vanishes when $U(\log U)^\nu=o(N_{\rm chk})$, where $N_{\rm chk}$ counts all measured $Z$ checks. The resource estimate depends on the chosen representatives through their union $U$; it can be written in terms of distance $d$ when those representatives additionally satisfy $U=O(d)$. The neighborhood counts and strength estimates are derived in \hyperref[sec:m-geometry]{Appendix}.

These resource savings concern a fixed number of selected logical qubits. Protecting the entire register of a constant-rate code is more demanding: a narrow neighborhood of each representative can collectively occupy an extensive part of the system. In fact, an extensive enhancement is necessary even to make the AQEC entropy per logical qubit vanish.

\begin{corollary}[Enhancement cost for a growing logical register]\label{cor:register}\label{thm:register}
Suppose each qubit meets at most $\Delta_Z$ measured checks and all strengths outside an enhanced set are bounded by $\beta_0<\infty$, uniformly in size. Let $F_{\mathrm{hi}}$ be the set of physical qubits touched by the enhanced checks. For $k\ge1$ selected logical qubits and independent flips at fixed $0<p<1$,
\begin{equation}
 \frac{\AQEC}{k}\ge2p(1-p)e^{-2\Delta_Z\beta_0}
 \left(1-\frac{|F_{\mathrm{hi}}|}{k}\right)_+,
 \label{eq:footprint}
\end{equation}
where $(z)_+=\max\{z,0\}$. Consequently, $\AQEC/k\to0$ requires $|F_{\mathrm{hi}}|\ge[1-o(1)]k$. With uniformly bounded check weight and $k=\Theta(n)$, $\Omega(n)$ enhanced checks are necessary.
\end{corollary}

Local noise on the qubits outside $F_{\mathrm{hi}}$ still probes at least $(k-|F_{\mathrm{hi}}|)_+$ independent logical parities. Each retains a finite leakage contribution at bounded bulk strength, giving the lower bound above (\hyperref[sec:m-leakage]{Appendix}). A small protection radius can thus coexist with an extensive total enhancement when many representatives are selected. Theorem~\ref{thm:enhancement} remains applicable to a growing register using its full support union $U$. Its implications for total AQEC entropy, entropy density, and full-register entanglement recovery are stated separately in \hyperref[sec:m-register]{Appendix}.

\subsection{Wilson restoration and coherent logical operations}
\label{sec:wilson}
Local enhancement protects an encoded input against local noise in both two and three dimensions. Whether distant representatives act identically on that input remains a separate question. We therefore enhance two straight, parallel representatives $\ell_1,\ell_2$ of the same logical class in a single preparation, while keeping the intervening bulk at fixed finite strength $\beta$. Their separation is at least $\alpha L$, with fixed $0<\alpha<1/2$. The state $|\psi_\beta\rangle$ includes both enhanced regions, as does the full strength profile abbreviated by its subscript.

Each representative defines an encoding isometry $V_i:\mathcal H_L\to\mathcal H_Q$ for one logical qubit. For $i=1,2$, its action in the logical basis of Eq.~\eqref{eq:encoding} is
\begin{equation}
 \begin{aligned}
 &V_i\bigl(a|0\rangle_L+b|1\rangle_L\bigr)\\
 &\qquad=a|\psi_\beta\rangle+bZ^{\ell_i}|\psi_\beta\rangle.
 \end{aligned}
 \label{eq:two-encodings}
\end{equation}
Thus $V_i$ maps an arbitrary logical input into the entire $n$-qubit physical system $Q$; the index $i$ specifies the representative used for the image of $|1\rangle_L$. Both encodings share the image of $|0\rangle_L$. Within each encoding, the two codewords are orthogonal because they have opposite eigenvalues of a complementary physical logical $X$ operator. In the projective code, $V_1=V_2$. At finite strength, their agreement is measured by the Wilson overlap
\begin{equation}
 w=\langle W(C)\rangle_\beta
 =\langle\psi_\beta|Z^{\ell_1}Z^{\ell_2}|\psi_\beta\rangle,
 \label{eq:wilson-overlap}
\end{equation}
where $C=\ell_1\oplus\ell_2$ and $0\leq w\leq1$. In particular, $w\to1$ makes the two physical encodings agree for every logical input (\hyperref[sec:m-operations]{Appendix}).

\begin{figure*}[!t]
 \centering\includegraphics[width=\textwidth]{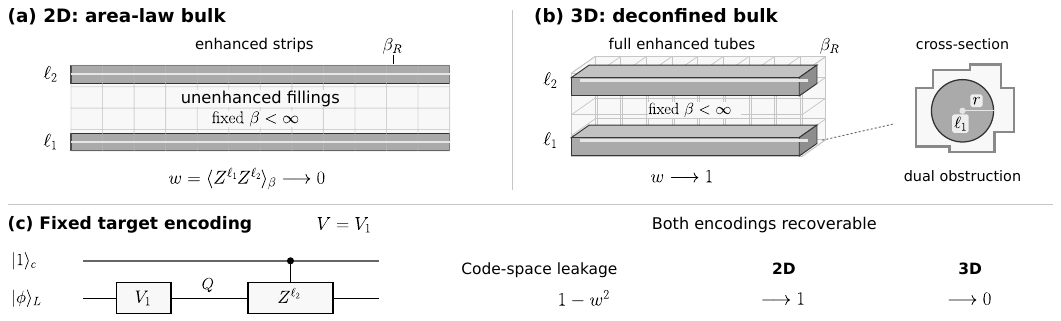}
 \caption[Wilson restoration and representative substitution]{\textbf{Wilson restoration and representative substitution.} Two distant representatives are enhanced in the same preparation, with the bulk held at fixed finite strength. \textbf{(a)} In 2D, macroscopic unenhanced areas in both complementary fillings retain the suppression of the Wilson overlap, giving $w\to0$. \textbf{(b)} In the strict 3D deconfined phase, a dual obstruction avoiding a full protected tube must extend over a distance of order its radius $r$. Subcritical dual-cluster tails and sufficiently strong checks give $w\to1$ for the logarithmic radius in Eq.~\eqref{eq:wilsonparameters}. The transverse inset shows a dual obstruction around one tube. \textbf{(c)} The circuit applies the distant representative to an arbitrary logical input $|\phi\rangle_L$ encoded by $V=V_1$. For control $|1\rangle_c$, replacing $\ell_1$ by $\ell_2$ produces code-space leakage $1-w^2$. The resulting limits differ despite the separate recoverability of both encodings under the stated low-noise conditions.}
 \label{fig:operations}
\end{figure*}

We now ask whether measurements confined near the two representatives can make $w$ approach one while the intervening bulk remains at finite strength. Recovery of the separate encodings is possible in either dimension; the bulk determines whether the same preparation also restores their agreement.

\begin{theorem}[Wilson restoration and dimensional separation]\label{thm:wilson}
In two dimensions, suppose the enhanced strips have width $r=o(L)$ and both complementary fillings contain at least $cL^2$ unenhanced plaquettes. At fixed $0<\beta<\infty$, even projective enhancement satisfies
\begin{equation}
 0\le w\le2\exp[-c|\log\tanh\beta|L^2]\longrightarrow0.
 \label{eq:2dno}
\end{equation}
In three dimensions, fix the bulk in the strict deconfined phase of the uniform model~\cite[Theorem~5]{DuncanSharp2025}, equivalently
\begin{equation}
 p^*(\beta)=\frac{2}{e^{2\beta}+1}<p_c^{\mathrm{FK}}(2;\mathbb Z^3),
 \label{eq:deconfined}
\end{equation}
where $p_c^{\mathrm{FK}}$ is the critical bond parameter of the Fortuin--Kasteleyn (FK) random-cluster model with cluster weight $q=2$~\cite{FortuinKasteleyn1972,Grimmett2006}. Full enhanced tubes around both representatives give $w\to1$ for
\begin{equation}
 \begin{aligned}
 r&=b\frac{\xi_\beta}{c_g}\log L,\qquad b>1,\\
 \beta_R&=(\tfrac12+\delta)\log L,\qquad\delta>0.
 \end{aligned}
 \label{eq:wilsonparameters}
\end{equation}
Here $\xi_\beta<\infty$ is a decay length for subcritical dual clusters and $c_g>0$ depends on the tube geometry. Both are independent of $L$ at fixed bulk strength.
\end{theorem}

The two-dimensional obstruction comes from the unenhanced filling area [Fig.~\ref{fig:operations}(a)]. Each unenhanced plaquette contributes a factor $\tanh\beta<1$, and a macroscopic number remains in each complementary filling. Sharpening checks near the boundary cannot remove this accumulated suppression.

In three dimensions, we adapt the random-cluster argument behind the perimeter law~\cite{DuncanSharp2025} to the locally enhanced preparation. The indicator coupling used in Refs.~\cite{DuncanSharp2025,DuncanTopological2025} equals $2\beta_p$ in the action convention of Eq.~\eqref{eq:general-wilson-action}. The plaquette random-cluster representation assigns the primal occupation parameter $\theta_p=1-e^{-2\beta_p}$ to a check of strength $\beta_p$. A dual edge is open when its primal plaquette is absent, and its FK bond parameter is
\begin{equation}
 p_e^*=\frac{2(1-\theta_p)}{\theta_p+2(1-\theta_p)}
      =\frac{2}{e^{2\beta_p}+1}.
 \label{eq:fk-parameters}
\end{equation}
These are random-cluster parameters, distinct from the physical noise probability $p$. Stronger checks therefore reduce the dual bond parameters. A dual obstruction avoiding a protected tube must extend over a distance of order $r$, or wind around the torus [Fig.~\ref{fig:operations}(b)]. At a fixed bulk point satisfying Eq.~\eqref{eq:deconfined}, the subcritical FK theorem of Ref.~\cite{Duminil2019} supplies exponential cluster tails. Combining these tails with the tube geometry and the finite-torus duality gives Theorem~\ref{thm:wilson}; the finite-strength and topological corrections are derived in \hyperref[sec:m-wilson]{Appendix} and Supplementary Note~4. Local enhancement thus restores $w\to1$, beyond the decay reduction implied by a bulk perimeter law.

The same preparation also restores AQEC for each encoding when the full tubes contain the ribbons used in Sec.~\ref{sec:enhancement} and the noise obeys $\lambda\leq\lambda_0<1$. With $d=L$, the enhanced checks number $O(L\log^2 L)$, a fraction $O(\log^2d/d^2)$ of all measured checks. The radius $O(\log d)$ here exceeds the sufficient $O(\sqrt{\log d})$ radius for AQEC alone in Table~\ref{tab:resources}; these are two sufficient constructions, and do not establish a necessary extra cost for Wilson restoration. In two dimensions, the encodings can likewise be separately recoverable while their Wilson overlap vanishes.

This distinction has a direct consequence for logical operations on an already encoded state. Fix the target encoding $V=V_1$. With the basis convention of Eq.~\eqref{eq:encoding}, the physical string $Z^{\ell_1}$ flips the logical labels, so $Z^{\ell_1}V=VX_L$, where $X_L=|0\rangle_L\langle1|_L+|1\rangle_L\langle0|_L$. For an ideal control qubit $c$, define
\begin{equation}
 \begin{aligned}
 G_j={}&|0\rangle_c\langle0|_c\otimes I_Q\\
 &+|1\rangle_c\langle1|_c\otimes Z^{\ell_j},\qquad j=1,2.
 \end{aligned}
 \label{eq:controlled-strings}
\end{equation}
Here $I_Q$ is the identity on the physical target. Changing the representative replaces $G_1$ by $G_2$ while leaving the encoded input fixed. The relevant error is therefore the difference between these two actions on the same code space. The following identity relates that error directly to the Wilson overlap.

\Needspace{6\baselineskip}
\begin{corollary}[Replacing a representative in an encoded CNOT]\label{cor:cnot}
With $V=V_1$ and $I_c$ the identity on the control,
\begin{align}
 G_1(I_c\otimes V)&=(I_c\otimes V)\operatorname{CNOT}_{c\to L},\nonumber\\
 \left\|(G_2-G_1)(I_c\otimes V)\right\|_\infty
 &=\sqrt{2(1-w)},
 \label{eq:cnot-error}
\end{align}
where $\operatorname{CNOT}_{c\to L}$ acts on the control and logical input, and $\|\cdot\|_\infty$ denotes the operator norm. For control $|1\rangle_c$, the second representative takes the target outside the fixed code space with probability $1-w^2$, for every logical target input.
\end{corollary}

In the enhanced three-dimensional preparation, either controlled string asymptotically implements the same logical CNOT on the fixed target encoding. In the two-dimensional setting of Theorem~\ref{thm:wilson}, the same substitution causes leakage approaching one for control $|1\rangle_c$, even when the target encoding is recoverable from the stated stochastic noise. The comparison concerns ideal controlled strings on a fixed encoding; implementing them with noisy local gates requires a separate circuit analysis. The overlap also determines the optimal common recovery when the representative used for encoding is undisclosed, as derived in \hyperref[sec:m-operations]{Appendix}.

\section{Discussion}\label{sec:discussion}
Finite-strength preparation is often judged by how much topological or long-range structure survives away from the projective limit. Our results show that this is not enough to determine whether the prepared state functions as a quantum code. What matters for recovery is how much information about the logical input remains accessible to local errors and, ultimately, to the noise environment. This changes the preparation problem from one of making the entire state uniformly close to an ideal code state to one of controlling the parts of the encoding map that expose logical information. For a fixed logical register, that control can be highly nonuniform: strengthening measurements only near selected logical supports is sufficient for recovery while the rest of the system stays at finite strength. In this sense, the relevant preparation resource is not simply global measurement fidelity, but where accuracy is placed relative to the logical information being protected.

The logarithmic strength scale gives this point a more general meaning. A local diagnostic can become small while a growing logical support still contains enough weakly distinguishable locations to reveal a finite amount of information collectively. Local indistinguishability at any one site is therefore not, by itself, a reliable proxy for recoverability in a growing code. The same mechanism survives beyond toric geometry whenever bounded check incidence allows local errors to probe selected logical parities. This suggests a useful way to think about imperfect code preparation: scalability is controlled not only by how strongly an individual check is measured, but also by how many locations carry the same logical information. The balance between these two quantities is what produces the logarithmic preparation scale in the toric protocols.

A second lesson is that finite-strength preparation separates two properties that coincide in an ideal stabilizer code. Recovering an unknown input does not automatically restore the equivalence of different physical representatives of a logical operator. Once the local leakage has been suppressed, the bulk can still matter for whether two representatives act coherently as the same logical operation on a fixed encoding. The Wilson overlap makes this statement operational: it is not merely a diagnostic of the prepared many-body state, but directly quantifies the error made when one physical representative is substituted for another. The contrast between two and three dimensions is therefore not simply another manifestation of area-law versus perimeter-law behavior. It shows that the bulk phase can become relevant to logical operation only after the more local requirement of recoverability has been satisfied. In the three-dimensional deconfined regime, suitable local enhancement can restore both properties; in two dimensions, recoverability can be restored while representative equivalence remains lost in the setting considered here.

These distinctions open several directions in which the present framework could become more general and more directly connected to fault-tolerant architectures. The first is to move beyond the fixed all-plus branch and ask how outcome-dependent preparation records should be incorporated into recovery and logical-operator equivalence. Doing so would turn the present conditioned analysis into a theory of imperfect measurement preparation with explicit record disorder and would clarify which information must be retained by a decoder. A second direction is to identify the analogue of Wilson restoration in general CSS and qLDPC codes, where two logical representatives are related by products of checks but no Euclidean filling surface is available. There, confinement, expansion, or other code-geometric properties may determine whether coherent representative equivalence can be restored together with recovery. Finally, incorporating efficient decoding and noisy implementations of logical operations would connect these static encoding criteria to circuit-level fault tolerance. That step is important because it would determine whether the separation found here between information recovery and logical-operator equivalence remains visible once preparation, decoding, and logical control are treated as one complete protocol.

\section{Appendix}\label{sec:methods}

\subsection{Finite-strength preparation and error overlaps}
\label{sec:m-encoding}
The overlap of a codeword with a physical error depends on the checks reversed by that error. This dependence can be evaluated without assuming that the measured checks are independent. The ancilla circuit in Fig.~\ref{fig:background}(a) gives the normalized Kraus operator $M_{j,+}=e^{\beta_jB_j/2}/\sqrt{2\cosh\beta_j}$, with $\tanh(\beta_j/2)=\tan t_j$, up to a global phase in the selected branch~\cite{Brun2002,Zhu2023}. Supplementary Note~1 derives the ancilla contraction and its normalization.

Write $F_\beta=\exp[\frac12\sum_j\beta_jB_j]$ and let $V_0|q\rangle_L=Z^{\ell(q)}|+\rangle^{\otimes n}$ be the zero-strength encoding. Complementary logical $X$ operators commute with $F_\beta$ and have eigenvalues $(-1)^{q_a}$ on its logical columns, making distinct columns orthogonal. Their norms are equal because $Z^{\ell(q)}$ is unitary and commutes with the filter. Consequently,
\begin{equation}
 V_0^\dagger F_\beta^2V_0=Z_\beta I_K,\qquad
 V_\beta=Z_\beta^{-1/2}F_\beta V_0.
 \label{meth:isometry}
\end{equation}
The all-plus success probability $Z_\beta/\prod_j(2\cosh\beta_j)$ is independent of the logical input, including inputs entangled with a reference. Normalizing the selected branch therefore yields a linear encoding isometry.

Every physical $X_x$ commutes with the complementary logical $X$ operators, so its matrix elements between distinct logical labels vanish. Commuting it through $Z^{\ell(q)}$ gives
\begin{equation}
 \begin{aligned}
 \langle q_\beta|X_x|q'_\beta\rangle
 &=\delta_{qq'}\chi_q(x)\kappa_\beta(x),\\
 \kappa_\beta(x)&=\langle\psi_\beta|X_x|\psi_\beta\rangle.
 \end{aligned}
 \label{meth:kernel}
\end{equation}
The logical dependence is entirely in the sign $\chi_q(x)$; the magnitude follows from the preparation filter. Moving $X_x$ between its two factors cancels precisely the checks in $\sigma(x)$. With $|+\rangle$ abbreviating $|+\rangle^{\otimes n}$,
\begin{equation}
 \kappa_\beta(x)=
 \frac{\langle+|e^{\sum_{j\notin\sigma(x)}\beta_jB_j}|+\rangle}
 {\langle+|e^{\sum_j\beta_jB_j}|+\rangle}.
 \label{meth:kernel-ratio}
\end{equation}
To bound this ratio, we use the high-temperature expansion of generalized Ising models~\cite{Wegner1971,Wegner2014}, $e^{\beta_jB_j}=\cosh\beta_j(1+\tanh\beta_jB_j)$. Only products whose binary check supports sum to zero survive the expectation. After extracting the hyperbolic-cosine factors, the numerator retains a subset of the denominator's nonnegative terms. This gives the upper bound below, including all check redundancies. For the lower bound, simultaneous diagonalization bounds the denominator of Eq.~\eqref{meth:kernel-ratio} by its numerator times $e^{\sum_{j\in\sigma(x)}\beta_j}$:
\begin{equation}
 e^{-\sum_{j\in\sigma(x)}\beta_j}\leq\kappa_\beta(x)
 \leq\prod_{j\in\sigma(x)}\sech\beta_j.
 \label{meth:local-kernel-bound}
\end{equation}
The lower bound leaves a nonzero overlap whenever an error reverses finitely many checks of bounded strength. The upper bound quantifies its suppression when those checks are strengthened. On the square torus, the only check redundancies are the empty and full plaquette sets. Deleting the two checks incident on one edge gives the exact expression in Eq.~\eqref{eq:2dkernel}. On the cubic torus, four checks meet each edge, giving the stated three-dimensional bounds. Supplementary Note~2 gives the exact redundancy sum and the full overlap calculation.

\subsection{Complementary channel and recovery}
\label{sec:m-recovery}
Recovery can be characterized by the logical information transferred to the noise environment~\cite{Schumacher2002,Kretschmann2008,Beny2010}. The same overlaps determine its complementary output. For any error distribution $P(x)$ in Eq.~\eqref{eq:noise}, a Stinespring dilation~\cite[Sec.~2.2.2]{Watrous2018} acts as $|\phi\rangle_Q\mapsto\sum_x\sqrt{P(x)}X_x|\phi\rangle_Q|x\rangle_E$. Encode one half of $|\Phi_K\rangle=K^{-1/2}\sum_q|q\rangle_R|q\rangle_L$ and trace out the physical output. Equation~\eqref{meth:kernel} gives
\begin{align}
 \rho_{RE}&=\frac1K\sum_q|q\rangle_R\langle q|_R\otimes\rho_E^q,\nonumber\\
 (\rho_E^q)_{xy}&=\sqrt{P(x)P(y)}\,\chi_q(x\oplus y)\kappa_\beta(x\oplus y).
 \label{meth:environment}
\end{align}
The diagonal $P(x)$ is input-independent; logical information resides in the error-label coherences. Although this state is classical on $q$, different conditional environment states suppress coherence between logical sectors. The recovery task still concerns a quantum input. The environment here belongs to the subsequent noise and is distinct from the fixed preparation record.

With rows $(q,x)$ and columns $(q',y)$, the normalized AQEC matrices of Ref.~\cite{Zhao2024}, defined in Eq.~\eqref{eq:matrixelements}, equal $\rho_{RE}^T$ and $(I_R/K\otimes\rho_E)^T$, respectively. Full transposition preserves relative entropy and trace norm, hence
\begin{equation}
 \begin{aligned}
 \AQEC&=I(R:E),\\
 \eps&=\tfrac12\|B\|_1
 =\tfrac12\left\|\rho_{RE}-\frac{I_R}{K}\otimes\rho_E\right\|_1.
 \end{aligned}
 \label{meth:information}
\end{equation}
We use $D(\rho\Vert\sigma)=\Tr[\rho(\log\rho-\log\sigma)]$ and natural logarithms. Recovery is possible as these correlations vanish. Writing $\cV_\beta(\rho)=V_\beta\rho V_\beta^\dagger$, the optimal squared entanglement fidelity is~\cite{Schumacher1996}
\begin{equation}
 F_e^\star=\max_{\cR}\langle\Phi_K|
 (\mathrm{id}_R\otimes\cR\cN\cV_\beta)
 (|\Phi_K\rangle\langle\Phi_K|)|\Phi_K\rangle.
 \label{eq:fe}
\end{equation}
This criterion uses the maximally mixed input on the entire selected register. The recovery acts on all physical qubits and returns that register; no efficiency or worst-case channel guarantee is assumed.

Uhlmann's theorem relates recovery to decoupling~\cite{Uhlmann1976,Schumacher2002,Beny2010}:
\begin{equation}
 \begin{aligned}
 F_e^\star&=\max_{\xi_E}F\!\left(\rho_{RE},\frac{I_R}{K}\otimes\xi_E\right),\\
 F(\rho,\sigma)&=\|\sqrt\rho\sqrt\sigma\|_1^2,
 \end{aligned}
 \label{meth:recovery-duality}
\end{equation}
where $\xi_E$ is an environment density matrix. Choosing $\xi_E=\rho_E$ and using the fidelity--trace-distance inequalities~\cite{Fuchs1999} gives $F_e^\star\ge(1-\eps)^2$. Conversely, recovery with infidelity $u$ implies $\eps\le2\sqrt u$. Together with quantum Pinsker's inequality~\cite[Theorem~5.38]{Watrous2018}, these give
\begin{equation}
 \frac{\eps^2}{4}\le1-F_e^\star\le2\eps,\qquad \AQEC\ge2\eps^2.
 \label{eq:recoverybounds}
\end{equation}
Supplementary Note~2 gives the full purification argument and the trace-distance estimates.

The conversion to total AQEC entropy also depends on the size of the logical register. Equation~\eqref{meth:environment} makes $\AQEC$ a Holevo information, bounded by $\log K$~\cite{Holevo1973}. Since the reference is classical in the same basis for $\rho_{RE}$ and its product comparator, the classical--quantum conditional-entropy continuity bound gives~\cite[Lemma~2]{Winter2016}
\begin{equation}
 \begin{aligned}
 \AQEC&\le\eps\log K+(1+\eps)h_2\!\left(\frac{\eps}{1+\eps}\right),\\
 h_2(t)&=-t\log t-(1-t)\log(1-t).
 \end{aligned}
 \label{eq:continuity}
\end{equation}
At fixed $k$, vanishing $\AQEC$, vanishing $\eps$, and $F_e^\star\to1$ are equivalent. If $k\ge1$ grows, $\eps\to0$ still guarantees recovery of the entire register and $\AQEC/k\to0$, whereas $k\eps\to0$ is sufficient for the total AQEC entropy to vanish. These information-theoretic conversions hold for arbitrary $P(x)$; independent flips and local stochasticity enter only the respective leakage and enhancement bounds.

\subsection{Local and collective leakage}
\label{sec:m-leakage}
The local obstruction in Theorem~\ref{thm:local} follows directly from the error-overlap matrix. Choose a physical site $e$ with $g(e)\ne0$. Matrix entries whose error labels differ by $e$ carry the logical sign $\chi_q(e)$ and the common overlap $\kappa_\beta(e)$. A signed permutation pairs $(q,x)$ with $(q,x\oplus e)$ and cancels this sign. Its operator norm is one, so trace-norm duality gives~\cite[Sec.~1.1.3]{Watrous2018}
\begin{equation}
 \begin{aligned}
 \|B\|_1&\ge\kappa_\beta(e)\sum_x\sqrt{P(x)P(x\oplus e)}\\
 &=2\sqrt{p(1-p)}\,\kappa_\beta(e)
 \end{aligned}
 \label{meth:witness}
\end{equation}
for independent flips. The sum includes every exterior error configuration; their probabilities sum to one. Thus the bound does not require the other qubits to remain error free. The same quantum Pinsker bound then yields
\begin{equation}
 \AQEC\ge\tfrac12\|B\|_1^2
 \ge2p(1-p)\kappa_\beta(e)^2.
 \label{meth:local-entropy}
\end{equation}
If each site meets at most $\Delta_Z$ measured checks and their strengths are bounded by $\beta_0$, Eq.~\eqref{meth:local-kernel-bound} gives $\AQEC\ge2p(1-p)e^{-2\Delta_Z\beta_0}$. At fixed $0<p<1$, a single site therefore leaves a positive lower bound independent of code size. The corresponding norm bound and Eq.~\eqref{eq:recoverybounds} also give the recovery obstruction in Corollary~\ref{cor:obstruction}. Its entropy-density statement follows by choosing sites that probe independent logical parities and applying the information chain rule, as proved in Supplementary Note~2.

Subsystem variance~\cite{Yi2024} identifies the local information responsible for this obstruction. The expectation of $X_e$ in a logical basis state is $\chi_q(e)\kappa_\beta(e)$, whereas its expectation in the maximally mixed code state is zero. Since $\|X_e\|_\infty=1$, the definition in Eq.~\eqref{eq:subsystem-variance} gives $\varepsilon_{\mathrm{SV}}(\{e\})\ge\kappa_\beta(e)$. In the toric codes, exact star constraints make the $Y_e$ and $Z_e$ expectations vanish for every encoded input. The entire single-edge marginal is therefore
\begin{equation}
 \rho_e=\frac12\left[I+\kappa_\beta(e)
             \sum_q(\rho_L)_{qq}\chi_q(e)X_e\right].
 \label{meth:toric-marginal}
\end{equation}
The code-average marginal is $I/2$. The coefficient of $X_e$ ranges between $-\kappa_\beta(e)$ and $\kappa_\beta(e)$, with both endpoints attained by logical basis states, so the full trace norm gives $\varepsilon_{\mathrm{SV}}(\{e\})=\kappa_\beta(e)$. This maximization includes arbitrary logical superpositions. The same residual overlap thus quantifies local distinguishability before noise and bounds information lost to the noise environment. For a general CSS encoding the variance need not equal the overlap, so the noise bound remains stated in terms of $\kappa_\beta(e)$.

The replica construction for imperfectly prepared toric codes~\cite[Sec.~V and Appendix~C]{Zhao2024} extends to the selected CSS register and nonuniform strengths. The resulting traces show how local contributions enter along the selected representatives. For integer $m\ge2$, multiplying the AQEC matrix around a replica cycle, with $x^{(m+1)}=x^{(1)}$, gives
\begin{equation}
 \begin{aligned}
 &K^{m-1}\Tr(\Lambda+B)^m\\
 &\quad=\sum_{x^{(1)},\ldots,x^{(m)}}
 \prod_{\alpha=1}^m
 P(x^{(\alpha)})\kappa_\beta(x^{(\alpha)}\oplus x^{(\alpha+1)}).
 \end{aligned}
 \label{meth:replica-sum}
\end{equation}
Each error probability appears in two neighboring square roots, and the logical signs cancel around the closed cycle. The expression for $K^{m-1}\Tr\Lambda^m$ has the same weights, with the additional constraint $g(x^{(\alpha)}\oplus x^{(\alpha+1)})=0$ at every step. Inserting the filter ratio in Eq.~\eqref{meth:kernel-ratio} gives the classical replica partition functions derived in Supplementary Note~2. The constraint excludes precisely the error differences that distinguish selected logical inputs.

Consider the first local-error configurations in this sum: one replica contains a single error $e$, and all other replicas contain no error. The two adjacent differences are both $e$, giving a factor $\kappa_\beta(e)^2$. This contribution occurs in the full trace but is excluded from the comparator whenever $g(e)\ne0$. Such a configuration can be placed in any replica and at any site on the union of selected representatives. A straight toric representative supplies $d$ such positions. The bounds on the local overlap involve only the two incident checks in 2D or four in 3D; at fixed finite strength, the overlap remains bounded below independently of the representative length. This counting is the origin of the accumulated leakage in the replica expansion.

At finite size, the AQEC entropy is recovered from these traces through
\begin{equation}
 \AQEC=\left.\partial_m\log
 \frac{\Tr(\Lambda+B)^m}{\Tr\Lambda^m}\right|_{m=1}.
 \label{meth:replica-derivative}
\end{equation}
The matrix powers near one are defined spectrally. The derivative equals the relative entropy because $\Tr(B\log\Lambda)=0$, a consequence of $\Tr_RB=0$. Supplementary Note~2 evaluates the local replica coefficient while retaining the terms needed in this limit. A truncation at fixed integer $m$ cannot be continued to $m=1$ by discarding those terms, nor does a finite-size dilute expansion by itself give a bound uniform in $d$.

The fixed-noise statement in Theorem~\ref{thm:local} follows from the collective estimate in the same Supplementary Note. For the $M$ sites specified there, the summed local signal is controlled by $\sum_i\kappa_\beta(e_i)$, while its second moment involves $\sum_{i\ne j}\kappa_\beta(e_i\oplus e_j)$. Disjoint syndrome supports bound the latter through Eq.~\eqref{meth:local-kernel-bound}, without assuming independent environment states. Data processing and Pinsker's inequality~\cite[Ch.~5]{Watrous2018}, together with a Cauchy--Schwarz estimate, give Eq.~\eqref{eq:collective}, whose denominator accounts for repeated signals about the same logical parity. With $M=d$, this proves Eq.~\eqref{eq:lognecessary} at fixed nonzero $p$. The same argument uses only the checks incident on those sites, so it also applies to their actual common enhanced strength $\beta_R$ when the bulk remains fixed.

\subsection{Connected-component recovery bound}
\label{sec:m-projective}
The finite-strength bound in Theorem~\ref{thm:enhancement} separates leakage from silent crossings from the error incurred by replacing the prepared state with its projection onto the enhanced-check sector. Write $|\psi_{\mathrm{soft}}\rangle=|\psi_{\boldsymbol\beta}\rangle$ for the finite-strength preparation. To bound the first contribution, consider
\begin{equation}
 \Pi_r=\prod_{j\in\mathcal P_r}\frac{I+B_j}{2},\qquad
 |\psi_r\rangle=\frac{\Pi_r|\psi_{\mathrm{soft}}\rangle}
 {\sqrt{\langle\psi_{\mathrm{soft}}|\Pi_r|\psi_{\mathrm{soft}}\rangle}}.
 \label{eq:hardstate}
\end{equation}
This comparison state has the same exterior strengths and selected representatives as the finite-strength preparation. The projection is used only in the proof. Writing $\kappa_r(x)=\langle\psi_r|X_x|\psi_r\rangle$, the identity $\Pi_rX_x\Pi_r=0$ makes this overlap vanish whenever $\sigma(x)\cap\mathcal P_r\ne\varnothing$.

Decompose the error difference $x=u\oplus v$ into connected components in $G_Z$. Distinct components touch disjoint sets of measured checks, so their syndromes cannot cancel. A nonzero contribution that distinguishes logical labels $q,q'$ must therefore contain silent components with different characters $\chi_q,\chi_{q'}$. Denote their class by $\mathcal C_r(q,q')$. Character multiplication gives an exact expansion of the conditional environment-state difference into families of these components:
\begin{equation}
 D^{q,q'}:=\rho_E^q-\rho_E^{q'}
 =\sum_{\substack{\varnothing\ne\mathcal F\subseteq\mathcal C_r(q,q')\\
                    \mathcal F\ \mathrm{compatible}}}D_{\mathcal F}^{q,q'}.
 \label{meth:family-expansion}
\end{equation}
Compatible families are mutually disconnected supports that can occur as complete components of one error difference. Each distinguishing component has character ratio $-1=1-2$, so expanding the product gives the signed coefficient $(-2)^{|\mathcal F|}$ in $D_{\mathcal F}^{q,q'}$. Supplementary Note~3 gives the full matrix elements.

For a fixed family with union $S$, the two error labels must be complementary on $S$ and equal on its external vertex boundary. After fixing these assignments, the remaining matrix is, up to signs and the family coefficient, a cross-Gram matrix of the vectors $\sqrt{P(u)}X_u|\psi_r\rangle$. The Schatten inequality $\|A^\dagger C\|_1\le\|A\|_2\|C\|_2$~\cite[Sec.~1.1.3]{Watrous2018} bounds its trace norm by the square root of the corresponding row and column probabilities. These probabilities already sum over exterior errors. Summing the common boundary assignments by Cauchy--Schwarz, then applying local stochasticity to the complementary occupations of $S$, gives a factor $\eta^{|S|/2}$. The $2^{|S|}$ assignments and $2^{|\mathcal F|}\le2^{|S|}$ family coefficient consequently yield
\begin{equation}
 \|D_{\mathcal F}^{q,q'}\|_1
 \le(4\sqrt\eta)^{|S|}
 =\prod_{\Gamma\in\mathcal F}(4\sqrt\eta)^{|\Gamma|}.
 \label{meth:family-norm}
\end{equation}
This estimate retains arbitrary noise correlations allowed by Eq.~\eqref{eq:localstochastic}; it is a bound on the full operator, rather than on individual matrix elements.

Compatibility can now be dropped when summing these nonnegative norm bounds. Every distinguishing component intersects the selected support union, with $U$ possible anchors, and has size at least $M(r)$. The exponential bound on the number of connected supports in a graph of bounded degree~\cite{Fawzi2018} therefore gives
\begin{equation}
 \sum_{\Gamma\in\mathcal C_r(q,q')}(4\sqrt\eta)^{|\Gamma|}
 \le\frac{CU\lambda^{M(r)}}{1-\lambda},
 \qquad \lambda=4\mu\sqrt\eta<1.
 \label{meth:component-tail}
\end{equation}
The family sum is bounded by the exponential of this single-component sum minus one. Averaging over logical labels then gives
\begin{equation}
 \begin{aligned}
 \|B_{\mathrm{hard}}\|_1
 &\le K^{-2}\sum_{q,q'}\|D^{q,q'}\|_1\\
 &\le\exp\!\left[\frac{CU\lambda^{M(r)}}{1-\lambda}\right]-1.
 \end{aligned}
 \label{eq:hardupper}
\end{equation}
The average introduces no factor growing with the logical dimension $K$. This proves the projective part of Theorem~\ref{thm:enhancement}.

\subsection{Finite-strength enhancement and resources}
\label{sec:m-finite}
For the unsigned CSS checks and nonnegative preparation strengths considered here, the probability of violating a measured check is at most $e^{-2\beta_j}$. To obtain this inequality, remove that check's coupling: the expansion in products of checks has nonnegative weights, hence a nonnegative expectation of $B_j$. Restoring its coupling reweights the two check eigenvalues by $e^{\pm\beta_j}$. A union bound therefore gives
\begin{equation}
 \begin{aligned}
 \Pr(B_j=-1)&\le e^{-2\beta_j},\\
 q_r:=1-\langle\Pi_r\rangle_{\mathrm{soft}}
 &\le|\mathcal P_r|e^{-2\beta_R}.
 \end{aligned}
 \label{meth:check-violation}
\end{equation}
No independence of check violations is required. Within the protected sector, each enhanced filter acts as a scalar, so the normalized projection is precisely the hard state with the exterior strengths unchanged.

The preparation states have overlap $\sqrt{1-q_r}$. Their purifications with the same maximally entangled logical input have this overlap as well, because the selected representatives are unchanged and commute with $\Pi_r$. The pure-state trace-distance formula~\cite{Fuchs1999} gives distance $\sqrt{q_r}$ between these purifications. A common noise dilation and partial trace cannot increase this distance~\cite[Sec.~3.3]{Watrous2018}, so the reference--environment states and their environment marginals obey the same bound. Thus
\begin{equation}
 \eps_{\mathrm{soft}}\le\eps_{\mathrm{hard}}+2\sqrt{q_r},
 \label{meth:soft-distance}
\end{equation}
and $\|B\|_1=2\eps$ gives
\begin{equation}
 \|B_{\mathrm{soft}}\|_1\le\|B_{\mathrm{hard}}\|_1
 +4\sqrt{|\mathcal P_r|}\,e^{-\beta_R}.
 \label{eq:softhard}
\end{equation}
Together with Eq.~\eqref{eq:hardupper}, this proves the finite-strength bound in Theorem~\ref{thm:enhancement}. The correction vanishes when $\sqrt{|\mathcal P_r|}e^{-\beta_R}\to0$, including the case of an empty protected set.

A complementary proof keeps finite measurement strength throughout the component expansion. For a fixed family, extracting the overlap factors associated with its syndrome leaves a positive Gram kernel, as shown in Supplementary Note~3. The cross-Gram argument then gives
\begin{equation}
 \|D_{\beta,\mathcal F}^{q,q'}\|_1\le
 \prod_{\Gamma\in\mathcal F}
 (4\sqrt\eta)^{|\Gamma|}(\sech\beta_R)^{|\sigma(\Gamma)\cap\mathcal P_r|}.
 \label{meth:soft-family}
\end{equation}
Here $D_{\beta,\mathcal F}^{q,q'}$ is the same family contribution evaluated in the finite-strength state. Every component smaller than $M(r)$ has a protected syndrome; the remaining components are bounded by size. Summing gives
\begin{equation}
 \|B_{\mathrm{soft}}\|_1\le\exp\!\left[\frac{CU}{1-\lambda}
 \bigl(\lambda\sech\beta_R+\lambda^{M(r)}\bigr)\right]-1.
 \label{eq:upper}
\end{equation}
The two terms describe suppression by measurement strength and by crossing size. Its simple sufficient strength condition is $U\sech\beta_R\to0$. The coefficients in Table~\ref{tab:resources} follow from the sharper soft--hard comparison, Eq.~\eqref{eq:softhard}.

\paragraph{Protection geometry.}
\label{sec:m-geometry}
For toric codes, take an integer $r\ge1$, set $r_0=\lfloor r/2\rfloor$, and assume $2r_0+1\le L$. Include parallel translates of each selected straight string with transverse coordinates between $-r_0$ and $r_0$, and protect the plaquette ribbons joining neighboring translates. Silence on a ribbon $A_y$ implies
\begin{equation}
 |\Gamma\cap\ell_y|+|\Gamma\cap\ell_{y+1}|
 =|\sigma(\Gamma)\cap A_y|=0\pmod2.
 \label{meth:ribbon-parity}
\end{equation}
Odd crossing of the selected string propagates to every translate. The $2r_0+1$ strings in a two-dimensional strip are disjoint, giving $|\Gamma|\ge r$. A full three-dimensional tube contains $(2r_0+1)^2$ such strings, giving $|\Gamma|\ge r^2$. This proves Eq.~\eqref{eq:toricM}.

For a family with linear confinement in the sense of Eq.~\eqref{eq:confinement}~\cite{Quintavalle2021,Hong2025Preparation}, protect every check touching the radius-$(r-1)$ neighborhood of the selected union. Consider an error with no protected syndrome and nonzero selected logical parity. If its weight is at least $T$, it already satisfies Eq.~\eqref{eq:confM}. For a lighter error, first minimize its full support modulo $X$ stabilizers, preserving syndrome and logical parity, and then choose a distinguishing component $\Gamma$. Every subset of a minimum-weight support is itself reduced, since a stabilizer reducing that subset would also reduce the full support. Let $\Gamma_t$ be its intersection with the radius-$t$ neighborhood of the selected union. These supports have weight below $T$, so Eq.~\eqref{eq:confinement} gives $|\sigma(\Gamma_t)|\ge\kappa|\Gamma_t|$. For $0\le t<r$, each such check is protected. Silence of the full component forces it to meet the next layer, where each added qubit can cancel at most $\Delta_Z$ checks. Hence
\begin{equation}
 |\Gamma_{t+1}|\ge(1+\kappa/\Delta_Z)|\Gamma_t|.
 \label{meth:layer-growth}
\end{equation}
Nonzero logical parity gives $|\Gamma_0|\ge1$, and iteration proves Eq.~\eqref{eq:confM}, with growth capped at $T$. Supplementary Note~3 gives both geometric proofs.

\paragraph{Resources at fixed logical-register size.}
For fixed straight toric registers, $U=O(d)$. Choose a fixed envelope $0<\lambda_0<1$ with $\lambda\le\lambda_0$. Taking $r\ge(1+\delta)\log d/|\log\lambda_0|$ in two dimensions, or $r^2$ at least that large in three, gives $\lambda^{M(r)}\le d^{-1-\delta}$ and hard leakage $O(d^{-\delta})$. Strips contain $O(dr)$ checks and full tubes $O(dr^2)$, so both constructions use $O(d\log d)$ enhanced checks. With $\beta_R=(\tfrac12+\delta)\log d$, Eq.~\eqref{eq:softhard} adds $O(d^{-\delta}\sqrt{\log d})$. Dividing by the total measured-check count, $\Theta(d^2)$ or $\Theta(d^3)$, gives the first two enhanced fractions in Table~\ref{tab:resources}. These choices work for every fixed finite bulk strength; widths are rounded up.

For confinement, write $\gamma=1+\kappa/\Delta_Z>1$. Assume $U\to\infty$, choose $c>1/|\log\lambda_0|$, and require $T\ge c\log U$. This cutoff follows from good linear confinement as defined in Results, since $U\le n$ and a uniform polynomial cutoff eventually exceeds $c\log U$~\cite{Quintavalle2021,Hong2025Preparation}. The radius
\begin{equation}
 r=\left\lceil\frac{\log(c\log U)}{\log\gamma}\right\rceil
 \label{meth:confinement-radius}
\end{equation}
then gives $M(r)\ge c\log U$ and $U\lambda^{M(r)}\to0$. If $D_G\ge2$ bounds the degree of $G_Z$, the neighborhood contains at most $CU(D_G+1)^r$ qubits. Bounded check incidence therefore gives
\begin{equation}
 \begin{aligned}
 |\mathcal P_r|&\le\min\{N_{\rm chk},CU(\log U)^\nu\},\\
 \nu&=\frac{\log(D_G+1)}{\log\gamma}.
 \end{aligned}
 \label{meth:confinement-footprint}
\end{equation}
At $\beta_R=(\tfrac12+\delta)\log U$, the finite-strength correction is $O(U^{-\delta}(\log U)^{\nu/2})\to0$. This proves the confinement row of Table~\ref{tab:resources} and completes Corollary~\ref{cor:resources}. The enhanced fraction vanishes under the additional condition $U(\log U)^\nu=o(N_{\rm chk})$; a small radius alone does not ensure a small fraction.

For the toric protocols, logarithmic strength is also necessary under independent flips at fixed $0<p<1$: all checks incident on the selected straight string lie in its protected neighborhood, so the collective bound of Theorem~\ref{thm:local} applies to their common actual strength $\beta_R$. This necessity statement uses the independent-noise model, whereas the sufficient construction allows uniformly weak local-stochastic noise. It concerns the actual strength used in the protocol; a lower bound $\beta_j\ge\beta_R$ alone cannot establish necessity for its parameter $\beta_R$.

\paragraph{Recovery and entropy for a growing logical register.}
\label{sec:m-register}
The cost of protecting many logical qubits is constrained by the physical sites left outside the enhanced checks. The binary matrix with rows $\ell_1,\ldots,\ell_k$ has rank $k$. Deleting columns in $F_{\mathrm{hi}}$, the qubits touched by the enhanced checks, leaves rank at least $(k-|F_{\mathrm{hi}}|)_+$. Choose sites $e_a$ with independent remaining columns $g(e_a)$. For a uniformly mixed logical register, the bits $g(e_a)\cdot q$ are independent and uniform. Let $E_{e_a}$ be the environment qubit recording the absence or presence of a physical flip at $e_a$. Under independent flips at fixed $0<p<1$, the information chain rule and data processing~\cite[Ch.~5]{Watrous2018}, together with the single-site environment bound in Supplementary Note~2, give
\begin{equation}
 \begin{aligned}
 \AQEC&\ge\sum_a I(g(e_a)\cdot q:E_{e_a})\\
      &\ge2p(1-p)\sum_a\kappa_\beta(e_a)^2.
 \end{aligned}
 \label{meth:independent-parities}
\end{equation}
If all exterior strengths are at most $\beta_0$ and each qubit touches at most $\Delta_Z$ measured checks, then $\kappa_\beta(e_a)\ge e^{-\Delta_Z\beta_0}$. Substitution proves the footprint bound in Corollary~\ref{cor:register}. Vanishing entropy density consequently requires $|F_{\mathrm{hi}}|\ge[1-o(1)]k$; bounded check weight makes an extensive number of enhanced checks necessary when $k=\Theta(n)$.

The upper bound in Theorem~\ref{thm:enhancement} also applies to a growing register, using its actual support union $U$ and uniform local constants. If its right-hand side tends to zero, Eqs.~\eqref{eq:recoverybounds} and \eqref{eq:continuity} give full-register entanglement recovery and $\AQEC/k\to0$. To make the total entropy vanish, the continuity bound gives the stronger sufficient conditions
\begin{equation}
 kU\lambda^{M(r)}\longrightarrow0,\qquad
 k\sqrt{|\mathcal P_r|}\,e^{-\beta_R}\longrightarrow0.
 \label{meth:register-total-entropy}
\end{equation}
Without the factors $k$, these conditions already suffice for recovery and vanishing entropy density. The extra factors ensure vanishing total entropy and are not claimed necessary for recovery. Supplementary Note~3 gives the register-dependent estimates.

\subsection{Wilson restoration on a finite torus}
\label{sec:m-wilson}
Theorem~\ref{thm:wilson} compares two representatives enhanced in the same preparation. Their Wilson overlap depends on the checks between them as well as those near their supports. Because $W(C)$ is diagonal, this dependence can be evaluated in the Born measure of the prepared state. Write $Z_e=(-1)^{s_e}$ with $s_e\in\mathbb F_2$; the plaquette parity $(\delta s)_p$ is the sum of these bits around $p$. Squaring the preparation amplitudes gives
\begin{equation}
 \pi_{\boldsymbol\beta}(s)=\frac{\exp[\sum_p\beta_p(-1)^{(\delta s)_p}]}
 {\sum_{s'}\exp[\sum_p\beta_p(-1)^{(\delta s')_p}]}.
 \label{meth:gauge}
\end{equation}
Each unoriented plaquette is counted once, and all chains and intersections below have $\mathbb F_2$ coefficients.

In two dimensions, the high-temperature expansion~\cite[Sec.~7]{Wegner2014}, $e^{\beta_pB_p}=\cosh\beta_p(1+u_pB_p)$, and the sum over edge bits retain plaquette sets with boundary zero in the denominator and boundary $C$ in the numerator. The only fillings are the two complementary regions $S$ and $S^c$, giving
\begin{equation}
 w=\frac{\prod_{p\in S}u_p+\prod_{p\in S^c}u_p}
          {1+\prod_pu_p},\qquad u_p=\tanh\beta_p.
 \label{meth:inhomogeneous-wilson}
\end{equation}
The denominator is at least one, and every enhanced-plaquette factor is at most one. Hence $w$ is bounded by $(\tanh\beta)^{|S\setminus\mathcal P_r|}+(\tanh\beta)^{|S^c\setminus\mathcal P_r|}$. When each filling retains at least $cL^2$ unenhanced plaquettes, this proves the two-dimensional part of Theorem~\ref{thm:wilson}, including the limit of projective enhancement.

In three dimensions, we adapt the random-cluster and dual-linking arguments used by Duncan and Schweinhart to prove the perimeter law~\cite[Secs.~5, 7, and 9]{DuncanSharp2025}. Our setting has nonuniform couplings and two winding representatives on a finite torus. The plaquette coupling of Ref.~\cite[Sec.~5.1]{DuncanSharp2025} gives a direct representation in this setting: conditional on the open plaquette set $A$, the gauge configurations are uniform subject to $B_p=+1$ for $p\in A$. Their Wilson character averages to one when $C$ bounds a chain of open plaquettes and to zero otherwise. Thus $w$ is the filling probability. Supplementary Note~4 derives this identity for the present couplings and bounds the obstructions that remain after local enhancement.

The parameter conventions matter in applying this representation. The identity $e^{\beta_pB_p}=e^{-\beta_p}e^{2\beta_p\mathbf1_{\{B_p=+1\}}}$ shows that the indicator-action coupling of Refs.~\cite{DuncanSharp2025,DuncanTopological2025} is $2\beta_p$. The primal plaquette parameter is $\theta_p=1-e^{-2\beta_p}$. A dual edge is open when its primal plaquette is absent, and its parameter is
\[
 p_e^*=\frac{2(1-\theta_p)}{\theta_p+2(1-\theta_p)}
       =\frac{2}{e^{2\beta_p}+1}.
\]
These are random-cluster weight parameters with cluster weight $q=2$, distinct from the physical noise probability. Stronger checks decrease $p_e^*$.

On the dual torus, write $D$ for the set of open edges. The Fortuin--Kasteleyn (FK) random-cluster measure is a probability distribution on these configurations~\cite{FortuinKasteleyn1972,Grimmett2006}, with weights
\begin{equation}
 \phi_{L,\boldsymbol p^*,2}(D)\propto
 2^{c(D)}\prod_{e\in D}p_e^*\prod_{e\notin D}(1-p_e^*).
 \label{meth:fk-definition}
\end{equation}
Here $c(D)$ counts connected components, including isolated vertices, on the graph with periodic boundaries; the weights are normalized over all $D$. Each component contributes a factor two, corresponding to its two spin assignments in the Ising representation. This cluster factor correlates the edge occupations, so $p_e^*$ is a weight parameter rather than the marginal probability that $e$ is open.

Plaquette-to-edge duality on a torus also retains winding information~\cite[Sec.~4.3]{DuncanTopological2025}. Keeping the ordinary gauge weights, the rank calculation in Supplementary Note~4 gives the induced dual law for our inhomogeneous couplings:
\begin{equation}
 \nu^*(D)=\frac{2^{-h(D)}\phi_{L,\boldsymbol p^*,2}(D)}
 {\mathbb E_{\phi_{L,\boldsymbol p^*,2}}[2^{-h(D)}]},
 \label{meth:torus-dual-law}
\end{equation}
where $h(D)$ counts the independent winding classes carried by $D$, with coefficients in $\mathbb F_2$. The factor $2^{-h(D)}$ is the topological correction to the standard FK distribution. Since the torus has three independent periodic directions, $0\leq h(D)\leq3$; both this factor and its FK average lie in $[1/8,1]$. Thus any event probability under $\nu^*$ is at most eight times its probability under periodic FK. This comparison allows the standard random-cluster estimates to be used for the gauge model.

At the fixed bulk point in Eq.~\eqref{eq:deconfined}, the dual FK model is subcritical. Theorem~1.2 of Ref.~\cite{Duminil2019} gives exponential decay of connections from the center to the wired boundary of a finite graph-distance ball. This is also the probability estimate used in the perimeter-law proof~\cite[Theorem~36]{DuncanSharp2025}. For unit bond couplings, Ref.~\cite{Duminil2019} uses an inverse-temperature parameter $b_{\mathrm{FK}}$ with bond parameter $1-e^{-b_{\mathrm{FK}}}$, so our bulk parameter corresponds to
\begin{equation}
 b_{\mathrm{FK}}=-\log[1-p^*(\beta)]=-\log\tanh\beta.
 \label{meth:fk-source-coupling}
\end{equation}
To use this theorem on the inhomogeneous torus, condition periodic FK on the edges outside a ball that lifts to $\mathbb Z^3$. Exterior connections identify some boundary vertices. The fully wired condition identifies all of them and therefore gives the largest connection probabilities. Together with monotonicity in the edge parameters~\cite[Chs.~3--4]{Grimmett2006}, the inequality $p_e^*\leq p^*(\beta)$ bounds the conditional connection probability by that of the homogeneous wired ball. Averaging over the exterior and including the factor-eight comparison yields
\begin{equation}
 \nu^*\{v\leftrightarrow\partial B_R(v)\}\leq C_\beta e^{-R/\xi_\beta},
 \quad 1\leq R\leq L/4,
 \label{meth:dual-tail}
\end{equation}
where $B_R(v)$ is the graph-distance ball about a dual vertex $v$. A union over vertices also bounds the probability of a cluster with periodic diameter at least $L/4$ by $C_\beta L^3e^{-c_\beta L}$. The constants are uniform in the enhanced parameters and system size at fixed bulk strength; no uniformity at the critical point is assumed.

Protect the plaquettes whose dual edges lie within cell-incidence distance $r$ of $C=\ell_1\oplus\ell_2$. For $1\leq r\leq c_\alpha L$, with $c_\alpha<\min\{\alpha/8,1/16\}$, these full tubes contain $|\mathcal P_r|=O(L(r+1)^2)$ checks. Projective enhancement closes every dual edge in the tubes. If an open filling fails to exist, linear duality supplies an Eulerian dual chain with odd intersection with a filling $S$. Exclude the macroscopic-cluster event just bounded. Each remaining cluster lifts to a contractible box, and at least one Eulerian part $\gamma$ has odd intersection with $S$. It bounds a dual surface $A_\gamma$ inside that box, so
\begin{equation}
 1=\gamma\cdot S=(\partial A_\gamma)\cdot S
   =A_\gamma\cdot C.
 \label{meth:boundary-intersection}
\end{equation}
The surface must meet $C$, and the cluster therefore approaches $C$ within a distance proportional to its diameter. A cluster avoiding the protected tube must consequently have diameter at least $c_gr$. At diameter scale $s$, it has an anchor among $O(L(s+1)^2)$ edges near $C$. Applying Eq.~\eqref{meth:dual-tail} at these anchors and summing over dyadic scales bounds the remaining filling-failure probability by $C_\beta L(r+1)^2e^{-c_gr/\xi_\beta}$, after adjusting geometric constants.

For finite-strength enhancement, a protected edge has conditional open probability at most $p_e^*\leq2e^{-2\beta_R}$ under ordinary FK. If $E_r$ is the event that any protected dual edge is open, the factor-eight comparison and a union bound give
\begin{equation}
 \nu^*(E_r)\leq16|\mathcal P_r|e^{-2\beta_R}.
 \label{meth:soft-wilson}
\end{equation}
Conditioning all protected edges to be closed makes their local weight factors constant and reproduces the projectively enhanced dual law exactly, including its topological weight. The Wilson deficit is thus bounded by the projective deficit plus $\nu^*(E_r)$:
\begin{equation}
 \begin{aligned}
 1-w\leq{}& C_\beta L(r+1)^2e^{-c_gr/\xi_\beta}\\
 &+C_\beta L^3e^{-c_\beta L}+C_\beta Lr^2e^{-2\beta_R}.
 \end{aligned}
 \label{eq:wilsonbound}
\end{equation}
The three terms respectively control local obstructing clusters, macroscopic clusters, and residual open edges in the protected tubes. For the choices in Eq.~\eqref{eq:wilsonparameters}, with $r$ rounded upward, they are $O(L^{1-b}\log^2L)$, $O(L^3e^{-c_\beta L})$, and $O(L^{-2\delta}\log^2L)$. All vanish, while $r\leq c_\alpha L$ and $\beta_R\geq\beta$ hold for sufficiently large $L$.

The same preparation restores AQEC under the separate low-noise assumption $\lambda\leq\lambda_0<1$. The cell-incidence and qubit check-graph distances differ only by fixed factors, so the full tubes contain the AQEC ribbons at width proportional to $r$. Their minimum silent distinguishing size is then proportional to $r^2$, making the projective activity $O(L e^{-c(\log L)^2})$ for some $c>0$. The $O(L\log^2L)$ enhanced checks give a soft--hard AQEC correction $O(L^{-\delta}\log L)$. Both contributions vanish for either chosen encoding. Supplementary Notes~3 and 4 give the geometric comparison and the complete recovery and Wilson estimates.

\subsection{Controlled strings and common recovery}
\label{sec:m-operations}
The operation in Corollary~\ref{cor:cnot} acts on the fixed target encoding $V=V_1$; only its physical representative is changed. The two encodings share their zero codeword, while their one codewords have overlap $w$. Opposite eigenvalues of the complementary physical logical $X$ symmetry make the cross-sector overlaps vanish. Thus, in the ordered logical basis $(|0\rangle_L,|1\rangle_L)$,
\begin{equation}
 V_1^\dagger V_2=\diag(1,w),\qquad
 \|V_1-V_2\|_\infty=\sqrt{2(1-w)}.
 \label{eq:encoder-distance}
\end{equation}
The norm follows from $(V_1-V_2)^\dagger(V_1-V_2)=\diag(0,2(1-w))$. In particular, $w\to1$ makes the encodings agree on arbitrary logical inputs, including inputs entangled with a reference.

For the fixed encoding, $Z^{\ell_1}V=VX_L$. Commutativity of the two physical strings and the same logical symmetry give
\begin{equation}
 V^\dagger Z^{\ell_2}V=wX_L,\qquad
 V^\dagger Z^{\ell_1}Z^{\ell_2}V=wI_L.
 \label{meth:string-compression}
\end{equation}
Consequently $[(Z^{\ell_2}-Z^{\ell_1})V]^\dagger[(Z^{\ell_2}-Z^{\ell_1})V]=2(1-w)I_L$. Since replacing $G_1$ by $G_2$ changes only the control-one block, its largest singular value gives Eq.~\eqref{eq:cnot-error}. The projection onto the fixed code space is $VV^\dagger$, and
\[
 V^\dagger Z^{\ell_2}VV^\dagger Z^{\ell_2}V=w^2I_L.
\]
For control $|1\rangle_c$, the target therefore remains in the code with probability $w^2$ for every logical input. This proves the leakage statement $1-w^2$. Tensoring the operator identities with a reference identity leaves these conclusions unchanged.

The overlap also constrains recovery when the encoder choice is undisclosed. With no subsequent noise, the optimal worst-choice fidelity of a single recovery $\cR$ from the full physical system $Q$ to the logical qubit is
\begin{equation}
 \max_{\cR}\min_{i=1,2}F_e(\cR\cV_i)
 =\frac{1+\sqrt{(1+w)/2}}{2},
 \label{meth:common-recovery}
\end{equation}
where $\cV_i(\rho_L)=V_i\rho_LV_i^\dagger$ and $F_e$ is the squared entanglement fidelity for a maximally mixed logical qubit~\cite{Schumacher1996}. The equal mixture of encoders retains coherence $\sqrt{(1+w)/2}$ between their shared zero codeword and the normalized symmetric combination of their one codewords. Cauchy--Schwarz bounds the recovered coherence, and a recovery using the symmetric and antisymmetric combinations attains the displayed fidelity for both encoders (Supplementary Note~4). Subsequent common noise cannot improve this optimum, since its composition with a recovery is already among the channels optimized above. Conversely, $w\to1$ transfers any successful fixed recovery under the same noise between the two encodings by channel contractivity~\cite[Sec.~3.3]{Watrous2018}. The common-recovery result relies on the shared zero codeword and concerns an unknown encoder choice; the controlled-string result keeps the encoding fixed throughout.

\section*{Acknowledge}
This work was supported by Quantum Science and Technology-National Science and Technology Major Project (Grant No. 2021ZD0302400), Shanghai Municipal Science and Technology (Grant No. 25LZ2600200), and National Natural Science Foundation of China (Grants No. 92365111).

\makeatletter
\let\auto@bib@innerbib\@empty
\makeatother


\clearpage
\onecolumngrid
\setcounter{section}{0}
\setcounter{subsection}{0}
\setcounter{equation}{0}
\setcounter{table}{0}
\setcounter{figure}{0}
\renewcommand{\thesection}{\arabic{section}}
\renewcommand{\theequation}{S\arabic{equation}}
\renewcommand{\thetable}{S\arabic{table}}
\renewcommand{\thefigure}{S\arabic{figure}}
\renewcommand{\theHsection}{SI.\arabic{section}}
\renewcommand{\theHequation}{SI.\arabic{equation}}
\renewcommand{\theHtable}{SI.\arabic{table}}
\renewcommand{\theHfigure}{SI.\arabic{figure}}
\let\lemma\supplementarylemma
\let\endlemma\endsupplementarylemma
\let\definition\supplementarydefinition
\let\enddefinition\endsupplementarydefinition
\let\remark\supplementaryremark
\let\endremark\endsupplementaryremark

\begin{center}
\large\bfseries
Detailed Derivations and Proofs of the Theorems
\end{center}
\startsupplementarycontents
\clearpage

\begingroup

\refstepcounter{section}
\section*{Supplementary Note 1: Finite-strength preparation and bare Wilson loops}\label{app:background}

The finite-strength measurement instrument fixes both the conditional encoding and
the classical measure used to evaluate its Wilson observables. We derive
this connection for general CSS codes before specializing to the square
and cubic toric codes. The two-dimensional Wilson expression is exact at
finite volume. In three dimensions, high- and low-temperature expansions
establish the area and perimeter mechanisms, and a sharp-transition
theorem extends the corresponding laws to the full strict phases. Wilson
restoration by local enhancement is proved in Supplementary Note~4.

\subsection{Ancilla circuit and measurement strength}\label{app:weak-circuit}

Let $B_j$ be a Hermitian Pauli check, $B_j^2=I$, and let $a$ denote a fresh
ancilla prepared in
$|+\rangle_a=(|0\rangle_a+|1\rangle_a)/\sqrt2$.
The circuit in the main text is the standard Pauli finite-strength measurement
construction~\cite{SI-Brun2002,SI-Zhu2023}:
\begin{equation}\label{bg:unitary}
 U_j(t_j)=e^{-it_jZ_a\otimes B_j}
 =\cos t_j\,I-i\sin t_j\,Z_a\otimes B_j,
 \qquad 0\le t_j\le\pi/4.
\end{equation}
The angle $t_j$ is the integrated coupling in units of $\hbar$; for an
interaction Hamiltonian $H_j(\tau)=g_j(\tau)Z_a\otimes B_j$,
$t_j=\hbar^{-1}\int g_j(\tau)\,d\tau$. It is not a duration by itself.
The displayed multiqubit gate specifies the measurement instrument; no
assumption of native multiqubit hardware is needed for the recovery results.

For an arbitrary data state $|\phi\rangle$, the joint state is
\begin{equation}\label{bg:jointstate}
 U_j(t_j)|+\rangle_a|\phi\rangle
 =\frac{|0\rangle_a e^{-it_jB_j}|\phi\rangle
       +|1\rangle_a e^{it_jB_j}|\phi\rangle}{\sqrt2}.
\end{equation}
Measure $Y_a$ and label its eigenstate
$|y_s\rangle=(|0\rangle+is|1\rangle)/\sqrt2$, $s=\pm1$.
The unnormalized data branch follows by taking the ancilla inner product:
\begin{align}
 \langle y_s|U_j(t_j)|+\rangle_a
 &=\frac{e^{-it_jB_j}-is e^{it_jB_j}}{2}\notag\\
 &=\frac{1-is}{2}(\cos t_j\,I+s\sin t_j\,B_j)\notag\\
 &=e^{-is\pi/4}\frac{\cos t_j\,I+s\sin t_j\,B_j}{\sqrt2}.
 \label{bg:contraction}
\end{align}
The phase $e^{-is\pi/4}$ is global within a fixed readout branch and has no
physical effect. Thus $s$ in the positive filter below is precisely the
$Y_a$ eigenvalue, with the same sign convention as the basic Pauli circuit
in Ref.~\cite{SI-Zhu2023}.

Define $\beta_j\ge0$ by $\tanh(\beta_j/2)=\tan t_j$.
Using $e^{s\beta_jB_j/2}=\cosh(\beta_j/2)I+s\sinh(\beta_j/2)B_j$
and $\cosh\beta_j=\cosh^2(\beta_j/2)+\sinh^2(\beta_j/2)$ yields
\begin{equation}\label{bg:kraus}
 M_{j,s}=\frac{\cos t_j\,I+s\sin t_j\,B_j}{\sqrt2}
       =\frac{e^{s\beta_jB_j/2}}{\sqrt{2\cosh\beta_j}}.
\end{equation}
The useful equivalent parameter conversions are
\begin{equation}\label{bg:beta-angle}
 \boxed{\ \beta_j=2\operatorname{artanh}(\tan t_j)
 =\log\tan(\pi/4+t_j),\quad
 \tanh\beta_j=\sin(2t_j),\quad
 \sech\beta_j=\cos(2t_j).\ }
\end{equation}
These identities hold for $0\le t_j<\pi/4$ and extend by a limit at
$\pi/4$. In particular, the amplitude exponent is $\beta_jB_j/2$,
whereas the squared-amplitude (classical gauge) exponent is $\beta_jB_j$.
The expansion weights used below are denoted by $u_j=\tanh\beta_j$ to
keep them distinct from the interaction angles $t_j$.

The POVM effects and probabilities for a normalized data density matrix
$\rho$ are
\begin{align}
 M_{j,s}^\dagger M_{j,s}
 &=\frac{I+s\tanh\beta_j B_j}{2},\qquad
 \sum_{s=\pm1}M_{j,s}^\dagger M_{j,s}=I,\label{bg:povm}\\
 p_j(s\mid\rho)&=\frac{1+s\tanh\beta_j\Tr(B_j\rho)}2,
 \qquad
 \rho_{j,s}=\frac{M_{j,s}\rho M_{j,s}^\dagger}{p_j(s\mid\rho)}.
 \label{bg:born}
\end{align}
The conditional state is defined when its probability is nonzero.
At $t_j=0$ both outcomes have probability $1/2$ and leave the state
unchanged. At $t_j\uparrow\pi/4$, $M_{j,s}\to(I+sB_j)/2$.
For a $B_j$ eigenstate with eigenvalue $b=\pm1$, an incorrect readout has
probability $(1-\sin2t_j)/2$. For a superposition, however, the finite
measurement also changes amplitudes coherently. It is not equivalent to a
projective measurement followed only by a classical readout error
~\cite{SI-Yang2022,SI-Zhao2024,SI-PNAS2025}.

\subsection{Postselection, normalization, and logical inputs}\label{app:postselection}

For the CSS preparation, the measured checks are unsigned $Z$ operators,
$B_j=Z^{h_j}$, where the binary vectors $h_j$ are the rows of
$H_Z\in\mathbb F_2^{m_Z\times n}$. They commute, and their rows need not
be independent. For a complete record
$\boldsymbol s=(s_1,\ldots,s_{m_Z})$, discarding only the irrelevant global
phase gives
\begin{equation}\label{bg:recordkraus}
 K_{\boldsymbol s}=\prod_jM_{j,s_j}
 =\frac{F_{\beta,\boldsymbol s}}{\sqrt{\prod_j2\cosh\beta_j}},
 \qquad
 F_{\beta,\boldsymbol s}=\exp\!\left(\frac12\sum_js_j\beta_jB_j\right).
\end{equation}
Hence $\Pr(\boldsymbol s\mid\rho)=\Tr(K_{\boldsymbol s}\rho
K_{\boldsymbol s}^\dagger)$ for an arbitrary input.
For $|+\rangle^{\otimes n}$, define
\begin{equation}\label{bg:generalrecord}
 Z_{\beta,\boldsymbol s}
 =\langle+|^{\otimes n}F_{\beta,\boldsymbol s}^2|+\rangle^{\otimes n},
 \qquad
 |\psi_{\beta,\boldsymbol s}\rangle
 =Z_{\beta,\boldsymbol s}^{-1/2}F_{\beta,\boldsymbol s}|+\rangle^{\otimes n}.
\end{equation}
A fixed record therefore gives a pure state. Postselection in this paper
means retaining precisely $s_j=+1$ for all $j$. Writing $F_\beta$ and
$Z_\beta$ for this record,
\begin{equation}\label{bg:allplus}
 K_+|+\rangle^{\otimes n}
 =\frac{F_\beta|+\rangle^{\otimes n}}{\sqrt{\prod_j2\cosh\beta_j}},
 \quad |\psi_\beta\rangle=\frac{F_\beta|+\rangle^{\otimes n}}{\sqrt{Z_\beta}},
 \quad p_+=\frac{Z_\beta}{\prod_j2\cosh\beta_j}.
\end{equation}
No independence of the measured checks has been assumed. To make the
postselection cost explicit even for redundant checks, expand
$e^{\beta_jB_j}=\cosh\beta_j(I+\tanh\beta_jB_j)$. A check product has a
nonzero expectation in $|+\rangle^{\otimes n}$ exactly when its binary
support vanishes. Consequently,
\begin{equation}\label{bg:success-redundancy}
 p_+=2^{-m_Z}\sum_{a\in\ker H_Z^T}
          \prod_{j:a_j=1}\tanh\beta_j.
\end{equation}
Independent rows give $p_+=2^{-m_Z}$; in the fully projective limit,
$p_+\to2^{-\operatorname{rank}H_Z}$. Thus the selected branch need not have
size-independent probability. Our enhanced-check counts quantify control
within this conditioned protocol, not the expected number of preparation
attempts. At finite strengths all records have nonzero probability; in
the projective limit, inconsistent records imposed by check redundancies
have zero probability.

Discarding the readout produces a different channel
~\cite{SI-Brun2002,SI-Zhu2023,SI-Lee2022}. For one check,
\begin{equation}\label{bg:unconditioned}
 \sum_sM_{j,s}\rho M_{j,s}^\dagger
 =\cos^2t_j\,\rho+\sin^2t_j\,B_j\rho B_j.
\end{equation}
For all checks the unconditioned output is
$\sum_{\boldsymbol s}K_{\boldsymbol s}\rho K_{\boldsymbol s}^\dagger$;
retaining a classical record instead gives
$\sum_{\boldsymbol s}|\boldsymbol s\rangle\langle\boldsymbol s|
\otimes K_{\boldsymbol s}\rho K_{\boldsymbol s}^\dagger$.
Neither object is the all-plus pure branch. Other fixed records put signs
$s_j\beta_j$ in the gauge action; the Born-weighted ensemble is the
disordered problem studied in Ref.~\cite{SI-Zhu2023}. The present all-plus
branch has a nonnegative, disorder-free gauge coupling.

The same conditioned operation encodes an unknown logical input. Choose
$k$ ideal logical $Z$ representatives $\ell_a$ that are independent modulo
$\operatorname{row}H_Z$, put $K=2^k$, and write
$\ell(q)=\bigoplus_{a=1}^kq_a\ell_a$ for $q\in\mathbb F_2^k$.
Here $\mathcal H_L$ is the $K$-dimensional logical input space and
$\mathcal H_Q=(\mathbb C^2)^{\otimes n}$ is the full physical space.
The map $V_0:\mathcal H_L\to\mathcal H_Q$ defined by
$V_0|q\rangle_L=Z^{\ell(q)}|+\rangle^{\otimes n}$ is an isometry;
$V_0$ denotes the zero-strength encoding. Since $F_\beta^2$ expands
entirely into $Z$ stabilizer products,
\begin{equation}\label{bg:logicalsuccess}
 {}_L\langle q|V_0^\dagger F_\beta^2V_0|q'\rangle_L
 =\langle+|^{\otimes n}Z^{\ell(q\oplus q')}F_\beta^2
 |+\rangle^{\otimes n}=Z_\beta\delta_{qq'}.
\end{equation}
For $q\ne q'$, the selected logical class cannot be canceled by a check
product; for $q=q'$, commutation makes the norm independent of $q$.
Therefore $K_+V_0=\sqrt{p_+}\,V_\beta$ with
$V_\beta=Z_\beta^{-1/2}F_\beta V_0$ and
$V_\beta^\dagger V_\beta=I_K$. Its physical codewords are
$|q_\beta\rangle=V_\beta|q\rangle_L=Z^{\ell(q)}|\psi_\beta\rangle$.
Tensoring this identity with $I_R$ shows that
success is input independent even for a logical input entangled with a
reference $R$. The normalized successful operation is a linear encoding
isometry. The same argument applies to every nonzero-probability fixed
record; only the all-plus branch is used below. Supplementary
Lemma~\ref{lem:a1} relates this encoding to the $X$-error overlap kernel.

\subsection{Why the \texorpdfstring{$B_j$}{B} sector is the relevant recovery problem}\label{app:one-sided}

CSS codes separate the two Pauli syndrome sectors~\cite{SI-CalderbankShor1996,SI-Steane1996}.
Our preparation treats them asymmetrically. Every $X$ check $A_\mu$
stabilizes $|+\rangle^{\otimes n}$ and commutes with both $F_\beta$ and
$Z^{\ell(q)}$, so
\begin{equation}\label{bg:exact-xchecks}
 A_\mu V_\beta=V_\beta.
\end{equation}
For toric codes this fixes every vertex charge, while the weakened
plaquette constraints allow flux fluctuations~\cite{SI-Kitaev2003,SI-Dennis2002,SI-Zhu2023}.
Equation~\eqref{bg:exact-xchecks} detects a $Z$ operator with nonzero
$X$ syndrome. Exact correction also requires that errors with the same
syndrome reveal no logical information, which follows from the matrix
elements below.

Let $d_Z$ be the minimum weight of a nontrivial ideal $Z$ logical operator,
and let $\mathcal E_Z=\{Z^z:z\in\mathcal S\}$ be any error set for which
$|z\oplus z'|<d_Z$ for every pair $z,z'\in\mathcal S$.
For $v=z\oplus z'$, commutation of $Z^v$ with the filter gives
\begin{equation}\label{bg:z-compression-elements}
 \langle q_\beta|Z^v|q'_\beta\rangle
 =\frac{\langle+|^{\otimes n}
       Z^{v\oplus\ell(q\oplus q')}F_\beta^2|+\rangle^{\otimes n}}
       {Z_\beta}.
\end{equation}
When $q\ne q'$, a nonzero numerator would require
$v\oplus\ell(q\oplus q')\in\operatorname{row}H_Z$, because
$F_\beta^2$ contains only products of measured checks. It would make $v$
a nontrivial ideal logical $Z$ operator, contradicting $|v|<d_Z$.
For $q=q'$, the result is independent of $q$. Hence
\begin{equation}\label{bg:z-kl}
 \boxed{\quad V_\beta^\dagger (Z^z)^\dagger Z^{z'}V_\beta
 =\langle\psi_\beta|Z^{z\oplus z'}|\psi_\beta\rangle\,I_K.
 \quad}
\end{equation}
This is the exact Knill--Laflamme condition~\cite{SI-Knill1997} for
$\mathcal E_Z$. In particular, all pure $Z$ errors of weight at most
$\lfloor(d_Z-1)/2\rfloor$ remain exactly correctable, for every finite
strength profile and in the projective limit. The result also covers
channels with Kraus operators in the linear span of this error set.

A product of $B_j$ may change the physical state at finite strength;
Eq.~\eqref{bg:z-kl} asserts a scalar \emph{compression}, not that every
$B_j$ acts as the identity on the imperfect subspace. Residual deformations
from degenerate $Z$ errors nevertheless carry no logical information in
this error set and can be removed by an unrestricted recovery. A usual
syndrome-dependent Pauli correction alone need not return the state to
the original imperfect subspace.

The $X$ sector has the opposite behavior. An $X_x$ error changes the signs
of the $B_j$ it anticommutes with; finite-strength filtering leaves a
nonzero overlap between error branches. As shown in Supplementary
Lemma~\ref{lem:a1}, that overlap is
$\langle q_\beta|X_x|q'_\beta\rangle
=\delta_{qq'}(-1)^{q\cdot g(x)}\kappa_\beta(x)$ and can depend on the
logical input even for a single-site error~\cite{SI-Zhao2024}.
This identifies the sector in which finite-strength preparation creates the recovery
obstruction studied here. Restricting the subsequent channel to $X$ noise
still tests preservation of arbitrary quantum superpositions and their
reference entanglement. Combining the sector-specific results into a
threshold for joint $X/Z$ noise, or allowing noisy syndrome-extraction
circuits, is a further task and is not assumed in the present bounds.

\subsection{Gauge measure and comparison of representatives}

For a general CSS code, the probability of a computational-basis string
$s\in\mathbb F_2^n$ in $|\psi_\beta\rangle$ is
\begin{equation}\label{bg:general-gauge}
 \mathcal S_\beta(s)=-\sum_j\beta_j(-1)^{(H_Zs)_j},\qquad
 \pi_\beta(s)=\frac{e^{-\mathcal S_\beta(s)}}{2^n Z_\beta}.
\end{equation}
The normalization is $\sum_s e^{-\mathcal S_\beta(s)}=2^nZ_\beta$;
$Z_\beta$ is the quantum normalization defined in
Eq.~\eqref{bg:allplus}. Diagonal quantum expectations therefore equal
averages in this classical multispin measure, whose generalized Ising form is described in Ref.~\cite{SI-Wegner1971}.

For representatives with difference $c=\ell_1\oplus\ell_2\in
\operatorname{im}H_Z^T$, put $W=Z^c$. Expanding the check factors as in the
high-temperature expansion~\cite[Sec.~4]{SI-Wegner2014} and using
$2^{-n}\sum_s(-1)^{v\cdot s}=\mathbf1_{\{v=0\}}$ gives
\begin{equation}\label{bg:general-wilson}
 \langle W\rangle_\beta=
 \frac{\displaystyle\sum_{\substack{a\in\mathbb F_2^{m_Z}\\H_Z^Ta=c}}
        \prod_{j:a_j=1}u_j}
      {\displaystyle\sum_{\substack{a\in\mathbb F_2^{m_Z}\\H_Z^Ta=0}}
        \prod_{j:a_j=1}u_j},\qquad u_j=\tanh\beta_j.
\end{equation}
The numerator sums over all products of measured checks implementing the
change of representative; the denominator sums over their relations.
This identity is exact for any check matrix, including redundant rows,
and any nonnegative strength profile. It also shows
$\langle W\rangle_\beta\geq0$ for the all-plus branch. For a signed record,
the same expansion replaces $u_j$ by $s_j u_j$; its weights need not be
nonnegative.

For toric codes this multispin action becomes the $\mathbb Z_2$ lattice gauge action~\cite{SI-Wegner1971,SI-Kogut1979}.
All chains, cochains, homology groups, and intersection numbers below have
coefficients in $\mathbb F_2$.  The boundary of a chain is denoted by
$\partial$, and the coboundary of a cochain by $\delta$.  On the cubic torus
$\mathbb T_L^D$, a computational-basis configuration is an edge cochain
$s\in C^1(\mathbb T_L^D;\mathbb F_2)$, with $Z_e=(-1)^{s_e}$.  Thus the
plaquette operator $B_p=\prod_{e\in\partial p}Z_e$ has eigenvalue
$(-1)^{(\delta s)_p}$.  The squared amplitudes of
$|\psi_{\beta}\rangle\propto
\exp[\frac12\sum_p\beta_pB_p]|+\rangle^{\otimes n}$ give the probability
measure
\begin{equation}\label{c:gauge}
 \pi_{\beta}(s)=
 \frac{\exp\!\left[\sum_p\beta_p(-1)^{(\delta s)_p}\right]}
 {\displaystyle\sum_{s'}
   \exp\!\left[\sum_p\beta_p(-1)^{(\delta s')_p}\right]}.
\end{equation}
Every diagonal Wilson expectation in the quantum state is therefore an
expectation in this classical gauge measure.  We count each unoriented
plaquette once.  The action $\beta_pB_p$ corresponds to
$\widetilde\beta_p\mathbf1_{\{B_p=+1\}}$ with
$\widetilde\beta_p=2\beta_p$ in the Potts-gauge convention of
Refs.~\cite{SI-DuncanSharp2025,SI-DuncanTopological2025}.  The convention that
counts both plaquette orientations in Ref.~\cite{SI-Forsstrom2025} instead
has coupling $\beta_{\mathrm{FV}}=\beta/2$ for our uniform action.

\begin{lemma}[Representative identity and comparison of encodings]\label{lem:c1}
Let $\ell_1$ and $\ell_2$ be two $Z$ representatives such that
$\ell_1\oplus\ell_2=\partial S$, and put $C=\ell_1\oplus\ell_2$.  Then
\begin{equation}\label{c:repidentity}
 Z^{\ell_1}Z^{\ell_2}=\prod_{p\in S}B_p=:W(C),
 \qquad
 \bigl\|(Z^{\ell_1}-Z^{\ell_2})|\varphi\rangle\bigr\|^2
 =2-2\langle W(C)\rangle_\varphi
\end{equation}
for every normalized state $|\varphi\rangle$.  Suppose that a normalized
state $|\psi\rangle$ is a $+1$ eigenstate of an $X$ logical operator that
anticommutes with both $Z^{\ell_i}$.  Define
$V_i|0\rangle_L=|\psi\rangle$ and
$V_i|1\rangle_L=Z^{\ell_i}|\psi\rangle$.  Both $V_i:\mathcal H_L\to\mathcal H_Q$ are isometries. In the common ordered logical basis $(|0\rangle_L,|1\rangle_L)$, with $w=\langle\psi|W(C)|\psi\rangle$, we have
\begin{equation}\label{c:encdistance}
 V_1^\dagger V_2=
 \begin{pmatrix}1&0\\0&w\end{pmatrix},
 \qquad
 \|V_1-V_2\|_\infty^2=2(1-w).
\end{equation}
\end{lemma}

\begin{proof}
Multiplying the two strings cancels their common edges.  Multiplying the
plaquette boundaries in $S$ cancels every interior edge, leaving precisely
$\partial S$.  This proves the operator identity in
Eq.~\eqref{c:repidentity}.  Since the two strings are commuting Hermitian
unitaries, the square of their difference is $2I-2W(C)$.

Let $\overline X$ be the logical operator in the hypothesis.  The identities
$\overline X|\psi\rangle=|\psi\rangle$ and
$\overline XZ^{\ell_i}\overline X=-Z^{\ell_i}$ imply
$\langle\psi|Z^{\ell_i}|\psi\rangle=0$.  The two columns of each $V_i$
are therefore orthonormal.  The same identities eliminate the off-diagonal
entries of $V_1^\dagger V_2$; its remaining entries are $1$ and $w$.
Consequently,
$(V_1-V_2)^\dagger(V_1-V_2)=\operatorname{diag}(0,2-2w)$, proving
Eq.~\eqref{c:encdistance}.
\end{proof}

For the states prepared with finite-strength measurements in this work, the specified $X$ logical
commutes with every preparation filter and stabilizes
$|+\rangle^{\otimes n}$.  Equation~\eqref{c:encdistance} thus applies
before and after the boundary enhancement.  It compares the two encoding
maps in a common logical basis.  In particular, its operator-norm bound
also holds after tensoring the maps with an identity on any reference
system.  Recovery after noise is a separate question, governed by the
reference--environment state.

The background figure also expresses this comparison as a state fidelity.
For $|\phi_i\rangle=Z^{\ell_i}|\psi_\beta\rangle$, the squared fidelity is
\begin{equation}\label{c:representative-fidelity}
 F_{\rm rep}:=|\langle\phi_1|\phi_2\rangle|^2
 =|\langle W(C)\rangle_\beta|^2.
\end{equation}
An area or perimeter law for the Wilson overlap therefore gives the
corresponding decay of this squared fidelity. For parallel winding strings
of length $L$ separated by $R=\Theta(L)$ on the square or cubic torus,
the smaller filling has area $A_{\min}=L\min(R,L-R)=\Theta(L^2)$,
while the total boundary length is $P=2L$, provided both complementary
separations are proportional to $L$. Thus a perimeter law can still give
$F_{\rm rep}\to0$ as $L\to\infty$ at fixed finite strength. The background
figure separates the representative comparison in panel (b) from the bare
large-loop laws in panel (c). At fixed $L$, changing $R$ changes the filling area but
leaves $P=2L$ unchanged; it is distinct from increasing $L$ at fixed $R/L$.
A perimeter law alone does not remove finite-separation corrections or
specify the full $R$ dependence.

\subsection{Exact area law in two dimensions}

In two dimensions, a closed boundary has only two plaquette fillings on the
torus.  This simple geometry makes the Wilson expectation exactly calculable
and exposes the obstruction to repairing it near the boundary alone. The uniform finite-torus expression is given in Ref.~\cite[Eq.~(65)]{SI-Wegner2014}; we retain distinct plaquette couplings below.

\begin{lemma}[Exact inhomogeneous Wilson expectation in two dimensions]\label{lem:c2}
On the connected square two-torus, let $\beta_p\geq0$ and
$u_p=\tanh\beta_p$.  For a null-homologous one-cycle $C=\partial S$,
\begin{equation}\label{c:2dexact}
 \langle W(C)\rangle_{\beta}
 =\frac{\displaystyle\prod_{p\in S}u_p+
                   \prod_{p\in S^c}u_p}
              {\displaystyle1+\prod_p u_p}.
\end{equation}
Here $S^c$ is the complementary plaquette chain.  The formula includes a
disconnected pair of homologous winding strings.  If every plaquette
outside a protected set $\mathcal P_r$ has the same coupling $0<\beta<\infty$,
then arbitrary nonnegative couplings inside $\mathcal P_r$ obey
\begin{equation}\label{c:2dnogo}
 0\leq\langle W(C)\rangle_{\beta}
 \leq(\tanh\beta)^{|S\setminus \mathcal P_r|}
      +(\tanh\beta)^{|S^c\setminus \mathcal P_r|}.
\end{equation}
If each unprotected filling contains at least $cL^2$ plaquettes,
\eqref{c:2dnogo} gives the exponentially small upper bound
\[
 2\exp[-c|\log\tanh\beta|L^2],
\]
also in the hard-protection limit.
\end{lemma}

\begin{proof}
Expand every local factor as
$e^{\beta_pB_p}=\cosh\beta_p(1+u_pB_p)$, then sum over the edge spins.
A product of plaquettes survives this sum precisely when its boundary
vanishes.  In the numerator with the insertion $W(C)$, it survives
precisely when its boundary is $C$.  On the connected two-torus the space
of two-cycles consists of the empty chain and the full surface.  Hence
the denominator has exactly these two surviving terms, and the numerator
has the two chains $S$ and $S^c$.  Their common factors cancel, giving
Eq.~\eqref{c:2dexact}.

The denominator is at least one and each protected factor satisfies
$0\leq u_p\leq1$.  Dropping these factors yields Eq.~\eqref{c:2dnogo}.
Taking $u_p\uparrow1$ on protected plaquettes proves the same bound for
hard protection.  At background $\beta=0$, the expectation is exactly
zero whenever each complementary filling contains an unprotected
plaquette.
\end{proof}

For two straight representatives separated by a distance proportional to
$L$, tubes of width $r=o(L)$ leave order $L^2$ plaquettes in both
complementary fillings.  This unprotected-area condition is part of the
statement: a protected set covering an entire filling is not covered by
the no-go bound with extensive unprotected area.

For uniform strength on the two-torus, Eq.~\eqref{c:2dexact} reduces to
\begin{equation}\label{bg:2duniform}
 \langle W(C)\rangle_\beta
 =\frac{u^{|S|}+u^{L^2-|S|}}{1+u^{L^2}},\qquad u=\tanh\beta.
\end{equation}
In a simply connected planar box there are no nonempty two-cycles.  The same
expansion consequently gives
\begin{equation}\label{bg:2dplane}
 \langle W(C)\rangle_\beta=(\tanh\beta)^{|S|}.
\end{equation}
This remains true in the infinite-plane limit for any fixed finite loop.
For every $0<\beta<\infty$, its area tension is
$-\log\tanh\beta>0$.  On the torus the smaller of the two complementary
areas controls the exponent.  In particular, loops with both areas
proportional to $L^2$ decay exponentially in $L^2$.  There is no finite
$\beta$ at which the bare two-dimensional Wilson observable acquires a
perimeter law.  These statements concern fixed $\beta$; their conclusion
must not be interchanged with a limit in which $\beta$ grows with $L$.

\subsection{Three dimensions: the high-temperature area bound}

The gauge representation also shows why a different behavior is possible
in three dimensions.  There are now many plaquette surfaces with the same
boundary.  On a finite cubic box with free boundary conditions, or on a finite
cubic torus for a null-homologous $C$, expansion of each factor
$e^{\beta B_p}$ gives the exact formula
\begin{equation}\label{bg:surfaceexpansion}
 \langle W(C)\rangle_\beta
 =\frac{\displaystyle\sum_{S:\,\partial S=C}u^{|S|}}
        {\displaystyle\sum_{T:\,\partial T=0}u^{|T|}},
 \qquad u=\tanh\beta.
\end{equation}
The common factors $2^{n}\prod_p\cosh\beta$ cancel.  The sums are over
binary plaquette chains; surfaces need not be embedded or orientable.

\begin{lemma}[Area bounds at sufficiently small strength]\label{lem:bg-area}
Let $C$ be a simple loop in a cubic box and let $A(C)$ be the smallest
number of plaquettes in a filling of $C$.  There are numerical constants
$c_0,c_1$, independent of the box and the loop, such that, when
$c_1\tanh\beta<1$,
\begin{equation}\label{bg:area-bound}
 (\tanh\beta)^{A(C)}
 \leq\langle W(C)\rangle_\beta
 \leq\frac{c_0[c_1\tanh\beta]^{A(C)}}{1-c_1\tanh\beta}.
\end{equation}
For a planar rectangle in the infinite lattice, $A(C)$ is its usual area.
\end{lemma}

\begin{proof}
Choose a minimum filling $S_0$.  Symmetric difference with $S_0$ bijects
closed chains $T$ with chains having boundary $C$.  For $0<u<1$, the inequality
$|S_0\oplus T|\leq|S_0|+|T|$ implies that the numerator of
Eq.~\eqref{bg:surfaceexpansion} is at least $u^{A(C)}$ times its denominator.
This proves the lower bound without an independence assumption.

For the upper bound, make plaquettes adjacent when they share an edge.
Each connected component of a plaquette chain has a boundary that is a
cycle, and distinct components have disjoint boundary edges. Each component
boundary is therefore a subcycle of $C$. Since a simple loop has no nonempty proper subcycle, exactly one component has boundary
$C$; every other component is closed.  Denote the former component by
$S_1$.  Removing it leaves a closed chain disjoint from $S_1$.  Dropping
this disjointness restriction bounds the numerator by
\[
 \left(\sum_{S_1\ \text{ connected}:\,\partial S_1=C}u^{|S_1|}\right)
 \left(\sum_{T:\,\partial T=0}u^{|T|}\right).
\]
The denominator cancels.  A component $S_1$ must contain a plaquette
incident on a fixed edge of $C$.  The plaquette adjacency graph has bounded
degree, so the number of such connected sets of size $m$ is at most
$c_0c_1^m$.  One way to obtain this bound is to encode a spanning tree by a
depth-first walk of length at most $2m$ on this graph.  Summing over
$m\geq A(C)$ gives Eq.~\eqref{bg:area-bound}; the case $\beta=0$
follows by continuity. Projection onto the plane
of a rectangular loop shows that every filling uses at least one
plaquette above each unit square of the rectangle, proving the final
assertion.
\end{proof}

The surface sum establishes an area-law interval: producing a Wilson
boundary costs a number of plaquettes proportional to its minimum filling
area.  Uniformity in the containing box carries this conclusion to the
thermodynamic limit.  The constants do not locate the transition.

\subsection{Three dimensions: the low-temperature perimeter bound}

At large $\beta$, it is more useful to count violated plaquettes.  Mark a
dual edge whenever its primal plaquette has $B_p=-1$.  The product of the
six plaquettes around every cube is one, so an even number of marked dual
edges meets each dual vertex.  The marked edges therefore form closed
networks.  Each violated plaquette carries the relative weight
$e^{-2\beta}$.

The elementary expansion below is taken in simply connected boxes with
flat exterior boundary conditions, followed by the infinite-volume limit.
Equivalently, it is the gas of finite closed dual networks with a cutoff
on their support.  The gauge-to-defect map has fibers of equal size, and
every such network has a filling in a simply connected box; hence these
weights give the gauge measure.  Winding networks on a finite torus impose
an additional global constraint.  Their contribution is treated in
Supplementary Note~4, rather than dropped here.

\begin{lemma}[Perimeter bounds at sufficiently large strength]\label{lem:bg-perimeter}
For all sufficiently large fixed $\beta$, there are constants
$0<c_\beta\leq C_\beta<\infty$ such that every simple finite loop in the
infinite cubic lattice satisfies
\begin{equation}\label{bg:perimeter-bound}
 e^{-C_\beta|C|}\leq\langle W(C)\rangle_\beta
 \leq e^{-c_\beta|C|}.
\end{equation}
The constants can be chosen with $c_\beta,C_\beta$ both of order
$e^{-8\beta}$ as $\beta\to\infty$.
\end{lemma}

\begin{proof}
First we prove the upper bound directly in the spin variables.  Choose a
subset $J$ of the edges of $C$ such that no plaquette contains two edges
of $J$.  An edge shares a plaquette with at most twelve other edges, so a
greedy choice gives $|J|\geq|C|/13$.  Conditional on all other spins, the
spins in $J$ are independent.  Each sees a field of magnitude at most
$4\beta$, because an edge belongs to four plaquettes.  Consequently,
\[
 |\langle W(C)\rangle_\beta|
 \leq(\tanh4\beta)^{|J|}
 \leq(\tanh4\beta)^{|C|/13}.
\]
The Wilson expectation is nonnegative: expansion in plaquette subsets
has nonnegative coefficients both for free boundaries and for exterior
edge spins fixed to $+1$. Thus one may take
$c_\beta=-\frac1{13}\log\tanh4\beta$.

For the lower bound, call a nonempty connected closed dual network
$\gamma$ a polymer.  Distinct polymers are compatible when they share no
dual vertex, and their activities are $z(\gamma)=e^{-2\beta|\gamma|}$.
There is a constant $a_0$ such that at most $a_0^m$ polymers of $m$ edges
contain a given dual vertex.  Every polymer has at least four edges and
at most $|\gamma|$ vertices.  For sufficiently large $\beta$,
\begin{equation}\label{bg:kpcheck}
 \sum_{\gamma':\,\gamma'\ \text{ incompatible with}\ \gamma}
 |z(\gamma')|e^{|\gamma'|}
 \leq |\gamma|\sum_{m\geq4}(a_0e^{1-2\beta})^m
 \leq|\gamma|.
\end{equation}
This is the standard polymer convergence criterion
\cite{SI-KoteckyPreiss1986}.  It guarantees absolute convergence of the
connected-cluster expansion of the logarithm of the partition function;
the sum of absolute cluster weights containing a specified polymer
$\gamma$ is bounded by $|z(\gamma)|e^{|\gamma|}$.  The same estimate holds
if any activities are multiplied by signs.

Choose a plaquette filling $S$ of $C$.  Inserting $W(C)$ multiplies each
polymer activity by $(-1)^{|\gamma\cap S^*|}$, where $S^*$ denotes the dual
edges crossing $S$.  Thus the logarithm of the Wilson expectation is the
difference of two convergent cluster expansions.  A contributing cluster
must contain a polymer with odd intersection with $S^*$.  Marking one such
polymer and applying the preceding bound gives
\begin{equation}\label{bg:linked-bound}
 |\log\langle W(C)\rangle_\beta|
 \leq2\sum_{\gamma:\,|\gamma\cap S^*|\ \text{ odd}}
             e^{-(2\beta-1)|\gamma|}.
\end{equation}
At $\beta=\infty$ the Wilson expectation is one. Absolute convergence
makes the twisted partition function nonzero, so continuity fixes a real
logarithm throughout this large-$\beta$ interval.

An odd-intersection polymer of size $m$ must approach $C$ within distance
of order $m$.  Indeed, its coordinate bounding box has diameter at most
of order $m$.  If this box were disjoint from $C$, the closed network would
bound inside the box, and its intersection with any filling of $C$ would
be even.  There are at most $a_1|C|(m+1)^3$ possible root vertices within
this distance of $C$.  Counting the polymers from these vertices in
Eq.~\eqref{bg:linked-bound} therefore gives
\[
 |\log\langle W(C)\rangle_\beta|
 \leq 2a_1|C|\sum_{m\geq4}(m+1)^3(a_0e^{1-2\beta})^m.
\]
The series converges under a sufficiently strong version of
Eq.~\eqref{bg:kpcheck}.  It supplies $C_\beta$ and is of order
$e^{-8\beta}$.  The direct upper bound has the same order since
$-\log\tanh4\beta\sim2e^{-8\beta}$.  Uniform convergence permits the
infinite-volume limit.
\end{proof}

This argument displays the geometric reason for the perimeter: a small
closed defect network can link a filling only near its boundary.  An
intersection deep in the interior requires a larger network.  More precise
low-temperature coefficients, including corner and bottleneck corrections,
are supplied by Forsstr\"om and Viklund~\cite{SI-Forsstrom2025}.  We use only
the coarse bounds above; an expansion controlled at large $\beta$ is not
being extrapolated to the transition.

\subsection{The full strict phases and the dual parameter}

The expansions explain the two decay mechanisms but leave an intermediate
range of measurement strengths unresolved. The sharp-transition theorem
supplies the laws throughout both strict phases.  Let $p_c^{\rm FK}(\mathbb Z^3,2)$ be the critical edge parameter
of the Fortuin--Kasteleyn random-cluster model on $\mathbb Z^3$ with cluster weight
two~\cite{SI-FortuinKasteleyn1972,SI-Grimmett2006}, and define $\beta_c$ by
\begin{equation}\label{bg:critical}
 \frac{2}{e^{2\beta_c}+1}=p_c^{\rm FK}(\mathbb Z^3,2).
\end{equation}
For axis-parallel rectangular loops $C_j$ whose two side lengths diverge
and whose minimum side divided by the logarithm of their maximum side
tends to infinity, Theorem~5 of Duncan and
Schweinhart~\cite{SI-DuncanSharp2025} implies
\begin{equation}\label{bg:sharp-phases}
 \begin{aligned}
 -\frac{\log\langle W(C_j)\rangle_\beta}{A(C_j)}
   &\longrightarrow\sigma_\beta>0, &&0<\beta<\beta_c,\\
 -\frac{\log\langle W(C_j)\rangle_\beta}{|C_j|}
   &\longrightarrow\tau_\beta>0, &&\beta>\beta_c.
 \end{aligned}
\end{equation}
Here the expectations are thermodynamic gauge expectations.  The $q=2$
statement is unconditional; the dual Ising surface-tension threshold equals
its critical threshold, as used in the cited theorem.  No assertion at
$\beta=\beta_c$ is required.

For clarity, the parameter conversion follows directly from
\[
 e^{\beta B_p}=e^{-\beta}e^{2\beta\mathbf1_{\{B_p=+1\}}}.
\]
Thus the Potts-gauge coupling in the reference is $2\beta$.  The primal
plaquette occupation parameter is $1-e^{-2\beta}$, and duality sends it to
$2/(e^{2\beta}+1)$, giving Eq.~\eqref{bg:critical}.  Increasing the
measurement strength decreases the dual parameter.  The strict
perimeter-law interval is consequently the subcritical interval of the
dual random-cluster model.

Equation~\eqref{bg:sharp-phases} concerns rectangular loops in an
infinite-volume measure.  The two distant winding representatives in our
restoration result live on a finite torus and form a different loop
family.  Their estimate is proved with the finite-torus dual law and its
topological correction in Supplementary Note~4.  This is the step
that allows the full strict condition $\beta>\beta_c$ to be used in the
restoration theorem.

\refstepcounter{section}
\section*{Supplementary Note 2: AQEC theory and finite-strength logical leakage}\label{app:leakage}

Finite-strength preparation leaves a residual overlap between a codeword
and a local error acting on it.  This overlap makes local reduced states
depend on the logical input and allows subsequent noise to transfer logical
information to its environment.  We first derive the overlap and the
corresponding single-site leakage.  Bounds for several sites then show why
vanishing leakage requires increasing measurement strength even when each
single-site signal tends to zero.

All logarithms are natural.  There are $n$ physical qubits and $m_Z$ measured
$Z$ checks, with binary check matrix $H_Z$.  Physical error supports are
identified with vectors in $\mathbb F_2^n$, and addition of supports means
symmetric difference.  For selected independent logical representatives
$\ell_1,\ldots,\ell_k$, set
\begin{equation}
 \sigma(x)=\supp(H_Zx),\qquad
 g(x)=\bigl(|x\cap\ell_1|,\ldots,|x\cap\ell_k|\bigr)\pmod 2,
 \qquad \chi_q(x)=(-1)^{q\cdot g(x)}.
 \label{a:syndrome-character}
\end{equation}
Thus $\sigma(x)$ records the measured checks that anticommute with $X_x$,
whereas $g(x)$ records its commutation signs with the selected logical
representatives.  Here $q\in\mathbb F_2^k$ labels a logical basis state and
$K=2^k$ is the logical Hilbert-space dimension.  System subscripts $L$, $Q$, $R$, and $E$ denote the logical input, the
physical qubits, an untouched reference, and the physical-noise environment,
respectively.  In
subscripts, $\beta$ abbreviates the full profile
$\boldsymbol\beta=(\beta_j)$.  A scalar $\beta$ in a bound is the common
strength of the checks specified in that statement.  Unless stated otherwise, all strengths are
finite and nonnegative.  The toric examples use the standard periodic
square or cubic lattice with side length $L\ge3$.

\subsection{The recovery task and normalized AQEC matrices}

An encoding isometry $V:\mathcal H_L\to\mathcal H_Q$, with
$\mathcal H_L\simeq\mathbb C^K$ and
$\mathcal H_Q=(\mathbb C^2)^{\otimes n}$, maps an arbitrary logical
input into the full physical system and fixes its code subspace.  For our
construction $V=V_\beta$.  Write
$\cV(\rho)=V\rho V^\dagger$.  A physical channel
$\cN(\rho)=\sum_xE_x\rho E_x^\dagger$ is exactly correctable when a
completely positive trace-preserving recovery $\cR$ satisfies
$\cR\cN\cV=\operatorname{id}_L$.  This equality protects arbitrary logical
inputs and their correlations with an untouched reference.  The
Knill--Laflamme condition~\cite{SI-Knill1997} is
\begin{equation}
 V^\dagger E_x^\dagger E_yV=\lambda_{xy}I_K.
 \label{a:kl}
\end{equation}
Thus overlaps between error branches may reveal an error syndrome, but
must not depend on the unknown logical input.  The condition includes
both equality of diagonal logical matrix elements and vanishing of the
off-diagonal elements.

For a physical input $|\phi\rangle_Q$, a Stinespring dilation of the noise~\cite[Sec.~2.2.2]{SI-Watrous2018}
$|\phi\rangle_Q\mapsto\sum_xE_x|\phi\rangle_Q|x\rangle_E$
has complementary output
\begin{equation}
 [\widehat\cN\cV(\rho)]_{xy}
 =\Tr(\rho V^\dagger E_y^\dagger E_xV).
 \label{a:complement-general}
\end{equation}
Equation~\eqref{a:kl} is therefore equivalent to an environment state
independent of $\rho$.  This is the information--disturbance formulation
of correction~\cite{SI-SchumacherNielsen1996,SI-Schumacher2002,SI-Kretschmann2008,SI-Beny2010}.
The environment here belongs to the subsequent physical noise.  It is
separate from the fixed all-plus preparation record.

To quantify approximate correction, prepare
$|\Phi_K\rangle=K^{-1/2}\sum_q|q\rangle_R|q\rangle_L$ and let
$\rho_{RE}$ be its reference--environment state after encoding and noise.
Its reference marginal is $I_R/K$.  Following Ref.~\cite{SI-Zhao2024}, define
normalized matrices on the product basis $|q\rangle_R|x\rangle_E$ by
\begin{align}
 (\Lambda+B)_{xy,qq'}
 &=\frac1K\langle q|_L V^\dagger E_x^\dagger E_yV|q'\rangle_L,\notag\\
 \Lambda_{xy,qq'}
 &=\frac{\delta_{qq'}}{K^2}\Tr(V^\dagger E_x^\dagger E_yV).
 \label{a:matrix-definitions}
\end{align}
The subscript $xy,qq'$ denotes row $(q,x)$ and column $(q',y)$.
Direct partial trace shows that these matrices are
$\rho_{RE}^T$ and $(I_R/K\otimes\rho_E)^T$, respectively.  The transpose
is a full transpose in this product basis; it preserves positivity,
relative entropy, and trace norm.  In particular, $\Lambda+B$ and
$\Lambda$ are density matrices, and $B$ is their traceless Hermitian
difference.  In particular, both partial traces of $B$ vanish.  The
normalization in Eq.~\eqref{a:matrix-definitions} is retained when the
number of selected logical qubits grows.

The standard AQEC entropy and its accompanying trace-distance defect are
\begin{equation}
 D_{\mathrm{AQEC}}=D(\Lambda+B\Vert\Lambda)=I(R:E),\qquad
 \eps=\tfrac12\|B\|_1.
 \label{a:metrics-definition}
\end{equation}
Here $D(\rho\Vert\sigma)=\Tr[\rho(\log\rho-\log\sigma)]$ and
$I(R:E)=S(\rho_R)+S(\rho_E)-S(\rho_{RE})$.  If $Q$ is the physical
output, purity of $RQE$ also gives
\begin{equation}
 D_{\mathrm{AQEC}}=\log K+S(\rho_E)-S(\rho_Q)
 =\log K-I_c(I_K/K,\cN\cV),
 \label{a:coherent-information}
\end{equation}
where the coherent information is $I_c=S(\rho_Q)-S(\rho_E)$.
The AQEC entropy is the coherent information lost relative to its
noiseless value~\cite{SI-SchumacherNielsen1996,SI-Schumacher2002}; it is not the
Shannon entropy of the error distribution.

Our recovery criterion is the optimal entanglement fidelity for the
maximally mixed input on the entire selected register~\cite{SI-Schumacher1996}:
\begin{equation}
 F_e^\star=\sup_{\cR}\langle\Phi_K|
 (\operatorname{id}_R\otimes\cR\cN\cV)
 (|\Phi_K\rangle\langle\Phi_K|)|\Phi_K\rangle.
 \label{a:entanglement-fidelity}
\end{equation}
The recovery acts on all of $Q$, outputs the logical register, and may
depend on the code and noise but not on the logical input.  Purification
relates this task to environmental decoupling:
\begin{equation}
 F_e^\star=\max_{\xi_E}
 F\!\left(\rho_{RE},\frac{I_R}{K}\otimes\xi_E\right),\qquad
 F(\rho,\sigma)=\|\sqrt\rho\sqrt\sigma\|_1^2.
 \label{eq:fixed-input-duality}
\end{equation}
The maximization is over environment density matrices, and fidelity is
squared throughout.  A proof of both directions and quantitative
trace-distance bounds is given in \cref{lem:a8}.  This establishes the
existence of recovery, without assuming an efficient decoding algorithm
or a recovery confined to the enhanced region.

The lower bounds below use independent $X$ noise with fixed
$0<p<1$.  The recovery upper bounds in \hyperref[app:enhancement]{Supplementary Note~3} require
only $\Pr[A\subseteq\supp(x)]\le\eta^{|A|}$ and therefore allow correlated
local stochastic noise.  A probability upper bound alone also allows
the identity channel; it cannot imply nonzero leakage.  This is why the
lower and upper bounds have different noise assumptions.

\subsection{The encoding and its overlap kernel}

The preparation filter commutes with complementary logical $X$ operators.
Their exact eigenvalues keep distinct logical basis states orthogonal,
although the measured $Z$ checks are no longer sharp.  Consequently every
$X$-error overlap reduces to a logical sign times one scalar kernel.

\begin{lemma}[Encoding isometry and logical matrix elements]\label{lem:a1}
Let $\ell_1,\ldots,\ell_k$ represent $Z$ logical operators that are
independent modulo the $Z$-stabilizer space.  The map
\begin{equation}
 V_\beta|q\rangle_L=|q_\beta\rangle:=Z^{\ell(q)}|\psi_\beta\rangle,
 \qquad \ell(q)=\bigoplus_{a=1}^k q_a\ell_a,
 \qquad q\in\mathbb F_2^k,
 \label{a:encoding}
\end{equation}
is an isometry from a register of dimension $K=2^k$.  Applying the preparation
filter to $V_0$ has success probability independent of the logical input,
including coherent superpositions and inputs entangled with a reference.
For every physical $X$ error,
\begin{equation}
 \langle q_\beta|X_x|q'_\beta\rangle
 =\delta_{qq'}\chi_q(x)\kappa_\beta(x),
 \qquad \kappa_\beta(x)=\langle\psi_\beta|X_x|\psi_\beta\rangle.
 \label{a:logical-matrix}
\end{equation}
\end{lemma}

\begin{proof}
Choose complementary logical operators $\overline X_a$ whose binary
commutation pairing with the selected $Z$ logicals is $\delta_{ab}$.  Each
$\overline X_a$ commutes with all measured checks and stabilizes
$|+\rangle^{\otimes n}$.  It therefore has eigenvalue $(-1)^{q_a}$ on
$|q_\beta\rangle$.  Different logical labels belong to orthogonal joint
eigenspaces.  All codewords have the same norm because $Z^{\ell(q)}$ is
unitary, proving the isometry claim.  Since $X_x$ commutes with every
$\overline X_a$, it cannot connect different logical labels.  Commuting it
through $Z^{\ell(q)}$ gives the diagonal matrix element in
Eq.~\eqref{a:logical-matrix}.

To check the normalization of postselection, define
\begin{equation}
 F_\beta=\exp\!\left(\frac12\sum_j\beta_jB_j\right),\qquad
 V_0|q\rangle_L=Z^{\ell(q)}|+\rangle^{\otimes n},\qquad
 Z_\beta=\langle+|^{\otimes n}F_\beta^\dagger F_\beta|+\rangle^{\otimes n}.
 \label{a:filter}
\end{equation}
Expanding $F_\beta^\dagger F_\beta$ into products of checks gives
\begin{equation}
 V_0^\dagger F_\beta^\dagger F_\beta V_0=Z_\beta I_K,
 \qquad V_\beta=Z_\beta^{-1/2}F_\beta V_0.
 \label{a:success-normalization}
\end{equation}
Every diagonal entry is $Z_\beta$.  For $q\ne q'$, the off-diagonal entry
vanishes: $\ell(q\oplus q')$ is a nontrivial logical class, so no product of
$Z$ checks can cancel it inside an expectation in $|+\rangle^{\otimes n}$.
Consequently a physically normalized Kraus operator $cF_\beta$ succeeds
with probability $|c|^2Z_\beta$ on every logical density operator.  Equation
\eqref{a:success-normalization}, tensored with the identity on a reference,
also proves the assertion for entangled inputs.
\end{proof}

The choice of representatives specifies the encoding map itself.  For
example, when two representatives of one ideal logical class use the same
$|\psi_\beta\rangle$ as their logical zero state, their encoding maps obey
\begin{equation}
 V_1^\dagger V_2=\operatorname{diag}
 \bigl(1,\langle Z^{\ell_1}Z^{\ell_2}\rangle_\beta\bigr).
 \label{a:encoding-overlap}
\end{equation}
Both maps act on the same logical input and have the same full physical
output space, but their image subspaces need not coincide at finite strength.
Recovery is therefore specified relative to a chosen encoding map.

\begin{lemma}[Exact overlap kernel]\label{lem:a2}
Let $\mathcal R_Z=\ker H_Z^T$ be the space of check redundancies and put
$u_j=\tanh\beta_j$.  Then
\begin{equation}
 \kappa_\beta(x)=
 \left(\prod_{j\in\sigma(x)}\sech\beta_j\right)
 \frac{\displaystyle
 \sum_{\substack{a\in\mathcal R_Z\\\supp(a)\cap\sigma(x)=\varnothing}}
       \prod_{j:a_j=1}u_j}
 {\displaystyle\sum_{a\in\mathcal R_Z}\prod_{j:a_j=1}u_j},
 \label{a:kernel-exact}
\end{equation}
and
\begin{equation}
 \exp\!\left(-\sum_{j\in\sigma(x)}\beta_j\right)
 \le \kappa_\beta(x)\le\prod_{j\in\sigma(x)}\sech\beta_j.
 \label{a:kernel-bounds}
\end{equation}
For a single edge in the standard periodic toric codes of side length
$L\ge3$, uniform preparation gives
\begin{equation}
 \kappa_\beta^{\mathrm{2D}}(e)
 =\frac{\sech^2\beta}{1+(\tanh\beta)^{L^2}},
 \qquad e^{-4\beta}\le\kappa_\beta^{\mathrm{3D}}(e)\le\sech^4\beta.
 \label{a:toric-edge-kernel}
\end{equation}
\end{lemma}

\begin{proof}
Write $|+\rangle=|+\rangle^{\otimes n}$ for this proof.  Moving $X_x$
through the preparation filter reverses the signs of precisely the checks
in $\sigma(x)$.  Those checks cancel between the two half-strength filters;
the others add.  Since $X_x|+\rangle=|+\rangle$,
\begin{equation}
 \kappa_\beta(x)=
 \frac{\langle+|\exp(\sum_{j\notin\sigma(x)}\beta_jB_j)|+\rangle}
 {\langle+|\exp(\sum_j\beta_jB_j)|+\rangle}.
 \label{a:kernel-ratio}
\end{equation}
In the expansion $e^{\beta_jB_j}=\cosh\beta_j(1+u_jB_j)$, a product of
checks has nonzero expectation exactly when its binary indicator belongs
to $\ker H_Z^T$.  This proves Eq.~\eqref{a:kernel-exact}.  All redundancy
weights are nonnegative, and restricting the redundancy sum can only
decrease it, proving the upper bound.

For the lower bound, simultaneous diagonalization of the commuting checks
gives
\begin{equation}
 \exp\!\left(\sum_j\beta_jB_j\right)
 \le \exp\!\left(\sum_{j\in\sigma(x)}\beta_j\right)
       \exp\!\left(\sum_{j\notin\sigma(x)}\beta_jB_j\right).
 \label{a:operator-comparison}
\end{equation}
Taking the $|+\rangle$ expectation and using Eq.~\eqref{a:kernel-ratio}
proves the lower bound.  In two dimensions the plaquette redundancies are
the empty set and the entire torus.  Removing the two plaquettes incident
on an edge leaves only the empty redundancy.  In three dimensions an edge
meets four plaquettes, so Eq.~\eqref{a:kernel-bounds} directly gives
Eq.~\eqref{a:toric-edge-kernel}.
\end{proof}

\subsection{Standard subsystem variance}

A local observer can detect imperfect encoding when the reduced state depends
on the logical input.  Subsystem variance measures the largest such dependence
relative to the code average; it is a property of the encoding before a noise
channel is specified.

Subsystem variance uses the reduced maximally mixed code state as its
comparator~\cite{SI-Yi2024}.  For a fixed physical region $A$, the definition
in the main text can equivalently be written
\begin{equation}
 \varepsilon_{\rm SV}(A)
 =\max_{\rho_L}\left\|
 \Tr_{A^c}(V_\beta\rho_LV_\beta^\dagger)
 -\Tr_{A^c}(P_\beta/K)\right\|_1.
 \label{a:sv-definition}
\end{equation}
Convexity makes maximization over pure inputs equivalent to maximization
over density matrices.  Here $P_\beta=V_\beta V_\beta^\dagger$ is the
code projector, as in the main text.  The convention uses the full trace
norm, without $1/2$.  Maximizing this quantity over the connected regions
of a prescribed maximum size gives the subsystem-variance parameter of
Ref.~\cite{SI-Yi2024}.  At fixed $A$, it measures the input dependence of
that region, including dependence on logical superpositions.  If it vanishes exactly, the channel
to $A$ is constant.  Since this channel is complementary to erasure of
$A$, exact correction from $A^c$ follows from
Eq.~\eqref{a:complement-general}.  Approximate bounds must retain the
specified noise channel: erasure and stochastic $X$ noise sample the
encoded state differently.

\begin{lemma}[Subsystem variance of a physical site]\label{lem:subsystem-variance}
For a site $e$ with $g(e)\ne0$, the finite-strength CSS encoding satisfies
\begin{equation}
 \varepsilon_{\rm SV}(\{e\})\ge\kappa_\beta(e).
 \label{a:sv-lower}
\end{equation}
For the edge-qubit toric codes, equality holds for every nonnegative
strength profile.  In particular, at uniform finite strength,
\begin{equation}
 \begin{split}
 \varepsilon_{\rm SV}^{\mathrm{2D}}(\{e\})
 &=\frac{\sech^2\beta}{1+\tanh^{L^2}\beta}\ge e^{-2\beta},\\
 e^{-4\beta}\le\varepsilon_{\rm SV}^{\mathrm{3D}}(\{e\})
 &=\kappa_\beta^{\mathrm{3D}}(e)\le\sech^4\beta.
 \end{split}
 \label{a:sv-toric}
\end{equation}
For a toric edge with $g(e)=0$, its subsystem variance is zero.
\end{lemma}

\begin{proof}
The character $\chi_q(e)$ takes both signs equally often when $g(e)\ne0$.
Equation~\eqref{a:logical-matrix} consequently gives
\begin{equation}
 V_\beta^\dagger X_eV_\beta
 =\kappa_\beta(e)\sum_q\chi_q(e)|q\rangle_L\langle q|_L,
 \qquad \Tr(X_eP_\beta/K)=0.
 \label{a:local-compression}
\end{equation}
Choose a logical basis state with $\chi_q(e)=1$.  The expectation of
$X_e$ in the difference entering Eq.~\eqref{a:sv-definition} is
$\kappa_\beta(e)$.  Since $\|X_e\|_\infty=1$, trace-norm duality proves
the general CSS lower bound.

For toric codes, every codeword is stabilized by every $X$ star.  A star
incident on $e$ anticommutes with $Y_e$ and $Z_e$, so their compressions
vanish on the entire code subspace:
\begin{equation}
 V_\beta^\dagger Y_eV_\beta=V_\beta^\dagger Z_eV_\beta=0.
 \label{a:star-compressions}
\end{equation}
This statement holds for arbitrary superpositions, not only for basis
codewords.  For any logical density matrix it gives
\begin{equation}
 \Tr_{\{e\}^c}(V_\beta\rho_LV_\beta^\dagger)=\frac12\left[
 I+\kappa_\beta(e)\sum_q(\rho_L)_{qq}\chi_q(e)X_e\right].
 \label{a:one-site-state}
\end{equation}
The code-average marginal is $I/2$ when $g(e)\ne0$, and
$\sum_q(\rho_L)_{qq}\chi_q(e)$ ranges over $[-1,1]$.  The trace norm of
the difference from $I/2$ is therefore at most $\kappa_\beta(e)$, with
equality at either endpoint.  Equation~\eqref{a:toric-edge-kernel} gives
the displayed uniform-strength results.  If $g(e)=0$, the sum is always
one, so every input has the same reduced state and the variance vanishes.
\end{proof}

\subsection{The complementary channel and replica identities}

The complementary channel determines how the local input dependence enters
the environment.  For stochastic $X$ noise, its matrix elements are given
directly by the same overlap kernel.  This form also yields exact replica
partition functions at finite size. The construction below extends the toric-code replica calculation of Ref.~\cite[Sec.~V and Appendix~C]{SI-Zhao2024} to the present CSS register and nonuniform strengths.

For stochastic $X$ noise with distribution $P(x)$, use the Stinespring
isometry
\begin{equation}
 |\phi\rangle_Q\longmapsto\sum_x\sqrt{P(x)}X_x|\phi\rangle_Q|x\rangle_E.
 \label{a:noise-dilation}
\end{equation}
Encoding half of the maximally entangled state of a $K$-dimensional
register gives
\begin{equation}
 |\Omega\rangle
 =\frac1{\sqrt K}\sum_{q,x}\sqrt{P(x)}
   |q\rangle_R X_x|q_\beta\rangle_Q|x\rangle_E.
 \label{a:purification}
\end{equation}
Tracing out the physical output $Q$ and using
Eq.~\eqref{a:logical-matrix} makes $\rho_{RE}$ classical on $R$ in the
$q$ basis.  We use the normalized AQEC matrices and entropy from
Eqs.~\eqref{a:matrix-definitions} and~\eqref{a:metrics-definition}.
Their entries, with the logical indices last, are
\begin{align}
 (\Lambda+B)_{xy,qq'}
 &=\frac{\delta_{qq'}}K\sqrt{P(x)P(y)}
       \chi_q(x\oplus y)\kappa_\beta(x\oplus y),
 \label{a:aqec-entry}\\
 \Lambda_{xy,qq'}
 &=\frac{\delta_{qq'}}K\sqrt{P(x)P(y)}\kappa_\beta(x\oplus y)
       \mathbf1_{\{g(x\oplus y)=0\}}.
 \label{a:product-entry}
\end{align}
The second line uses orthogonality of the characters under uniform $q$.
In particular $\Tr_R B=0$.  Equivalently,
\begin{equation}
 \rho_{RE}=\frac1K\sum_q|q\rangle_R\langle q|_R\otimes\rho_E^q,
 \quad \rho_E=\frac1K\sum_q\rho_E^q,
 \quad D_{\mathrm{AQEC}}=S(\rho_E)-\frac1K\sum_qS(\rho_E^q).
 \label{a:cq-holevo}
\end{equation}
This is the Holevo information of the conditional environment
ensemble~\cite{SI-Holevo1973}, hence $D_{\mathrm{AQEC}}\le\log K$ for this
model.  The error-label diagonal is $P(x)$ for every $q$, and all logical
dependence resides in the off-diagonal entries.  Classicality on $R$ does
not make the recovery task classical: different $\rho_E^q$ suppress
coherence between the corresponding physical logical sectors.
Although the physical channel is a stochastic
mixture of Pauli errors, its complementary output retains coherences
between error labels.  Dephasing those labels would discard the very
information measured by these bounds.

\begin{lemma}[Replica identities at finite size]\label{lem:a3}
For an integer $m\ge2$, put $x^{(m+1)}=x^{(1)}$ and
$d_\alpha=x^{(\alpha)}\oplus x^{(\alpha+1)}$.  Then
\begin{align}
 K^{m-1}\Tr(\Lambda+B)^m
 &=\sum_{x^{(1)},\ldots,x^{(m)}}
   \prod_{\alpha=1}^mP(x^{(\alpha)})\kappa_\beta(d_\alpha),
 \label{a:replica-full}\\
 K^{m-1}\Tr\Lambda^m
 &=\sum_{x^{(1)},\ldots,x^{(m)}}
   \prod_{\alpha=1}^mP(x^{(\alpha)})\kappa_\beta(d_\alpha)
      \mathbf1_{\{g(d_\alpha)=0\}}.
 \label{a:replica-product}
\end{align}
At every finite system size,
\begin{equation}
 D_{\mathrm{AQEC}}
 =\left.\frac{d}{dm}\log\frac{\Tr(\Lambda+B)^m}{\Tr\Lambda^m}
 \right|_{m=1}.
 \label{a:replica-derivative}
\end{equation}
\end{lemma}

\begin{proof}
Multiply the matrix blocks around the replica cycle.  Their character
product equals one because $\bigoplus_\alpha d_\alpha=0$.  The logical sum
then contributes $K$, and every error probability appears as two square
roots, giving one $P(x^{(\alpha)})$ per replica.  This proves the integer
identities.

For real $m$ near one, define the powers through the spectra of the finite
matrices.  Since both matrices have trace one, the derivative in
Eq.~\eqref{a:replica-derivative} is
$\Tr[(\Lambda+B)\log(\Lambda+B)]-\Tr(\Lambda\log\Lambda)$.
The support of $\rho_{RE}$ lies in
$\mathcal H_R\otimes\supp\rho_E$.  On this support,
$\log\Lambda=-(\log K)I+I_R\otimes\log\rho_E^T$.
Both $\Tr B=0$ and $\Tr_RB=0$, so $\Tr(B\log\Lambda)=0$.
The derivative is therefore exactly the relative entropy in
Eq.~\eqref{a:metrics-definition}.  This spectral definition supplies the
real-$m$ derivative; the integer replica sums by themselves do not justify
an analytic continuation after a thermodynamic or dilute-noise limit.
\end{proof}

The replica sums have a local classical-spin representation.  For
$z=(z_1,\ldots,z_n)\in\{\pm1\}^n$, let
$b_j(z)=\prod_{i:(H_Z)_{ji}=1}z_i$ be the classical value of check $j$.
The normalization in Eq.~\eqref{a:filter} is
$Z_\beta=2^{-n}\sum_z\exp[\sum_j\beta_jb_j(z)]$.  Since a check survives
in Eq.~\eqref{a:kernel-ratio} precisely when $(H_Zd_\alpha)_j=0$,
Eq.~\eqref{a:replica-full} becomes
\begin{align}
 K^{m-1}\Tr(\Lambda+B)^m
 &=\frac{1}{(2^nZ_\beta)^m}
 \sum_{\{x^{(\alpha)},z^{(\alpha)}\}}
 \prod_{\alpha=1}^m P(x^{(\alpha)})\notag\\[-2pt]
 &\qquad\times
 \exp\!\left\{\sum_{\alpha,j}\frac{\beta_j}{2}
 \left[1+(-1)^{(H_Zd_\alpha)_j}\right]
 b_j(z^{(\alpha)})\right\}.
 \label{a:replica-classical}
\end{align}
The expression for $K^{m-1}\Tr\Lambda^m$ inserts
$\prod_\alpha\mathbf1_{\{g(d_\alpha)=0\}}$ into the same sum.
For independent noise, $P(x^{(\alpha)})$ also factors over physical sites.
Thus Eq.~\eqref{a:replica-classical} is a classical partition function
with local check interactions between adjacent replicas.  In toric
codes, the variables $z_i$ live on edges and $b_j$ is a plaquette product,
giving the gauge-theory form.  This is the direct extension of the
replica mapping in Ref.~\cite{SI-Zhao2024} to the check structure and
nonuniform strengths considered here.  The two partition functions
share all local weights; the product comparator additionally forbids
replica differences with nonzero logical parity.

The simplest such contribution has a single flip $e$ in one replica and
no errors in the others.  The two adjacent error differences then both
equal $e$, giving the weight
$P(e)P(0)^{m-1}\kappa_\beta(e)^2$ in
Eq.~\eqref{a:replica-full}.  If $g(e)\ne0$, this configuration is excluded
from Eq.~\eqref{a:replica-product}.  Such terms occur at every site in
$\bigcup_a\ell_a$, so the replica difference includes contributions from
local errors along the entire selected support.  The single-flag calculation
below determines the local dilute coefficient while retaining the terms
needed for the $m\to1$ limit.  Accumulation at fixed nonzero noise strength
is established by the subsequent collective bounds, with correlations
between different sites included.

\subsection{A local witness and the dilute coefficient}

A lower bound does not require diagonalizing the environment state.  It is
enough to identify an observable whose mean depends on the logical input.
The following witness compares error branches differing by a fixed support;
its expectation separates the probability supplied by the noise from the
overlap left by finite-strength preparation.

\begin{lemma}[A norm-one witness for leakage]\label{lem:a4}
If $g(u)\ne0$, define the Hellinger overlap of an error distribution with
its translate~\cite{SI-Fuchs1999} by
\begin{equation}
 H_P(u)=\sum_x\sqrt{P(x)P(x\oplus u)}.
 \label{a:hellinger}
\end{equation}
Then
\begin{equation}
 \|B\|_1\ge H_P(u)\kappa_\beta(u),\qquad
 D_{\mathrm{AQEC}}\ge\frac12H_P(u)^2\kappa_\beta(u)^2.
 \label{a:single-witness}
\end{equation}
For independent bit flips of probability $p$,
$H_P(u)=[2\sqrt{p(1-p)}]^{|u|}$.
\end{lemma}

\begin{proof}
On the matrix-label space, the operator defined by
$W_u(|q\rangle_R|x\rangle_E)=\chi_q(u)|q\rangle_R|x\oplus u\rangle_E$
is unitary and has operator norm one.  Since $g(u)\ne0$, the corresponding
entries of $\Lambda$ vanish.  Equation~\eqref{a:aqec-entry} gives
\begin{equation}
 \Tr(W_uB)=\frac1K\sum_{q,x}\chi_q(u)^2
 \sqrt{P(x)P(x\oplus u)}\kappa_\beta(u)
 =H_P(u)\kappa_\beta(u).
 \label{a:witness-expectation}
\end{equation}
Trace-norm duality~\cite[Sec.~1.1.3]{SI-Watrous2018} gives the first bound, and quantum Pinsker with natural logarithms~\cite[Theorem~5.38]{SI-Watrous2018} gives $D_{\mathrm{AQEC}}\ge\|B\|_1^2/2$.
For independent noise, the overlap factors over the sites of $u$; sites
outside $u$ contribute one.
\end{proof}

For a toric edge, \cref{lem:subsystem-variance,lem:a4} give the direct
connection to the complete noise channel:
\begin{equation}
 \|B\|_1\ge2\sqrt{p(1-p)}\,\varepsilon_{\rm SV}(\{e\}),\qquad
 D_{\mathrm{AQEC}}\ge2p(1-p)\varepsilon_{\rm SV}(\{e\})^2.
 \label{a:sv-aqec}
\end{equation}
For a general CSS encoding the common lower bounds remain those in terms
of $\kappa_\beta(e)$; an upper bound on subsystem variance would be needed
to replace $\kappa_\beta(e)$ by the variance itself in that more general case.
For bounded check incidence and uniformly bounded strengths, any site
with $g(e)\ne0$ has a positive, size-independent kernel lower bound.
Equations~\eqref{a:kernel-bounds} and~\eqref{a:single-witness} then give
positive norm and entropy defects at every fixed $0<p<1$.  The norm defect obstructs recovery directly through
\cref{lem:a8}, even when the number of selected logical qubits grows.

An exact environmental calculation makes the dilute expansion transparent.
For $g(e)\ne0$, let $S=g(e)\cdot q$ be the corresponding uniformly
random reference parity and retain the environment flag $E_e$.  This is
a partial trace of the environment of the original full noise channel.
In the error-flag basis $\{|0\rangle_{E_e},|1\rangle_{E_e}\}$, the
conditional density matrices are
\begin{equation}
 \rho_{E_e}^{\pm}=
 \begin{pmatrix}
 1-p&\pm\sqrt{p(1-p)}\,\kappa_\beta(e)\\
 \pm\sqrt{p(1-p)}\,\kappa_\beta(e)&p
 \end{pmatrix}.
 \label{a:flag-state}
\end{equation}
The superscripts $+$ and $-$ correspond to $S=0$ and $S=1$, respectively;
the flag basis records absence or presence of the physical flip and is
not a logical basis.  Tracing all other flags sets their two error indices
equal and sums their probabilities to one.  The average is $\operatorname{diag}(1-p,p)$.
Diagonalizing these two matrices and applying data processing~\cite[Theorem~5.35]{SI-Watrous2018} gives
\begin{equation}
 D_{\mathrm{AQEC}}\ge I(S:E_e)
 =h_2(p)-h_2\!\left(
 \frac{1-\sqrt{(1-2p)^2+4p(1-p)\kappa_\beta(e)^2}}2\right)
 \ge2p(1-p)\kappa_\beta(e)^2,
 \label{a:flag-information}
\end{equation}
where $h_2(z)=-z\log z-(1-z)\log(1-z)$.  All entropy bounds refer to the
same $D_{\mathrm{AQEC}}$ of the full channel; $I(S:E_e)$ is only the
intermediate mutual information produced by discarding other systems.
The final inequality follows from Pinsker applied to each conditional
matrix and its average.

This calculation also fixes the local coefficient in a dilute replica
expansion.  At fixed $0<p<1/2$, expansion in $\kappa_\beta(e)$ gives
\begin{equation}
 I(S:E_e)=
 \frac{p(1-p)}{1-2p}\log\frac{1-p}{p}\,\kappa_\beta(e)^2
 +O_p(\kappa_\beta(e)^4).
 \label{a:flag-t-expansion}
\end{equation}
Indeed, the two conditional eigenvalues are
$1-p+\frac{p(1-p)}{1-2p}\kappa_\beta(e)^2+O_p(\kappa_\beta(e)^4)$
and $p-\frac{p(1-p)}{1-2p}\kappa_\beta(e)^2+O_p(\kappa_\beta(e)^4)$.
The corresponding replica entropy difference is
$(m-1)^{-1}\log\{\Tr[(\rho_{E_e}^{+})^m]/[(1-p)^m+p^m]\}$.
Its coefficient of $\kappa_\beta(e)^2$ is
\begin{equation}
 \frac{mp(1-p)}{(m-1)(1-2p)}
 \frac{(1-p)^{m-1}-p^{m-1}}{(1-p)^m+p^m}
 \xrightarrow[m\to1]{}
 \frac{p(1-p)}{1-2p}\log\frac{1-p}{p}.
 \label{a:replica-coefficient}
\end{equation}
Both $p$ and $p^m$ occur before the $m\to1$ limit.  Dropping $p^m$ at
fixed integer $m>1$ creates a spurious $m/(m-1)$ singularity.  Keeping
both terms gives the finite coefficient proportional to $p\log(1/p)$.
Equivalently, at fixed overlap,
\begin{align}
 I(S:E_e)
 &=p\kappa_\beta(e)^2\log(1/p)\notag\\
 &\quad+p\left\{\kappa_\beta(e)^2+
 [1-\kappa_\beta(e)^2]\log[1-\kappa_\beta(e)^2]\right\}
 +O(p^2\log(1/p)).
 \label{a:flag-p-expansion}
\end{align}
We use $0\log0=0$.  At $p=1/2$, the quadratic coefficient in
Eq.~\eqref{a:flag-t-expansion} is its continuous value $1/2$.
A local expansion of this kind identifies the leakage mechanism but
cannot be summed over an increasing number of positions without
controlling correlations and information saturation.  The following
collective bounds perform that step nonperturbatively.

\subsection{Independent parities and collective anchors}

There are two ways for local leakage to accumulate.  Different locations may
reveal independent logical parities, or they may provide repeated evidence
about the same parity.  The first allows an additive information bound.  The
second requires a collective witness that retains correlations and respects
the finite information available in each logical degree of freedom.

\begin{lemma}[Independent logical parities]\label{lem:a5}
For a physical set $A$, choose $e_1,\ldots,e_r\in A$ whose columns
$g(e_a)$ form a basis of the span of $\{g(e):e\in A\}$.
Under independent bit-flip noise,
\begin{equation}
 D_{\mathrm{AQEC}}\ge2p(1-p)\sum_{a=1}^r\kappa_\beta(e_a)^2.
 \label{a:independent-parities}
\end{equation}
If $k\ge1$, every physical qubit meets at most $\Delta_Z$ measured
checks, and $0\le\beta_j\le\beta_0$ for all $j$, choosing $A=[n]$ gives
\begin{equation}
 \frac{D_{\mathrm{AQEC}}}{k}\ge2p(1-p)e^{-2\Delta_Z\beta_0}.
 \label{a:density-lower}
\end{equation}
\end{lemma}

\begin{proof}
The bits $S_a=g(e_a)\cdot q$ are independent and uniform because the
columns are linearly independent and $q$ is uniform. The mutual-information chain rule, data processing, and nonnegativity of conditional mutual information~\cite[Ch.~5]{SI-Watrous2018} give
\begin{align}
 I(R:E)&\ge I(S_1\cdots S_r:E)\notag\\
 &=\sum_a\bigl[I(S_a:E)+I(S_a:S_1\cdots S_{a-1}\mid E)\bigr]
 \ge\sum_a I(S_a:E_{e_a}).
 \label{a:parity-chain}
\end{align}
Here independence removes $I(S_a:S_1\cdots S_{a-1})$, and the conditional
mutual information is nonnegative.  Equation~\eqref{a:flag-information}
bounds each term.  The environment flags themselves need not be
independent.

The physical binary rows $\ell_a$ are linearly independent: a relation
giving the empty physical support would contradict independence of their
logical classes.  The full matrix with columns $g(e)$ therefore has rank
$k$.  For $A=[n]$ one may choose $r=k$, and
Eq.~\eqref{a:kernel-bounds} gives
$\kappa_\beta(e_a)\ge
\exp[-\sum_{j\in\sigma(e_a)}\beta_j]\ge e^{-\Delta_Z\beta_0}$.
The same argument applies to any subset whose incident checks obey the
strength bound, as used for the enhanced-footprint converse in
Eq.~\eqref{b:register-footprint-lower}.
\end{proof}

\begin{lemma}[Many anchors for one logical parity]\label{lem:a6}
Suppose $M\ge1$ sites $e_1,\ldots,e_M$ have the same nonzero column $g(e_i)=g_0$,
their syndrome supports are pairwise disjoint, and each syndrome contains
$s$ checks.  Suppose all checks in these syndromes have strength $\beta$;
checks elsewhere may have arbitrary nonnegative strengths.  Under independent noise with
$a=2\sqrt{p(1-p)}$,
\begin{equation}
 D_{\mathrm{AQEC}}\ge
 \frac{a^2Me^{-2s\beta}}{2[1+2^{2s}a^2Me^{-2s\beta}]}.
 \label{a:collective-entropy}
\end{equation}
The constants are independent of system size, logical dimension, and
preparation strength.
\end{lemma}

\begin{proof}
Measure Pauli $X$ on the selected environment flags, obtaining
$Y_i=\pm1$, and set $S=(-1)^{g_0\cdot q}$.  Because $g_0\ne0$ and $q$
is uniform, the two signs of $S$ are equally probable.  In this proof,
$\langle\cdot\rangle$ denotes a classical average over the joint measurement
distribution $P(q,Y)=K^{-1}P(Y\mid q)$, while
$\langle f(Y)\rangle_q=\sum_Y P(Y\mid q)f(Y)$ fixes the logical label.
For any subset $A$ of the selected physical sites, direct summation of
Eq.~\eqref{a:aqec-entry} yields
\begin{equation}
 \left\langle\prod_{e_i\in A}Y_i\right\rangle_q
 =a^{|A|}\chi_q(A)\kappa_\beta(A).
 \label{a:flag-moments}
\end{equation}
Each flipped environment index contributes $2\sqrt{p(1-p)}$; every
unflipped index contributes one.  With $T=\sum_iY_i$,
\begin{align}
 \langle ST\rangle&=a\sum_i\kappa_\beta(e_i)\ge aMe^{-s\beta},
 \label{a:first-moment}\\
 \langle T^2\rangle&=M+a^2\sum_{i\ne j}\kappa_\beta(e_i\oplus e_j)
 \le M+a^2M(M-1)\sech^{2s}\beta.
 \label{a:second-moment}
\end{align}
The pair bound follows from disjoint syndrome supports and
Eq.~\eqref{a:kernel-bounds}; it does not factor the joint environment state.

The uniform binary prior gives
\begin{align}
 I(S:Y)
 &=\sum_Y P(Y)\left[\log2-h_2\!\left(P(S=1\mid Y)\right)\right]\notag\\
 &\ge\frac12\sum_Y P(Y)
       [P(S=1\mid Y)-P(S=-1\mid Y)]^2\notag\\
 &\ge\frac{\langle ST\rangle^2}{2\langle T^2\rangle}.
 \label{a:posterior-bound}
\end{align}
The first inequality is the binary Pinsker bound~\cite[Theorem~5.38]{SI-Watrous2018}.  The second is
Cauchy--Schwarz applied to
$\langle ST\rangle=\sum_Y T(Y)[P(S=1,Y)-P(S=-1,Y)]$.
Data processing gives $D_{\mathrm{AQEC}}\ge I(S:Y)$.
Substitution of the two moments, followed
by $\sech\beta\le2e^{-\beta}$ and $M-1\le M$, proves the bound.
\end{proof}

For standard straight toric representatives, distinct collinear edges do
not share a plaquette.  Thus $M=L$, with $s=2$ in two dimensions and $s=4$
in three dimensions.  The same anchors work when all two or three toric
logical qubits are selected: the other coordinate-direction representatives
do not use these physical edges.  At any fixed $0<p<1$,
Eq.~\eqref{a:collective-entropy} shows that $D_{\mathrm{AQEC}}\to0$
requires respectively
\begin{equation}
 \beta-\frac14\log L\longrightarrow+\infty\quad(\mathrm{2D}),
 \qquad
 \beta-\frac18\log L\longrightarrow+\infty\quad(\mathrm{3D}).
 \label{a:necessary-toric}
\end{equation}
These are necessary conditions; the coefficients are not asserted to be
optimal.  The denominator in Eq.~\eqref{a:collective-entropy} is essential:
many anchors observe the same binary variable, whose information cannot
grow without bound.

The same argument supplies a necessary strength for local enhancement.
If each selected edge lies on the enhanced representative and all checks
incident on these edges have strength $\beta_R$, then both
Eq.~\eqref{a:first-moment} and Eq.~\eqref{a:second-moment} use only
$\beta_R$.  The kernel bounds on these syndromes do not involve checks in the
bulk.  Taking the same $M=L$ disjoint collinear anchors gives
\begin{equation}
 \beta_R-\frac14\log L\longrightarrow+\infty\quad(\mathrm{2D}),
 \qquad
 \beta_R-\frac18\log L\longrightarrow+\infty\quad(\mathrm{3D})
 \label{a:necessary-toric-enhanced}
\end{equation}
as necessary conditions for vanishing total AQEC entropy, even when the
bulk strength is kept fixed.  Thus the necessary and sufficient local
enhancement strengths agree at logarithmic order; no matching of their
leading coefficients is asserted.

A concrete example explains why the collective estimate is necessary.
For the two-dimensional code, take $\beta(L)=\frac14\log L$.
Equation~\eqref{a:sv-toric} gives
\begin{equation}
 \max_e\varepsilon_{\rm SV}(\{e\})
 =\frac{\sech^2\beta(L)}{1+\tanh^{L^2}\beta(L)}
 \sim\frac4{\sqrt L}\longrightarrow0.
 \label{a:sv-vanishing-example}
\end{equation}
Nevertheless $Le^{-4\beta(L)}=1$, so
Eq.~\eqref{a:collective-entropy} gives
\begin{equation}
 D_{\mathrm{AQEC}}\ge
 \frac{2p(1-p)}{1+64p(1-p)}>0
 \qquad(0<p<1\text{ fixed}).
 \label{a:sv-global-counterexample}
\end{equation}
Thus no size-independent upper bound on the full AQEC entropy can be a
function only of single-site subsystem variance that vanishes when that
variance vanishes.  Many individually weak local signals can leave a
finite total correlation with the environment.  This example applies
also at any sufficiently small fixed noise probability, within the
regime where recovery upper bounds are relevant.

\begin{lemma}[Many anchors at arbitrary logical dimension]\label{lem:a7}
The common-column requirement can be removed for a trace-distance bound.
For $M\ge1$ sites satisfying $g(e_i)\ne0$, pairwise disjoint syndrome supports,
and $|\sigma(e_i)|\le\Delta_Z$, uniform strength and independent noise give
\begin{equation}
 \eps=\frac12\|B\|_1\ge
 \frac{a^2Me^{-2\Delta_Z\beta}}
 {4[1+2^{2\Delta_Z}a^2Me^{-2\Delta_Z\beta}]},
 \qquad a=2\sqrt{p(1-p)}.
 \label{a:register-collective}
\end{equation}
\end{lemma}

\begin{proof}
Let $P(q,Y)$ denote the joint classical distribution obtained by measuring
the reference in its classical basis and the chosen environment flags in
the $X$ basis.  Its product comparator is
$Q(q,Y)=K^{-1}P(Y)$, where $P(Y)=\sum_qP(q,Y)$.  Use the statistic
\begin{equation}
 F(q,Y)=\sum_i\chi_q(e_i)Y_i,\qquad
 b_i=e^{-\beta|\sigma(e_i)|}.
 \label{a:register-statistic}
\end{equation}
We write $\langle F\rangle_P=\sum_{q,Y}P(q,Y)F(q,Y)$, and similarly
for $F^2$ and the distribution $Q$.  The first moments and character
orthogonality give
\begin{equation}
 \langle F\rangle_P=a\sum_i\kappa_\beta(e_i)\ge a\sum_ib_i,
 \qquad\langle F\rangle_Q=0.
 \label{a:register-first}
\end{equation}
For the second moments,
\begin{align}
 \langle F^2\rangle_Q
 &=M+a^2\!\sum_{\substack{i\ne j\\g(e_i)=g(e_j)}}
       \kappa_\beta(e_i\oplus e_j)\notag\\
 &\le\langle F^2\rangle_P
 =M+a^2\sum_{i\ne j}\kappa_\beta(e_i\oplus e_j)
 \le M+2^{2\Delta_Z}a^2\left(\sum_ib_i\right)^2.
 \label{a:register-second}
\end{align}
Nonnegativity of the kernel proves the first inequality.  Disjoint
syndromes and $\sech\beta\le2e^{-\beta}$ bound each pair kernel by
$2^{2\Delta_Z}b_ib_j$.

For any two distributions and real statistic, Cauchy--Schwarz with weight
$|P-Q|$ gives
\begin{equation}
 |\langle F\rangle_P-\langle F\rangle_Q|^2
 \le2\operatorname{TV}(P,Q)
          (\langle F^2\rangle_P+\langle F^2\rangle_Q).
 \label{a:tv-second-moment}
\end{equation}
Here $\operatorname{TV}(P,Q)=\tfrac12\sum_{q,Y}|P(q,Y)-Q(q,Y)|$.
Trace-distance contractivity under the measurement channel~\cite[Sec.~3.3]{SI-Watrous2018} then yields
\begin{equation}
 \eps\ge\operatorname{TV}(P,Q)
 \ge\frac{a^2(\sum_ib_i)^2}
 {4[M+2^{2\Delta_Z}a^2(\sum_ib_i)^2]}.
 \label{a:register-tv}
\end{equation}
The right-hand side increases with $(\sum_ib_i)^2$.  Substituting
$\sum_ib_i\ge Me^{-\Delta_Z\beta}$ proves the claim.
\end{proof}

Let $G_Z$ join physical qubits sharing a measured check, and suppose its
maximum degree is $D$.  Set $U=|\bigcup_a\ell_a|$, as in the main text.
The induced graph on the union of selected representatives has an independent
set of size at least $U/(D+1)$.  Its vertices have disjoint syndrome supports, and
all have nonzero logical column.  For fixed $0<p<1$ and size-independent bounds $D$ and $\Delta_Z>0$,
vanishing $\eps$ therefore requires
\begin{equation}
 \beta-\frac{\log U}{2\Delta_Z}\longrightarrow+\infty.
 \label{a:necessary-register}
\end{equation}
This argument makes no fixed-$k$ assumption.  At constant encoding rate,
$U\ge k=\Theta(n)$, so the necessary strength is logarithmic
in $n$.  For general selected encodings, $U$ cannot be replaced
by $n$: adding ancillary qubits outside the logical supports does not add
vulnerable anchors.  This obstruction concerns full-register trace-distance
decoupling; it is not a necessary condition for vanishing entropy density
alone.

\subsection{Conversion to recovery and entropy density}

Trace-distance decoupling controls recovery of the entire selected register
uniformly in its dimension.  Conversion to total AQEC entropy depends on
that dimension, so total entropy and entropy per logical qubit must be
kept distinct when $k$ grows.

\begin{lemma}[Recovery and entropy at arbitrary rate]\label{lem:a8}
Let $F_e^\star$ be the optimal entanglement fidelity for the maximally mixed
input on the entire selected register, with squared fidelity convention.
The entropy-density statements assume $k\ge1$. For an empty register,
$K=1$ and $B=D_{\mathrm{AQEC}}=0$, so no density is defined.
The fidelity obeys the fixed-input duality in \cref{eq:fixed-input-duality} and the bounds
\begin{equation}
 \frac{\eps^2}{4}\le1-F_e^\star
 \le2\eps-\eps^2\le2\eps.
 \label{a:recovery-bounds}
\end{equation}
For the classical--quantum reference--environment state of this model,
\begin{equation}
 D_{\mathrm{AQEC}}\le\eps\log K
 +(1+\eps)h_2\!\left(\frac{\eps}{1+\eps}\right).
 \label{a:entropy-upper}
\end{equation}
Consequently $\eps\to0$ is equivalent to $F_e^\star\to1$ and implies
$D_{\mathrm{AQEC}}/k\to0$.  The stronger condition $k\eps\to0$ implies
$D_{\mathrm{AQEC}}\to0$.  For fixed $k$, vanishing AQEC entropy,
vanishing $\eps$, and $F_e^\star\to1$ are therefore equivalent.
\end{lemma}

\begin{proof}
First prove \cref{eq:fixed-input-duality}.  Fix an environment state
$\xi_E$ and a purification $|\chi\rangle_{FE}$.  Uhlmann's
theorem~\cite{SI-Uhlmann1976} optimizes the overlap of a purification of
$\rho_{RE}$ with $|\Phi_K\rangle_{RL}|\chi\rangle_{FE}$ by an
isometry from the physical output $Q$ to $LF$, enlarging $F$ when
necessary.  Regard this isometry as a recovery dilation.  The resulting
entanglement fidelity, which sums over all states of $FE$, is at least
this one squared overlap.  Maximizing over $\xi_E$ proves the
$\ge$ direction.  Conversely, for any recovery dilation let
$|\Psi\rangle_{RLFE}$ be its pure output and put
$|v\rangle_{FE}=(\langle\Phi_K|_{RL}\otimes I_{FE})|\Psi\rangle$.
Its squared norm is that recovery's entanglement fidelity.  If it is
nonzero, normalize $|v\rangle$ to $|\chi\rangle$ and set
$\xi_E=\Tr_F|\chi\rangle\langle\chi|$.  The overlap with
$|\Phi_K\rangle|\chi\rangle$ is then exactly $\|v\|$, so Uhlmann's
theorem bounds this recovery fidelity by the right-hand side of
\cref{eq:fixed-input-duality}.  A zero fidelity is immediate.  Taking
the supremum over recoveries proves equality.

For squared fidelity
$F(\rho,\sigma)=\|\sqrt\rho\sqrt\sigma\|_1^2$, the
Fuchs--van de Graaf inequality~\cite{SI-Fuchs1999} gives
\begin{equation}
 F\!\left(\rho_{RE},\frac{I_R}{K}\otimes\rho_E\right)
 \ge(1-\eps)^2.
 \label{a:decoupling-fidelity}
\end{equation}
Choosing $\xi_E=\rho_E$ in Eq.~\eqref{eq:fixed-input-duality} gives
$F_e^\star\ge(1-\eps)^2$ and hence the upper infidelity bound.

Conversely, let a recovery have entanglement infidelity $r$.
Uhlmann's theorem applied to its output and the desired maximally entangled
state gives some environment state $\xi_E$ for which
\begin{equation}
 \frac12\left\|\rho_{RE}-\frac{I_R}{K}\otimes\xi_E\right\|_1
 \le\sqrt r,\qquad
 \frac12\|\rho_E-\xi_E\|_1\le\sqrt r.
 \label{a:converse-decoupling}
\end{equation}
The second inequality follows by partial trace.  The triangle inequality
then gives $\eps\le2\sqrt r$, proving the lower bound on optimal
infidelity.

Finally, both $\rho_{RE}$ and $I_R/K\otimes\rho_E$ are classical on the
same $K$-dimensional reference.  The classical--quantum case of Lemma~2 in
Ref.~\cite{SI-Winter2016}, with its systems $A=R$ and $B=E$, bounds their
conditional-entropy difference by
$\eps\log K+(1+\eps)h_2(\eps/(1+\eps))$.  Using
$I(R:E)=\log K-S(R|E)$ gives Eq.~\eqref{a:entropy-upper}.
Here the classical reference permits the coefficient $\log K$, rather
than the $2\log K$ of the general quantum bound.  Since
$\log K=k\log2$, division by $k$ proves the density statement;
$k\eps\to0$ controls the total entropy as well.
\end{proof}

These bounds concern entanglement fidelity for the full maximally mixed
logical input; the proof does not replace it with a worst-case recovery
criterion.  Entropy density alone is weaker even than this full-register
criterion.  For example, complete dephasing of one logical qubit and the
identity channel on the other $k-1$ qubits give
\begin{equation}
 D_{\mathrm{AQEC}}=\log2,\qquad
 \frac{D_{\mathrm{AQEC}}}{k}\longrightarrow0,\qquad
 F_e^\star=\frac12.
 \label{a:density-counterexample}
\end{equation}
This example explains why the rate-dependent statements retain separate
conditions for entropy density, total leakage, and recovery of the entire
logical register.

\refstepcounter{section}
\section*{Supplementary Note 3: Local enhancement: operator estimates and geometry}
\label{app:enhancement}

Local enhancement suppresses the local leakage identified in
Supplementary Note~2. Errors with syndrome on an enhanced check have a small
residual overlap, while silent errors with nonzero logical parity must grow
across the protected neighborhood. We first bound their contributions to the
environment state under projective enhancement, and then treat finite
measurement strength. The estimates apply to correlated noise; the code
geometry enters through the minimum size of a silent component.

All qubit supports are identified with vectors over $\mathbb F_2$; addition of
supports is symmetric difference, denoted by $\oplus$. The probability of the
physical $X$-error support $x$ is $P(x)$. We assume local stochasticity~\cite{SI-Fawzi2018},
\begin{equation}
 \Pr[A\subseteq x]\le \eta^{|A|}
 \quad\text{for every fixed qubit set }A,
 \qquad 0\le\eta\le1.
 \label{b:local-stochastic}
\end{equation}
The distribution $P$ may contain correlations.  Local stochasticity controls
the probability of a prescribed support while allowing arbitrary dependence
among errors elsewhere.

Let $G_Z$ be the graph whose vertices are physical qubits, with an edge between
two distinct qubits whenever they belong to a common measured $Z$ check. We use
connectedness and graph distance in this graph. Its degree is bounded uniformly
for the code families considered here. Write
\begin{equation}
 \sigma(x)=\supp(H_Zx),\qquad
 g(x)=\bigl(|x\cap\ell_1|,\ldots,|x\cap\ell_k|\bigr)\pmod2,
 \qquad
 \chi_q(x)=(-1)^{q\cdot g(x)}.
 \label{b:notation}
\end{equation}
Here $\ell_1,\ldots,\ell_k$ are the selected independent $Z$-logical
representatives, $q\in\mathbb F_2^k$, $K=2^k$, and
$\ell(q)=\bigoplus_a q_a\ell_a$. The protected measured checks form a set
$\cP_r$. A support is \emph{silent in the protected region} when
$\sigma(x)\cap\cP_r=\varnothing$. This condition need not make its syndrome
vanish outside that region.

\subsection{Projection and the exact connected-component expansion}

Hard projection isolates the enhancement mechanism in its simplest form:
any error with a protected syndrome has zero overlap within the encoded
subspace.  Decomposing the remaining errors into connected components then
separates those that can carry logical information from those that cannot.

\begin{lemma}[Hard projection and logical sectors]
\label{lem:b1}
For the finite-strength state $|\psi_{\beta}\rangle$, define
\begin{equation}
 \Pi_r=\prod_{j\in\cP_r}\frac{I+B_j}{2},\qquad
 |\psi_r\rangle=
 \frac{\Pi_r|\psi_{\beta}\rangle}
 {\sqrt{\langle\psi_{\beta}|\Pi_r|
                 \psi_{\beta}\rangle}},\qquad
 |q\rangle_r=Z^{\ell(q)}|\psi_r\rangle.
 \label{b:hard-state}
\end{equation}
These states are well defined and the states $|q\rangle_r$ are orthonormal.
With $\kappa_r(x)=\langle\psi_r|X_x|\psi_r\rangle$,
\begin{align}
 {}_r\langle q|X_x|q'\rangle_r
 &=\delta_{q,q'}\chi_q(x)\kappa_r(x),
 \label{b:hard-sector}\\
 \sigma(x)\cap\cP_r\ne\varnothing
 &\quad\Longrightarrow\quad\kappa_r(x)=0.
 \label{b:hard-cancellation}
\end{align}
\end{lemma}

\begin{proof}
All measured checks are positive-phase $Z$ operators. The computational-basis
state $|0\rangle^{\otimes n}$ therefore has eigenvalue $+1$ for every $B_j$.
Its amplitude in the finite-strength state is nonzero, so the denominator
in \eqref{b:hard-state} is strictly positive. On the range of $\Pi_r$, each
protected filter $\exp(\beta_jB_j/2)$ acts as the scalar
$\exp(\beta_j/2)$. Consequently $|\psi_r\rangle$ depends only on the exterior
measurement strengths and the protected projectors, not on the finite
strengths used on protected checks before projection.

Choose logical $X$ operators dual to the selected $Z$ representatives. They
commute with every $Z$-check filter and with $\Pi_r$, and they stabilize the
initial state $|+\rangle^{\otimes n}$. Their eigenvalues on $|q\rangle_r$ are
$(-1)^{q_a}$. Distinct logical labels thus lie in orthogonal joint eigenspaces.
Every physical $X_x$ commutes with these logical $X$ operators, so its
matrix elements between distinct $q$ sectors vanish. For the diagonal matrix
element, commute $X_x$ through $Z^{\ell(q)}$ to obtain the sign
$\chi_q(x)$ in \eqref{b:hard-sector}.

The same argument covers a selected subset of all logical qubits. The
unnormalized filtered states belonging to its basis labels have the same norm,
because $Z^{\ell(q)}$ is unitary and commutes with the filter and projector.
Their orthogonality then makes the branch success probability independent of
an arbitrary logical superposition. Hence the normalized encoding is an
isometry, rather than an input-dependent postselection map.

Finally, if $j\in\sigma(x)\cap\cP_r$, then $B_j$ anticommutes with $X_x$,
and
\begin{equation}
 (I+B_j)X_x(I+B_j)=0.
 \label{b:projector-zero}
\end{equation}
All protected projectors commute. Their product therefore satisfies
$\Pi_rX_x\Pi_r=0$, which proves \eqref{b:hard-cancellation}.
\end{proof}

\begin{lemma}[Syndrome separation of connected components]
\label{lem:b2}
If $x=\bigsqcup_i\Gamma_i$ is its connected-component decomposition in $G_Z$,
then
\begin{equation}
 \sigma(x)=\bigsqcup_i\sigma(\Gamma_i),\qquad
 \chi_q(x)=\prod_i\chi_q(\Gamma_i).
 \label{b:component-separation}
\end{equation}
Thus a protected-syndrome-silent error has only silent components. If it
distinguishes $q$ from $q'$, an odd number of its components distinguish that
same pair.
\end{lemma}

\begin{proof}
A measured check cannot touch two distinct components: a qubit of one and a
qubit of the other would then be adjacent in $G_Z$. The sets of checks touched
by distinct components are consequently disjoint, so their syndromes cannot
cancel each other. This proves the first identity. Intersection parity adds
on disjoint supports, proving the second. In particular,
$\chi_q(x)\chi_{q'}(x)$ is the product of a factor $-1$ for each component
that distinguishes the two labels and a factor $+1$ for each other component.
\end{proof}

For a logical pair $q,q'$, define
\begin{align}
 \cC(q,q')&=\{\Gamma:\ \Gamma\text{ is connected in }G_Z,
                         \ \chi_q(\Gamma)\ne\chi_{q'}(\Gamma)\},\\
 \cC_r(q,q')&=\{\Gamma\in\cC(q,q'):
                         \sigma(\Gamma)\cap\cP_r=\varnothing\}.
 \label{b:component-classes}
\end{align}
A family of connected supports is \emph{compatible} if its members are
pairwise disjoint and no edge of $G_Z$ joins distinct members. This is precisely
the condition for its members to occur as complete connected components of one
error difference. We write $\mathrm{Comp}(x)$ for the set of connected
components of $x$.

\begin{lemma}[Parity-family identity]
\label{lem:b3}
Define the pairwise environment matrix
\begin{equation}
 (D_r^{q,q'})_{uv}=\sqrt{P(u)P(v)}
 [\chi_q(u\oplus v)-\chi_{q'}(u\oplus v)]\kappa_r(u\oplus v).
 \label{b:pair-matrix}
\end{equation}
For each compatible, nonempty family $F\subseteq\cC_r(q,q')$, define
\begin{equation}
 (D_{r,F}^{q,q'})_{uv}
 =(-2)^{|F|}\sqrt{P(u)P(v)}\,
   \chi_{q'}(u\oplus v)\kappa_r(u\oplus v)
   \mathbf1\{F\subseteq\mathrm{Comp}(u\oplus v)\}.
 \label{b:family-matrix}
\end{equation}
Then the following is an exact matrix identity:
\begin{equation}
 D_r^{q,q'}=
 \sum_{\substack{\varnothing\ne F\subseteq\cC_r(q,q')\\
                  F\text{ compatible}}}D_{r,F}^{q,q'}.
 \label{b:family-identity}
\end{equation}
\end{lemma}

\begin{proof}
For a fixed error difference $x$, let $A(x)$ be the components of $x$ that
distinguish $q$ from $q'$. By character multiplication,
\begin{align}
 \chi_q(x)-\chi_{q'}(x)
 &=\chi_{q'}(x)
      \left[\prod_{\Gamma\in A(x)}(1-2)-1\right]\\
 &=\chi_{q'}(x)
      \sum_{\varnothing\ne F\subseteq A(x)}(-2)^{|F|}.
 \label{b:parity-expansion}
\end{align}
When $\kappa_r(x)\ne0$, Lemmas~\ref{lem:b1} and~\ref{lem:b2} imply that
all components of $x$ are silent in $\cP_r$. Every member of $A(x)$ then
belongs to $\cC_r(q,q')$. Actual components are compatible, and the indicator
in \eqref{b:family-matrix} selects exactly the nonempty subsets of $A(x)$.
Multiplying \eqref{b:parity-expansion} by
$\sqrt{P(u)P(v)}\kappa_r(x)$ with $x=u\oplus v$ proves the claimed identity
entry by entry. If $\kappa_r(x)=0$, both sides vanish. The signed
inclusion--exclusion coefficients are retained throughout this step.
\end{proof}

\subsection{A trace-norm estimate for each family}

For each family, a cross-Gram representation bounds the trace norm after
summing over all exterior errors. Local stochasticity then controls the
probability of the prescribed support, without conditioning the noise on the
exterior configuration.

\begin{lemma}[Cross-Gram estimate]
\label{lem:b4}
Every compatible, nonempty family satisfies
\begin{equation}
 \|D_{r,F}^{q,q'}\|_1
 \le \prod_{\Gamma\in F}(C_0\sqrt\eta)^{|\Gamma|},
 \qquad C_0=4.
 \label{b:cross-gram-bound}
\end{equation}
The constant is independent of system size and logical dimension.
\end{lemma}

\begin{proof}
For $\eta=0$, \eqref{b:local-stochastic} implies that the empty error has
probability one, and the claim is immediate. Assume $\eta>0$. Put
$S=\bigcup_{\Gamma\in F}\Gamma$ and define its external vertex boundary by
\begin{equation}
 \partial_GS=\{e\notin S:\ e\text{ has a neighbor in }S\}.
 \label{b:external-boundary}
\end{equation}
Compatibility gives the exact equivalence
\begin{equation}
 F\subseteq\mathrm{Comp}(u\oplus v)
 \quad\Longleftrightarrow\quad
 (u\oplus v)|_S=\mathbf1,\qquad
 (u\oplus v)|_{\partial_GS}=0.
 \label{b:component-event}
\end{equation}
No condition is imposed outside $S\cup\partial_GS$.

Partition the permitted matrix entries into assignments
\begin{equation}
 u|_S=z,\quad v|_S=\mathbf1-z,\qquad
 u|_{\partial_GS}=v|_{\partial_GS}=w.
 \label{b:local-assignments}
\end{equation}
For fixed $z,w$, denote the allowed row labels by $\mathcal U_{z,w}$ and
column labels by $\mathcal V_{z,w}$. The character
$\chi_{q'}(u\oplus v)=\chi_{q'}(u)\chi_{q'}(v)$ merely multiplies rows
and columns by signs and therefore does not change singular values. After
removing these signs, the submatrix is $(-2)^{|F|}A^\dagger C$, where
\begin{equation}
 A=\sum_{u\in\mathcal U_{z,w}}\sqrt{P(u)}
       |X_u\psi_r\rangle\langle u|,\qquad
 C=\sum_{v\in\mathcal V_{z,w}}\sqrt{P(v)}
       |X_v\psi_r\rangle\langle v|.
 \label{b:cross-gram-operators}
\end{equation}
Indeed, $\kappa_r(u\oplus v)=\langle X_u\psi_r|X_v\psi_r\rangle$.
The vectors in these operators have unit norm, giving
$\|A\|_2^2=\Pr[x\in\mathcal U_{z,w}]$ and
$\|C\|_2^2=\Pr[x\in\mathcal V_{z,w}]$. Schatten H\"older~\cite[Sec.~1.1.3]{SI-Watrous2018} thus yields
\begin{equation}
 \|(-2)^{|F|}A^\dagger C\|_1
 \le2^{|F|}\sqrt{\Pr[x\in\mathcal U_{z,w}]
                         \Pr[x\in\mathcal V_{z,w}]}.
 \label{b:schatten-holder}
\end{equation}
These probabilities already include the sum over every exterior error
configuration.

Sum the boundary assignments before applying local stochasticity. For fixed
$z$, Cauchy--Schwarz gives
\begin{align}
 &\sum_w\sqrt{\Pr[x\in\mathcal U_{z,w}]
                    \Pr[x\in\mathcal V_{z,w}]}\nonumber\\
 &\qquad\le
 \sqrt{\Pr[x|_S=z]\Pr[x|_S=\mathbf1-z]}
 \le\sqrt{\eta^{|z|}\eta^{|S|-|z|}}
 =\eta^{|S|/2}.
 \label{b:boundary-sum}
\end{align}
Here the event $x|_S=z$ requires occupation of the $|z|$ specified sites, so
\eqref{b:local-stochastic} bounds its probability by $\eta^{|z|}$; requiring
additional unoccupied sites can only reduce that probability. The complementary
assignment is treated in the same way. This argument neither conditions on the
boundary nor assumes that boundary and interior errors are independent.

Use the triangle inequality to sum the assignment submatrices, sum all
$2^{|S|}$ choices of $z$, and use $|F|\le|S|$:
\begin{equation}
 \|D_{r,F}^{q,q'}\|_1
 \le2^{|F|+|S|}\eta^{|S|/2}
 \le(4\sqrt\eta)^{|S|}.
 \label{b:family-weight}
\end{equation}
The supports in $F$ are disjoint, so $|S|=\sum_{\Gamma\in F}|\Gamma|$,
which proves \eqref{b:cross-gram-bound}. Summing $w$ by
\eqref{b:boundary-sum} is also why no factor exponential in the boundary size
is necessary.
\end{proof}

\begin{lemma}[Summing the families]
\label{lem:b5}
Set
\begin{equation}
 \Delta_r(q,q')=
 \sum_{\Gamma\in\cC_r(q,q')}(C_0\sqrt\eta)^{|\Gamma|}.
 \label{b:hard-activity}
\end{equation}
Then
\begin{equation}
 \|D_r^{q,q'}\|_1\le e^{\Delta_r(q,q')}-1,
 \qquad
 \frac12\|B_r\|_1
 \le\frac1{2K^2}\sum_{q,q'}
                \bigl[e^{\Delta_r(q,q')}-1\bigr].
 \label{b:trace-average}
\end{equation}
Here $B_r$ is the normalized AQEC matrix deviation for the hard-enhanced
encoding and noise distribution $P$.
\end{lemma}

\begin{proof}
Apply the triangle inequality to the exact operator decomposition
\eqref{b:family-identity} and then use Lemma~\ref{lem:b4}:
\begin{align}
 \|D_r^{q,q'}\|_1
 &\le\sum_{\substack{\varnothing\ne F\subseteq\cC_r(q,q')\\
                      F\text{ compatible}}}
       \prod_{\Gamma\in F}(C_0\sqrt\eta)^{|\Gamma|}\nonumber\\
 &\le\prod_{\Gamma\in\cC_r(q,q')}
       \left[1+(C_0\sqrt\eta)^{|\Gamma|}\right]-1
 \le e^{\Delta_r(q,q')}-1.
 \label{b:family-resummation}
\end{align}
Compatibility is dropped only in the second line, after the contribution of
each compatible family has been bounded in trace norm. The final inequality is
$1+t\le e^t$ for $t\ge0$.

For completeness, let $\rho_E^q$ be the conditional environment state at
logical label $q$. Equation~\eqref{b:hard-sector} makes the reference--environment
state block diagonal in $q$, with conditional matrices
\begin{equation}
 (\rho_E^q)_{uv}=\sqrt{P(u)P(v)}\,
                 \chi_q(u\oplus v)\kappa_r(u\oplus v).
 \label{b:conditional-environment}
\end{equation}
Thus $D_r^{q,q'}=\rho_E^q-\rho_E^{q'}$. Since
$\rho_E=K^{-1}\sum_{q'}\rho_E^{q'}$, the deviation from the product of
the marginals is
\begin{equation}
 B_r=\bigoplus_q\frac1K(\rho_E^q-\rho_E)
    =\bigoplus_q\frac1{K^2}\sum_{q'}D_r^{q,q'}.
 \label{b:logical-blocks}
\end{equation}
The full-transpose convention for the AQEC matrices leaves all trace norms
unchanged; here the conditional matrices are also real symmetric. Trace-norm
additivity over the logical blocks, followed by the triangle inequality within
each block, proves the second bound in \eqref{b:trace-average}. It involves an
average over pairs of logical labels, not a multiplicative penalty growing
with $K$.
\end{proof}

For the geometric estimates, write the actual union of selected
representatives and its size as
\begin{equation}
 R=\bigcup_{a=1}^k\ell_a,\qquad U=|R|.
 \label{b:anchor-union}
\end{equation}
Suppose at most $\mu^m$ connected qubit sets of size $m$ contain a fixed
qubit, uniformly in system size. Such an exponential bound follows from bounded degree~\cite{SI-Fawzi2018}, with a constant $\mu$ chosen for the family. Set
\begin{equation}
 \lambda=\mu C_0\sqrt\eta<1.
 \label{b:lambda}
\end{equation}
The operator estimates themselves do not require $\lambda<1$; this
restriction makes the subsequent sum over connected supports converge
uniformly.

Every component distinguishing $q$ from $q'$ intersects
$\ell(q\oplus q')\subseteq R$. If every such protected-silent
component has size at least $M(r)$, anchoring it at a qubit of $R$
therefore gives
\begin{equation}
 \Delta_r(q,q')
 \le U\sum_{m\ge M(r)}\lambda^m
 \le\frac{U}{1-\lambda}\lambda^{M(r)}.
 \label{b:hard-geometric-sum}
\end{equation}
The sum runs over positive integers and is empty when $M(r)=\infty$.
Substituting into \eqref{b:trace-average} gives the projective part of the
main-text enhancement theorem.

\subsection{Finite measurement strength and the residual kernel}

Finite enhancement replaces exact cancellation by a small residual overlap.
The same component expansion applies because the residual kernel admits a
Gram representation. Each protected syndrome check contributes a suppression
factor to the family bound.

Let every strength obey $0\le\beta_j<\infty$, with
$\beta_j\ge\beta_R\ge0$ on $\cP_r$.

\begin{lemma}[Residual kernel and finite-strength bound]
\label{lem:b6}
Define $D_{\beta}^{q,q'}$ and
$D_{\beta,F}^{q,q'}$ by
\eqref{b:pair-matrix} and \eqref{b:family-matrix}, replacing $\kappa_r$ by
$\kappa_{\beta}$ and allowing compatible nonempty
$F\subseteq\cC(q,q')$. Then
\begin{equation}
 \|D_{\beta,F}^{q,q'}\|_1
 \le\prod_{\Gamma\in F}
 \left[(C_0\sqrt\eta)^{|\Gamma|}
       (\sech\beta_R)^{|\sigma(\Gamma)\cap\cP_r|}\right].
 \label{b:soft-family-bound}
\end{equation}
Suppose every silent distinguishing component has size at least $M(r)>0$.
If no such component exists, take $M(r)=\infty$ and $\lambda^{M(r)}=0$.
Then
\begin{align}
 \Delta_{r,\beta_R}(q,q')
 &:=\sum_{\Gamma\in\cC(q,q')}
      (C_0\sqrt\eta)^{|\Gamma|}
      (\sech\beta_R)^{|\sigma(\Gamma)\cap\cP_r|}\nonumber\\
 &\le\frac{U}{1-\lambda}
             \left[\lambda\sech\beta_R+\lambda^{M(r)}\right].
 \label{b:soft-activity}
\end{align}
The bounds in \eqref{b:trace-average} hold for the finite-strength state with
$\Delta_r$ replaced by $\Delta_{r,\beta_R}$.
\end{lemma}

\begin{proof}
The parity expansion \eqref{b:parity-expansion} holds without projection. It
gives the exact family identity over all of $\cC(q,q')$. It remains to
justify the factors involving $\sech\beta_R$ in the operator norm bound,
rather than infer them from an elementwise estimate.

Fix a compatible family $F$, and write
\begin{equation}
 S=\bigcup_{\Gamma\in F}\Gamma,\qquad
 J=\bigsqcup_{\Gamma\in F}\sigma(\Gamma).
 \label{b:family-syndrome}
\end{equation}
Lemma~\ref{lem:b2} makes the syndrome union disjoint. Fix local assignments
on $S\cup\partial_GS$ as in \eqref{b:local-assignments}. Remove those
assignments from $u$ and $v$, leaving supports $\bar u,\bar v$ outside
$S\cup\partial_GS$. No remaining qubit shares a check with $S$. Moreover the
prescribed boundary assignments cancel in $u\oplus v$. Consequently
\begin{equation}
 \sigma(u\oplus v)=J\sqcup T,\qquad
 T=\sigma(\bar u\oplus\bar v),\qquad T\cap J=\varnothing.
 \label{b:residual-syndrome}
\end{equation}
In fact each of $\sigma(\bar u)$ and $\sigma(\bar v)$ separately avoids
all checks touching $S$.

The finite-strength-state overlap depends only on the syndrome. Let
$k_{\beta}(s)$ denote its expression for a check subset $s$, so
$\kappa_{\beta}(x)=k_{\beta}(\sigma(x))$ for physical
supports $x$. Put $u_j=\tanh\beta_j$ and
$\mathcal R_Z=\ker H_Z^{\mathsf T}$. The redundancy expansion gives
\begin{equation}
 k_{\beta}(s)
 =\left(\prod_{j\in s}\sech\beta_j\right)
   \frac{\displaystyle
    \sum_{\substack{a\in \mathcal R_Z\\\supp(a)\cap s=\varnothing}}
             \prod_ju_j^{a_j}}
        {\displaystyle\sum_{a\in \mathcal R_Z}\prod_ju_j^{a_j}}.
 \label{b:redundancy-kernel}
\end{equation}
Commuting $X_x$ between the two amplitude filters gives this formula. A check
in $\sigma(x)$ cancels out of their combined exponent,
whereas every other check retains its full coefficient $\beta_j$. Expanding
$e^{\beta_jB_j}=\cosh\beta_j(I+u_jB_j)$ against
$|+\rangle^{\otimes n}$ retains only products with
$H_Z^{\mathsf T}a=0$. Dividing by the normalization gives
\eqref{b:redundancy-kernel}. Both sums are positive because they contain the
empty redundancy, including when some $u_j$ vanish.

For fixed $J$, define on all check subsets outside $J$
\begin{align}
 f_J(T)
 &:=\frac{k_{\beta}(J\sqcup T)}
           {k_{\beta}(J)}\nonumber\\
 &=\left(\prod_{j\in T}\sech\beta_j\right)
 \frac{\displaystyle
    \sum_{\substack{a\in \mathcal R_Z\\\supp(a)\cap J=\varnothing}}
       \left(\prod_ju_j^{a_j}\right)
       \mathbf1\{\supp(a)\cap T=\varnothing\}}
      {\displaystyle
    \sum_{\substack{a\in \mathcal R_Z\\\supp(a)\cap J=\varnothing}}
                   \prod_ju_j^{a_j}}.
 \label{b:residual-kernel}
\end{align}
This defines an extension to the full Boolean group of those check subsets,
even if some subsets are not attainable physical syndromes. The function has
nonnegative Fourier coefficients on that group. Indeed, for $b\in\{0,1\}$
and $c=\sech\beta_j\in(0,1]$,
\begin{equation}
 c^b=\frac{1+c}{2}+\frac{1-c}{2}(-1)^b,
 \qquad
 \mathbf1\{\supp(a)\cap T=\varnothing\}
 =\prod_{j\in\supp(a)}
   \frac{1+(-1)^{\mathbf1\{j\in T\}}}{2}.
 \label{b:nonnegative-fourier}
\end{equation}
Every displayed coefficient is nonnegative. Products correspond to convolution
of Fourier coefficients, and the weights $\prod_ju_j^{a_j}$ in
\eqref{b:residual-kernel} are nonnegative. Products and weighted sums thus
preserve nonnegativity. After division by the positive denominator, write
\begin{equation}
 f_J(T)=\sum_b c_b(-1)^{b\cdot T},\qquad
 c_b\ge0,\qquad \sum_bc_b=f_J(\varnothing)=1.
 \label{b:fourier-probability}
\end{equation}
The index $b$ runs over binary strings on the measured checks outside $J$.
The vectors with components $\sqrt{c_b}(-1)^{b\cdot s}$ have unit norm
and inner product $f_J(s\oplus s')$. Use the vectors indexed by
$s=\sigma(\bar u)$ and $s'=\sigma(\bar v)$ in place of
$|X_u\psi_r\rangle$ and $|X_v\psi_r\rangle$ in
\eqref{b:cross-gram-operators}. This expresses each residual assignment
submatrix as a cross-Gram matrix of unit vectors. The same Schatten estimate
and boundary sum apply, with the single additional scalar $k_{\beta}(J)$. From \eqref{b:redundancy-kernel},
\begin{equation}
 0<k_{\beta}(J)
 \le\prod_{j\in J}\sech\beta_j
 \le(\sech\beta_R)^{|J\cap\cP_r|}.
 \label{b:syndrome-scalar}
\end{equation}
The numerator in the redundancy ratio is a sub-sum of the denominator.
Disjointness of the syndromes in \eqref{b:family-syndrome} makes
$|J\cap\cP_r|=\sum_{\Gamma\in F}|\sigma(\Gamma)\cap\cP_r|$.
Together with \eqref{b:family-weight}, this proves
\eqref{b:soft-family-bound}.

For the geometric sum, each distinguishing component intersects
$\ell(q\oplus q')\subseteq R$, and therefore has an anchor in the actual
union $R$. There are at most $U\mu^m$ such connected supports of size $m$.
If $m<M(r)$, the component cannot be silent, so its weight contains at least
one $\sech\beta_R$ factor. For $m\ge M(r)$ we simply bound all such factors
by one. Hence
\begin{align}
 \Delta_{r,\beta_R}(q,q')
 &\le U\left[
  \sech\beta_R\sum_{1\le m<M(r)}\lambda^m
       +\sum_{m\ge M(r)}\lambda^m\right]\nonumber\\
 &\le\frac{U}{1-\lambda}
        [\lambda\sech\beta_R+\lambda^{M(r)}].
 \label{b:soft-geometric-sum}
\end{align}
Both sums run over positive integers. If $M(r)$ is not an integer, the second
sum starts at $\lceil M(r)\rceil$, and
$\lambda^{\lceil M(r)\rceil}\le\lambda^{M(r)}$.
The same family resummation and logical-block average as in
Lemma~\ref{lem:b5} finish the proof.
\end{proof}

The two terms in \eqref{b:soft-activity} express different mechanisms.
Increasing $\beta_R$ suppresses crossings that leave a protected syndrome.
Increasing $r$ raises the minimum size of a silent crossing in the geometries
below, reducing its contribution through the probability of its support.

\subsection{The minimum size of a silent component}

The operator estimate is common to all the code families considered here.
Their differences enter through geometry: a silent component that changes a
logical parity must cross enough parallel representatives, or satisfy enough
successive syndrome constraints, to acquire a large support.

\begin{lemma}[Toric-code tube geometry]
\label{lem:b7}
Choose a fixed number of straight, minimum-weight $Z$-logical strings in the
two- or three-dimensional edge-qubit toric code of linear size $L$. Let $r$
be a positive integer, set $r_0=\lfloor r/2\rfloor$, and assume
$2r_0+1\le L$. Around each chosen string include parallel translates with
displacement at most $r_0$ in each transverse coordinate, and protect all
plaquette ribbons joining neighboring translates in this array. Every
protected-syndrome-silent distinguishing component satisfies
\begin{align}
 |\Gamma|&\ge2r_0+1\ge r
          &&\text{in two dimensions},\label{b:toric-2d-core}\\
 |\Gamma|&\ge(2r_0+1)^2\ge r^2
          &&\text{in three dimensions}.\label{b:toric-3d-core}
\end{align}
The protected ribbons lie in a check tube of radius $O(r)$ around each selected
string. Protecting further checks in that tube preserves both bounds.
\end{lemma}

\begin{proof}
A distinguishing component has odd intersection with at least one selected
string. Indeed, $\chi_q(\Gamma)\ne\chi_{q'}(\Gamma)$ requires a nonzero
parity among the selected representatives. Fix one such string.

In two dimensions, label its parallel translates by $\ell_y$,
$-r_0\le y\le r_0$. Let $A_y$ be the plaquette ribbon joining neighboring
translates. The stabilizer relation
$\ell_y\oplus\ell_{y+1}=H_Z^{\mathsf T}\mathbf1_{A_y}$ implies
\begin{equation}
 |\Gamma\cap\ell_y|+|\Gamma\cap\ell_{y+1}|
 =|\sigma(\Gamma)\cap A_y|=0\pmod2.
 \label{b:ribbon-parity}
\end{equation}
The last equality holds because every plaquette in the ribbon is protected.
Odd intersection with the central string therefore propagates to every one
of its $2r_0+1$ translates. They are edge-disjoint, so $\Gamma$ must contain
at least one qubit on each, proving \eqref{b:toric-2d-core}.

In three dimensions label the parallel translates by $\ell_{y,z}$ with
$-r_0\le y,z\le r_0$. Apply the same parity relation to the protected
ribbons in each transverse direction. The transverse square is connected by
these neighboring relations, so odd intersection propagates throughout its
$(2r_0+1)^2$ edge-disjoint strings. This proves
\eqref{b:toric-3d-core}. The transverse square has $\ell_\infty$ radius
$r_0$ and transverse graph distance at most $2r_0\le r$ from its center;
the ribbons therefore lie in a tube of radius $O(r)$. The condition
$2r_0+1\le L$ ensures that the listed translates are distinct on the torus.
If tubes around different selected strings overlap, the argument is unchanged:
all ribbons used for the chosen central string remain protected.
\end{proof}

For a fixed number of straight representatives, $U\le CL$ for a constant $C$.
Using the minimum sizes in Lemma~\ref{lem:b7} and counting through anchors
therefore yields, uniformly in the logical pair,
\begin{equation}
 \Delta_r^{\mathrm{2D}}(q,q')
 \le\frac{CL}{1-\lambda}\lambda^{2r_0+1},\qquad
 \Delta_r^{\mathrm{3D}}(q,q')
 \le\frac{CL}{1-\lambda}\lambda^{(2r_0+1)^2}.
 \label{b:toric-hard-activity}
\end{equation}
For finite-strength enhancement, substitute these same minimum sizes into
\eqref{b:soft-activity}. The separate Wilson-restoration conditions are proved in
\hyperref[app:wilson]{Supplementary Note~4}.

For more general CSS codes, a confinement inequality can replace the ribbon
construction. The linear syndrome bound is the robustness property used in
quantum expander decoding~\cite{SI-Leverrier2015,SI-Fawzi2018}. Confinement entered
single-shot correction through subsystem codes~\cite{SI-Bombin2015}; good linear
confinement was subsequently shown to suffice under local stochastic
noise~\cite{SI-Quintavalle2021}. It also enters dimension-jump state
preparation~\cite{SI-Hong2025Preparation} and passive quantum memory at nonzero
temperature~\cite{SI-Hong2025Memory}. Its classical counterpart has been used to
establish eigenstate localization in quantum-perturbed code
Hamiltonians~\cite{SI-Yin2024Localization}. Here we use the inequality to bound
the growth of an error that remains silent throughout the enhanced region.
The cutoff in the following definition specifies the range of error weights
to which the argument applies.

\begin{definition}[Linear confinement in the $X$-error sector]
\label{def:linear-confinement}
Let $A_\mu=X^{a_\mu}$ be the $X$ checks and
$S_X=\operatorname{span}_{\mathbb F_2}\{a_\mu\}$ their support space. For an error $X_e$, its measured syndrome is the
set $\sigma(e)=\supp(H_Ze)$, and its reduced weight is
\begin{equation}
 |e|_{\mathrm{red}}=\min_{s\in S_X}|e\oplus s|.
 \label{b:reduced-weight}
\end{equation}
The code has linear confinement up to cutoff $T>0$ if
\begin{equation}
 |\sigma(e)|\ge\kappa|e|_{\mathrm{red}}
 \qquad\text{whenever }|e|_{\mathrm{red}}<T,
 \label{b:confinement}
\end{equation}
for some $\kappa>0$. A family of $n$-qubit codes has \emph{good linear
confinement} in this sector if the inequality holds with
$\kappa\ge\kappa_0>0$ and $T\ge c_Tn^\alpha$, where
$\kappa_0,c_T,\alpha>0$ are independent of $n$~\cite{SI-Quintavalle2021,SI-Hong2025Preparation}.
\end{definition}

Reduced weight here is taken only modulo stabilizers, preserving logical
parity as well as syndrome. Errors with the same syndrome but different
logical actions are not identified. The adjective ``good'' specifies the
uniform polynomial cutoff; it imposes no code-rate assumption. We require
confinement only for $X$ errors detected by the measured $Z$ checks. The
constant $\kappa$ is distinct from the preparation overlap $\kappa_\beta(x)$.

\begin{lemma}[Growth of a silent core under confinement]
\label{lem:b8}
Assume Eq.~\eqref{b:confinement} with constants $\kappa>0$ and $T>0$.
Let $N_t(R)$ be the radius-$t$ qubit neighborhood of the actual union $R$ in
$G_Z$, let $r\ge1$ be an integer, and assume $\cP_r$ contains every measured
check touching $N_{r-1}(R)$. If each qubit touches at most $\Delta_Z>0$
measured checks, then every protected-syndrome-silent error with $g(e)\ne0$
satisfies
\begin{equation}
 |e|\ge\min\left\{T,
                 \left(1+\frac{\kappa}{\Delta_Z}\right)^r\right\}.
 \label{b:confinement-core}
\end{equation}
In particular the right-hand side is a valid choice of the minimum-size lower
bound $M(r)$ in Lemma~\ref{lem:b6}.
\end{lemma}

\begin{proof}
Every subset of a minimum-weight representative modulo $S_X$ is itself
minimum weight in its stabilizer coset. Indeed, let $a\subseteq e$ with $e$
minimum weight modulo $S_X$. If some $s\in S_X$ reduced the weight of $a$,
then
\begin{equation}
 |a\oplus s|<|a|
 \quad\Longleftrightarrow\quad
 |a\cap s|>\frac{|s|}{2}.
 \label{b:reduced-subset}
\end{equation}
It would follow that $|e\cap s|>|s|/2$ and hence
$|e\oplus s|<|e|$, a contradiction.

If $|e|\ge T$, \eqref{b:confinement-core} is immediate. Otherwise choose a
minimum-weight representative $e_{\min}=e\oplus s$ modulo $S_X$, so
$|e_{\min}|\le|e|<T$. Multiplication by an $X$ stabilizer changes neither
the $Z$ syndrome nor intersection parity with any selected $Z$ logical:
$H_Zs=0$ by the CSS commutation relation, and $s\cdot\ell_a=0$ because
$\ell_a$ commutes with every $X$ stabilizer. Thus $e_{\min}$ is still silent
in $\cP_r$ and has $g(e_{\min})\ne0$.

By Lemma~\ref{lem:b2}, each component of $e_{\min}$ is silent in $\cP_r$,
and at least one component $\Gamma$ has $g(\Gamma)\ne0$. Being a subset of
the reduced support $e_{\min}$, $\Gamma$ and each of its subsets are reduced.
In particular, $|\Gamma|<T$. Set
\begin{equation}
 \Gamma_t=\Gamma\cap N_t(R),\qquad 0\le t\le r.
 \label{b:layers}
\end{equation}
Nonzero logical parity forces $\Gamma$ to intersect $R$, so
$|\Gamma_0|\ge1$. For $0\le t\le r-1$, reducedness and the cutoff allow
\eqref{b:confinement} to be applied to $\Gamma_t$:
\begin{equation}
 |\sigma(\Gamma_t)|\ge \kappa|\Gamma_t|.
 \label{b:layer-confinement}
\end{equation}
Every check in this syndrome touches $N_t(R)\subseteq N_{r-1}(R)$ and is
therefore protected. The full $\Gamma$ is silent on that check, whereas
$\Gamma_t$ has odd overlap with it. The check must consequently also touch a
qubit of $\Gamma\setminus\Gamma_t$ to cancel that parity. It shares the
check with a qubit of $\Gamma_t$, so common-check adjacency places the added
qubit in $\Gamma_{t+1}\setminus\Gamma_t$.

Each added qubit can account for at most $\Delta_Z$ such checks. Counting
check--qubit incidences gives
\begin{equation}
 \Delta_Z|\Gamma_{t+1}\setminus\Gamma_t|
 \ge|\sigma(\Gamma_t)|\ge \kappa|\Gamma_t|,
 \qquad
 |\Gamma_{t+1}|\ge
      \left(1+\frac{\kappa}{\Delta_Z}\right)|\Gamma_t|.
 \label{b:layer-growth}
\end{equation}
Iterating $r$ times and using $|\Gamma_0|\ge1$ yields
\begin{equation}
 |e|\ge|e_{\min}|\ge|\Gamma_r|
 \ge\left(1+\frac{\kappa}{\Delta_Z}\right)^r.
 \label{b:iterated-growth}
\end{equation}
This proves the claim in the case $|e|<T$, and together with the first case
proves \eqref{b:confinement-core}. Reducing the full error before choosing its
component is essential: the confinement hypothesis controls reduced weight,
whereas the connected-family sum ranges over arbitrary physical supports.
\end{proof}

Combining the appropriate minimum-size bound with Lemma~\ref{lem:b6} gives
the width and measurement-strength conditions used in the main text. For
uniform asymptotic claims, the graph-counting constant, confinement constant,
check incidence, and cutoff must be controlled at the stated scales. Growth
in \eqref{b:confinement-core} is guaranteed only up to the cutoff $T$.
Throughout, the anchor factor is the actual union size $U=|\bigcup_a\ell_a|$;
it should not be replaced by the size of a single representative when the
selected logical register grows.

\subsection{A soft--hard comparison for finite-strength CSS filters}

A second route to finite-strength recovery compares the prepared state
directly with its projection onto the protected sector.  This separates the
quality of that projection from the geometric recovery argument.  The
comparison uses nonnegative couplings and unsigned $Z$ checks, and remains
valid when the protected checks are dependent.

\begin{lemma}[Protected-sector probability and AQEC comparison]\label{lem:c8}
Consider the normalized CSS-filter state
\begin{equation}\label{c:csssoft}
 |\psi_{\mathrm{soft}}\rangle\propto
 \exp\!\left(\frac12\sum_j\beta_jB_j\right)|+\rangle^{\otimes n},
 \qquad B_j=Z^{h_j},\quad\beta_j\geq0.
\end{equation}
Let $\cP_r$ be a set of checks with $\beta_j\geq\beta_R$, and set
$\Pi_r=\prod_{j\in\cP_r}(I+B_j)/2$.  The probability of a violation
of at least one protected check satisfies
\begin{equation}\label{c:protectedprob}
 q_r:=1-\langle\psi_{\mathrm{soft}}|\Pi_r|
                     \psi_{\mathrm{soft}}\rangle
 \leq\sum_{j\in\cP_r}\frac{1}{1+e^{2\beta_j}}
 \leq |\cP_r|e^{-2\beta_R}.
\end{equation}
Define
$|\psi_{\mathrm{hard}}\rangle
=\Pi_r|\psi_{\mathrm{soft}}\rangle/\sqrt{1-q_r}$.
Construct the soft and hard logical bases from the same selected $Z$
representatives, under the logical orthonormality assumptions of the
encoding construction, and apply the same physical noise channel.
Then their reference--environment states obey
\begin{equation}\label{c:aqeccomparison}
 \frac12\|\rho_{RE}^{\mathrm{soft}}-
                 \rho_{RE}^{\mathrm{hard}}\|_1\leq\sqrt{q_r},
 \qquad
 \eps_{\mathrm{soft}}\leq\eps_{\mathrm{hard}}+2\sqrt{q_r},
\end{equation}
where
$\eps=\frac12\|\rho_{RE}-(I_R/K)\otimes\rho_E\|_1$
is the normalized AQEC trace-distance deviation and $K$ is the logical
dimension.
\end{lemma}

\begin{proof}
Remove only the coupling $\beta_j$, and let
$m_j=\langle B_j\rangle_{\beta_j=0}$ in the resulting Born measure.
Its high-temperature expansion has nonnegative coefficients:
each coefficient is a product of nonnegative $\tanh\beta_i$, and
\begin{equation}\label{c:csspositivity}
 \langle+|^{\otimes n}B_j\prod_{i\in A}B_i
                         |+\rangle^{\otimes n}\in\{0,1\}.
\end{equation}
The denominator has the same positivity and contains an empty term.
Therefore $m_j\geq0$.  Restoring $\beta_j$ reweights the two values
of $B_j$ by $e^{\pm\beta_j}$, giving the exact formula
\begin{equation}\label{c:singleviolation}
 \Pr(B_j=-1)
 =\frac{1-m_j}{e^{2\beta_j}(1+m_j)+(1-m_j)}
 \leq\frac{1}{1+e^{2\beta_j}}.
\end{equation}
The event $1-\Pi_r$ is the union of the protected-check violation
events, so a union bound proves Eq.~\eqref{c:protectedprob}.  No
independence assumption is involved.  Since all checks are unsigned,
the simultaneous $+1$ sector has nonzero weight; the normalized
projection is well-defined.  On this sector each protected filter is
scalar.  Consequently $|\psi_{\mathrm{hard}}\rangle$ is precisely
the hard-protected state with all outside couplings unchanged.

The overlap of the soft and hard preparation states is
$\sqrt{1-q_r}$.  The map
\begin{equation}\label{c:referenceisometry}
 |\varphi\rangle\longmapsto
 \frac1{\sqrt K}\sum_q|q\rangle_R
                      Z^{\ell(q)}|\varphi\rangle
\end{equation}
is an isometry: its squared norm is the average of $K$ identical
squared norms, because the reference labels are orthogonal and every
$Z^{\ell(q)}$ is unitary.  Each representative also commutes with
$\Pi_r$.  The two reference--code purifications therefore have overlap
$\sqrt{1-q_r}$, and the pure-state trace-distance formula~\cite{SI-Fuchs1999} gives distance $\sqrt{q_r}$. Apply a
common Stinespring dilation of the noise and trace out the physical
output. Contractivity~\cite[Sec.~3.3]{SI-Watrous2018} gives the first inequality in
Eq.~\eqref{c:aqeccomparison}, and the same bound holds for the
environment marginals.

Finally, the triangle inequality gives
\begin{equation}\label{c:epsdifference}
 \begin{split}
 \eps_{\mathrm{soft}}-\eps_{\mathrm{hard}}
 &\leq\frac12\|\rho_{RE}^{\mathrm{soft}}
                   -\rho_{RE}^{\mathrm{hard}}\|_1
       +\frac12\|\rho_E^{\mathrm{soft}}
                   -\rho_E^{\mathrm{hard}}\|_1\\
 &\leq2\sqrt{q_r}.
 \end{split}
\end{equation}
A full transpose, if used to identify these states with the AQEC
matrix convention, leaves all trace norms unchanged.
\end{proof}

Combining Lemma~\ref{lem:c8} with \eqref{b:trace-average} and
\eqref{b:hard-geometric-sum}, and using $\|B\|_1=2\eps$, gives
\begin{equation}
 \|B_{\mathrm{soft}}\|_1
 \le \exp\!\left[\frac{U\lambda^{M(r)}}{1-\lambda}\right]-1
       +4\sqrt{|\cP_r|}\,e^{-\beta_R}.
 \label{b:combined-recovery}
\end{equation}
This proves the main-text finite-strength enhancement theorem. The
counting convention used here absorbs the local constant into $\mu$.
At fixed $k$, $U\lambda^{M(r)}\to0$ and
$\sqrt{|\cP_r|}e^{-\beta_R}\to0$ imply both
$D_{\mathrm{AQEC}}\to0$ and $F_e^\star\to1$ by
Lemma~\ref{lem:a8}. The direct estimate in Lemma~\ref{lem:b6} remains a
separate bound. Its sufficient condition $U\sech\beta_R\to0$ is stronger
than the soft--hard condition for the toric neighborhoods below, and does
not supply their stated strength coefficient for every $\delta>0$.
The dependence on a growing logical register is treated at the end of this
Note.

\subsection{Resource estimates and the logarithmic strength scale}
\label{app:resources}

The two suppression mechanisms lead to separate choices of protection width
and measurement strength.  Their combination gives the main-text corollary on
toric resources and logarithmic strength. The distance $d$ in these statements is the distance of the ideal code; in the two toric
families under consideration, $d=L$. The exact Knill--Laflamme distance of
the finite-strength subspace is not being used as a growing parameter.
Take a fixed number of straight representatives, so $U\le Cd$, and choose
a fixed positive noise envelope $0<\lambda_0<1$ with
$\lambda\le\lambda_0$. This also includes zero physical noise and keeps
the explicit width formulas below well defined.

Fix $\delta>0$. Lemma~\ref{lem:b7} permits the choices
\begin{equation}
 \begin{split}
 r_{\mathrm{2D}}
 &=\left\lceil\frac{(1+\delta)\log d}{|\log\lambda_0|}\right\rceil,
 \\
 r_{\mathrm{3D}}
 &=\left\lceil\sqrt{
          \frac{(1+\delta)\log d}{|\log\lambda_0|}}\right\rceil.
 \end{split}
 \label{b:explicit-widths}
\end{equation}
The corresponding minimum silent sizes obey
$\lambda^{M(r)}\le d^{-1-\delta}$. Therefore
\eqref{b:toric-hard-activity} and Lemma~\ref{lem:b5} give
\begin{equation}
 \eps_{\mathrm{hard}}=O(d^{-\delta}).
 \label{b:hard-resource-error}
\end{equation}
All tube widths are sublinear in $L$, so the distinct-translate assumption
of Lemma~\ref{lem:b7} holds for sufficiently large $d$.

A two-dimensional strip contains $O(dr)$ protected plaquettes. A
three-dimensional tube contains $O(dr^2)$: there are $O(r^2)$ transverse
positions along a string of length $d$, and each position touches a bounded
number of plaquettes. Both choices in \eqref{b:explicit-widths} consequently
use
\begin{equation}
 |\cP_r|=O(d\log d).
 \label{b:toric-footprint}
\end{equation}
With a common enhanced strength
\begin{equation}
 \beta_R=\left(\frac12+\delta\right)\log d,
 \label{b:explicit-strength}
\end{equation}
Lemma~\ref{lem:c8} gives
\begin{equation}
 q_r=O(d^{-2\delta}\log d),\qquad
 \eps_{\mathrm{soft}}
       =O\!\left(d^{-\delta}\sqrt{\log d}\right)\longrightarrow0.
 \label{b:soft-resource-error}
\end{equation}
The recovery and entropy bounds in Lemma~\ref{lem:a8} then imply
$F_e^\star\to1$ and $D_{\mathrm{AQEC}}\to0$ for fixed $k$.
The bulk measurement strength can remain fixed and finite throughout this
construction. The widths differ, although the two constructions have the same
asymptotic enhanced-check count. Counting all measured plaquette checks gives
$N_{\rm chk}=d^2$ in two dimensions and $N_{\rm chk}=3d^3$ in three,
including the checks related by stabilizer redundancies. Hence
\begin{equation}
 \frac{|\cP_r|}{N_{\rm chk}}
 =\begin{cases}
 O(\log d/d),&\text{two dimensions},\\
 O(\log d/d^2),&\text{three dimensions}.
 \end{cases}
 \label{b:toric-enhanced-fractions}
\end{equation}
These are the first two fractions in the main-text resource table.

The logarithmic strength scale is also necessary within these protocols.
To compare the lower and upper statements, take every enhanced check to have
strength exactly $\beta_R$ and keep the exterior strength fixed. Along one
straight selected representative, its $d$ edges have the same nonzero logical
membership column and pairwise disjoint syndromes. Every syndrome lies inside
the protected tube and contains two checks in two dimensions or four in three
dimensions. The proof of Lemma~\ref{lem:a6} thus applies with the common
incident strength $\beta_R$: it uses the lower kernel bound for a single
edge and the upper kernel bound for a pair of these edges, and both involve
only their incident checks. The exterior couplings do not enter these bounds.
Consequently, vanishing total AQEC entropy requires
\begin{equation}
 \begin{split}
 \beta_R-\tfrac14\log d&\longrightarrow+\infty
       \quad\text{in two dimensions},\\
 \beta_R-\tfrac18\log d&\longrightarrow+\infty
       \quad\text{in three dimensions}.
 \end{split}
 \label{b:enhanced-strength-necessary}
\end{equation}
For fixed $k$, these are necessary for asymptotically faithful recovery as
well, by Lemma~\ref{lem:a8}. Together with
\eqref{b:explicit-strength}, they establish necessity and sufficiency of the
logarithmic order. They do not identify an optimal leading coefficient or a
necessary protection width. If the upper-bound hypothesis is stated only as
$\beta_j\ge\beta_R$, its lower parameter $\beta_R$ alone does not constrain
the actual strengths from above; the necessity statement concerns the common
strength used by the specified protocol.

For a family satisfying the confinement hypothesis of Lemma~\ref{lem:b8},
write $\gamma=1+\kappa/\Delta_Z>1$. The sufficient hard-state condition is
\begin{equation}
 |\log\lambda_0|\min(T,\gamma^r)-\log U
       \longrightarrow+\infty.
 \label{b:confinement-resource-condition}
\end{equation}
Assume $U\to\infty$. It suffices to have $T\ge c\log U$ and
$\gamma^r\ge c\log U$, with $c>1/|\log\lambda_0|$.
Good linear confinement provides this cutoff because $U\le n$ and its
polynomial cutoff eventually exceeds $c\log U$. Taking
\begin{equation}
 r=\left\lceil\frac{\log(c\log U)}{\log\gamma}\right\rceil
 \label{b:confinement-radius}
\end{equation}
gives $r=O(\log\log U)$ and
$U\lambda^{M(r)}\le U^{1-c|\log\lambda_0|}\to0$.

The enhanced-check count follows from the bounded degree of $G_Z$.
Let $D_G\ge2$ be a uniform degree bound. The union of radius-$r$
neighborhoods about the $U$ anchor qubits contains at most
$CU(D_G+1)^r$ qubits. Each qubit touches at most $\Delta_Z$ measured
checks, so enhancing exactly the checks touching $N_{r-1}(R)$ requires
\begin{equation}
 \begin{split}
 |\cP_r|&\le\min\{N_{\rm chk},CU(\log U)^\nu\},\\
 \nu&=\frac{\log(D_G+1)}{\log\gamma}.
 \end{split}
 \label{b:confinement-footprint}
\end{equation}
Here $C$ is independent of code size, and $N_{\rm chk}$ counts the same
measured $Z$ checks as the numerator. For
$\beta_R=(\tfrac12+\delta)\log U$, the soft--hard correction is bounded by
\begin{equation}
 4\sqrt{|\cP_r|}e^{-\beta_R}
 =O\!\left(U^{-\delta}(\log U)^{\nu/2}\right)\longrightarrow0.
 \label{b:confinement-soft-resource}
\end{equation}
Together with the hard-state estimate, this proves the confinement row in
the main-text resource table for fixed $k$. Its enhanced fraction vanishes
under the additional condition $U(\log U)^\nu=o(N_{\rm chk})$.

More generally, for finite strengths and nonempty protected sets,
Lemma~\ref{lem:c8} adds the condition
\begin{equation}
 \beta_R-\tfrac12\log|\cP_r|\longrightarrow+\infty.
 \label{b:general-soft-resource-condition}
\end{equation}
If the protected set is empty, the soft--hard correction is exactly zero;
the equivalent product condition
$\sqrt{|\cP_r|}e^{-\beta_R}\to0$ remains well defined in either case.
Alternatively, the direct finite-strength bound in Lemma~\ref{lem:b6}
uses $\beta_R-\log U\to+\infty$. Which estimate is smaller depends on the
actual enhanced-check count. The confinement inequality controls the minimum
silent support only up to $T$; it gives neither an unrestricted exponential
growth law nor a small aggregate footprint by itself.

\subsection{A growing logical register}
\label{app:register}

Protecting a growing register requires accounting for all vulnerable logical
supports and for the distinction between total leakage and leakage density.
The recovery theorem also applies when the selected number $k$ and the
union size $U$ grow, provided the local constants remain uniform. The average
over logical pairs in \eqref{b:trace-average}
prevents an additional factor $K=2^k$ from appearing in the trace-norm
estimate. In particular, the direct finite-strength bound gives
\begin{equation}
 \eps\le\frac12\left\{
 \exp\!\left[\frac{U}{1-\lambda}
       \bigl(\lambda\sech\beta_R+\lambda^{M(r)}\bigr)\right]-1
 \right\}.
 \label{b:register-direct}
\end{equation}
With a fixed $0<\lambda_0<1$ and $\lambda\le\lambda_0$, it therefore suffices that
\begin{equation}
 \beta_R-\log U\to+\infty,\qquad
 |\log\lambda_0|M(r)-\log U\to+\infty
 \label{b:register-decoupling-conditions}
\end{equation}
to obtain $\eps\to0$ and faithful entanglement recovery of the entire
selected register. Its entropy per logical qubit vanishes as well. Indeed,
Lemma~\ref{lem:a8} implies
\begin{equation}
 \frac{D_{\mathrm{AQEC}}}{k}
 \le \eps\log2+
     \frac{1+\eps}{k}
          h_2\!\left(\frac{\eps}{1+\eps}\right)
 \longrightarrow0.
 \label{b:register-entropy-density}
\end{equation}
To force the total entropy to vanish, it suffices instead that
$k\eps\to0$. Applied to \eqref{b:register-direct}, this amounts to replacing
$U$ by $kU$ in both sufficient conditions
\eqref{b:register-decoupling-conditions}. Under confinement, the minimum
$M(r)\ge\min(T,\gamma^r)$ is used in either version; the cutoff must reach
the logarithm of $U$ or of $kU$, respectively.

The soft--hard route has a corresponding distinction. Full-register
recovery follows from a vanishing hard-state bound and
$\sqrt{|\cP_r|}e^{-\beta_R}\to0$. A sufficient condition for vanishing
total entropy is
\begin{equation}
 kU\lambda^{M(r)}\to0,
 \qquad k\sqrt{|\cP_r|}\,e^{-\beta_R}\to0.
 \label{b:register-softhard-conditions}
\end{equation}
These explicit factors matter when many representatives are protected. The
anchor count is the size of their actual union, not the length of one
representative.

Finally, consider the aggregate lower bound under independent noise at fixed
$0<p<1$. Suppose checks outside an enhanced set $\cP_{\mathrm{hi}}$ have
strength at most fixed $\beta_0$, and let
$F_{\mathrm{hi}}=\bigcup_{j\in\cP_{\mathrm{hi}}}\supp B_j$ be the qubits
touched by that set. The binary matrix with rows $\ell_1,\ldots,\ell_k$ has
rank $k$. Deleting its columns in $F_{\mathrm{hi}}$ reduces rank by at most
$|F_{\mathrm{hi}}|$. Among the remaining columns, one can therefore select
at least $(k-|F_{\mathrm{hi}}|)_+$ independent logical parities. Their qubits
touch only checks with strength at most $\beta_0$, giving
$\kappa_{\beta}(e)\ge e^{-\beta_0\Delta_Z}$. Applying
Lemma~\ref{lem:a5} to these independent parities yields
\begin{equation}
 \frac{D_{\mathrm{AQEC}}}{k}
 \ge 2p(1-p)e^{-2\beta_0\Delta_Z}
            \left(1-\frac{|F_{\mathrm{hi}}|}{k}\right)_+.
 \label{b:register-footprint-lower}
\end{equation}
Here $(x)_+=\max\{x,0\}$. If each measured check has weight at most a
constant $w_Z$, then $|F_{\mathrm{hi}}|\le w_Z|\cP_{\mathrm{hi}}|$.
Vanishing entropy density requires
$|F_{\mathrm{hi}}|\ge[1-o(1)]k$. At constant encoding rate
$k=\Theta(n)$, this entails $\Omega(n)$ enhanced checks, as stated in the
main-text corollary on the cost of protecting a growing logical register. An
$O(\log\log n)$ protection radius
can coexist with this extensive footprint. Conversely, a vanishing encoding
rate alone does not ensure that the union of selected representatives is
subextensive.

\refstepcounter{section}
\section*{Supplementary Note 4: Enhanced Wilson loops and logical operations}\label{app:wilson}

Recovering each selected encoding does not yet establish that two distant
representatives implement the same logical action.  This Supplementary Note addresses
that remaining question through the main-text Wilson-restoration theorem and
the encoded-CNOT corollary.  The bulk coupling
remains fixed while only checks near the representatives are strengthened.
In three dimensions, the finite-torus gauge measure in Eq.~\eqref{c:gauge}
turns Wilson restoration into a bound on dual clusters that obstruct a
plaquette filling.
All chains and intersection numbers have coefficients in $\mathbb F_2$;
$\partial$ and $\delta$ denote boundary and coboundary.

\subsection{The two-dimensional obstruction}

Take two straight representatives separated by $a$, with
$\alpha L\leq a\leq(1-\alpha)L$ for fixed $0<\alpha<1/2$. Enhanced
strips of width $r=o(L)$ leave order $L^2$ unprotected plaquettes in
each complementary filling. Lemma~\ref{lem:c2} therefore gives
\[
 0\leq\langle W\rangle
 \leq 2\exp[-c|\log\tanh\beta|L^2]\longrightarrow0
\]
for fixed $0<\beta<\infty$, whatever the nonnegative couplings inside the
tubes, including infinite couplings.  At $\beta=0$, the expectation is
exactly zero as long as each complementary filling contains an
unprotected plaquette.  This proves the two-dimensional part of the main-text
Wilson-restoration theorem; its condition on unprotected area is essential.

\subsection{Finite-torus duality and the subcritical probability bound}

The three-dimensional argument adapts the random-cluster and dual-linking
methods used in the perimeter-law proof of Duncan and
Schweinhart~\cite[Secs.~5, 7, and 9]{SI-DuncanSharp2025}. Here we need
finite-volume estimates for two winding representatives with enhanced
couplings near their supports. Since
$W(C)$ is diagonal in the computational basis, its expectation in the
prepared quantum state is exactly its average in the gauge measure
of Eq.~\eqref{c:gauge}.  The plaquette random-cluster representation
expresses this average as the probability that an open filling of $C$
exists. We derive this identity below from the coupling construction
of Ref.~\cite[Sec.~5.1]{SI-DuncanSharp2025}, allowing each plaquette its
own coupling. On the torus, winding dual clusters
modify the induced dual measure through a bounded topological weight.

The local gauge action satisfies
$\beta_pB_p=2\beta_p\mathbf1_{\{B_p=+1\}}-\beta_p$, so the
indicator-action coupling used in
Refs.~\cite{SI-DuncanSharp2025,SI-DuncanTopological2025} is $2\beta_p$.
The resulting plaquette random-cluster model has cluster weight $q=2$.
Let $A$ consist of its open primal plaquettes and the full primal
one-skeleton.  A dual edge $e=p^*$ is open in the complementary graph
$D$ precisely when the corresponding primal plaquette $p$ is absent
from $A$.  The graph $D$ includes all dual vertices, including isolated
ones. The configurations $D$ describe the auxiliary statistical model
used to evaluate the prepared-state overlap.

\begin{lemma}[Exact dual law and the Wilson filling event]\label{lem:c3}
For each primal plaquette $p$ and its dual edge $e=p^*$, set
\begin{equation}\label{c:dualparameters}
 \theta_p=1-e^{-2\beta_p},
 \qquad
 p_e^*=\frac{2(1-\theta_p)}{\theta_p+2(1-\theta_p)}
           =\frac{2}{e^{2\beta_p}+1}.
\end{equation}
Here $\theta_p$ and $p_e^*$ are parameters of the primal plaquette and
dual edge weights, respectively. They are not the marginal occupation
probabilities: the cluster weights correlate the occupations. Neither is
the physical noise probability used in the AQEC calculation.
The Fortuin--Kasteleyn (FK) random-cluster measure on the periodic dual
graph is the probability distribution~\cite{SI-FortuinKasteleyn1972,SI-Grimmett2006}
\begin{equation}\label{c:fk-definition}
 \phi_{L,\boldsymbol p^*,2}(D)\propto
 2^{c(D)}\prod_{e\in D}p_e^*\prod_{e\notin D}(1-p_e^*),
\end{equation}
normalized by summing these weights over all edge configurations.
Here $c(D)$ counts the connected components of the periodic graph,
including isolated vertices. The cluster weight is $q=2$; summing
over the two Ising spin assignments on each component gives the factor
$2^{c(D)}$. For a dual configuration $D$, define
\begin{equation}\label{c:homologyrank}
 h(D)=\operatorname{rank}\operatorname{im}
 \bigl[H_1(D)\longrightarrow H_1(\mathbb T_L^3)\bigr]
 \in\{0,1,2,3\}.
\end{equation}
The induced dual law is
\begin{equation}\label{c:duallaw}
 \nu^*(D)=
 \frac{2^{-h(D)}\phi_{L,\boldsymbol p^*,2}(D)}{
       \displaystyle\sum_{D'}2^{-h(D')}\phi_{L,\boldsymbol p^*,2}(D')}.
\end{equation}
For every $C=\partial S$,
\begin{equation}\label{c:filling}
 \langle W(C)\rangle
 =\Pr\{C\text{ is the boundary of a chain of open primal plaquettes}\}.
\end{equation}
The probability is taken over $A$ in the gauge/random-cluster coupling.
In particular, for every dual-edge event $E$,
\begin{equation}\label{c:factor8}
 \frac18\phi_{L,\boldsymbol p^*,2}(E)
 \leq\nu^*(E)\leq8\phi_{L,\boldsymbol p^*,2}(E),
\end{equation}
uniformly in the local couplings.
\end{lemma}

\begin{proof}
The finite-volume coupling in Ref.~\cite[Sec.~5.1]{SI-DuncanSharp2025}
is stated for uniform parameters. Its product construction gives the
inhomogeneous coupling used here, as the calculation below shows.
For torus duality, Sec.~4.3 of Ref.~\cite{SI-DuncanTopological2025}
introduces a homologically reweighted, or balanced, measure and proves
its exact duality in Theorem~18. We retain the ordinary gauge weights
and compute their dual explicitly for $q=2$, obtaining the factor
$2^{-h(D)}$ in Eq.~\eqref{c:duallaw}.

The identity
$e^{\beta_pB_p}=e^{\beta_p}
[(1-\theta_p)+\theta_p\mathbf1_{\{B_p=+1\}}]$ gives the marginal weight
\begin{equation}\label{c:primalrc}
 \Pr(A)\propto
 \prod_{p\in A}\theta_p\prod_{p\notin A}(1-\theta_p)\,2^{b_1(A)},
\end{equation}
where $b_i(A)=\dim H_i(A;\mathbb F_2)$.  To see the power of two,
write $N_e$ and $N_v$ for the numbers of primal edges and vertices.
The number of edge cochains satisfying the checks selected by $A$ is
$2^{N_e-\operatorname{rank}(\partial_2|_A)}
=2^{N_v-1+b_1(A)}$.  The factor $2^{N_v-1}$ is independent of $A$.
Conditional on $A$, the allowed cochains are uniform in this linear
space.  The character $(-1)^{s(C)}$ averages to one if
$C\in\operatorname{im}(\partial_2|_A)$ and to zero otherwise.  This is
linear duality over $\mathbb F_2$, and proves Eq.~\eqref{c:filling}.

Put $N=L^3$ and retain $c(D)$ for the number of dual connected components.  The
ranks needed to dualize Eq.~\eqref{c:primalrc} are
\begin{equation}\label{c:rankidentities}
 \begin{split}
 b_2(A)&=c(D)-1+3-h(D),\\
 b_1(A)&=|D|-N+c(D)+3-h(D).
 \end{split}
\end{equation}
For the first identity, dualize primal two-cycles to dual one-cocycles
and restrict them to $D$.  Exact dual one-cocycles, restricted to this
graph, form a space of dimension $N-c(D)$.  Modulo this space, the
restrictions of the three global cohomology classes have rank $h(D)$,
by the homology--cohomology pairing.  The full space of primal two-cycles
has dimension $(N-1)+3$.  The kernel of restriction therefore has
dimension
$(N-1)+3-[N-c(D)+h(D)]$.  This kernel consists precisely of the
two-cycles supported in $A$.  Since the complex $A$ has no three-cells,
its dimension is $b_2(A)$, proving the first identity.  The full
one-skeleton is connected.  Applying Euler's identity with
$N_e=3N$ and $|A|=3N-|D|$ gives the second identity.

Substituting Eq.~\eqref{c:rankidentities} into Eq.~\eqref{c:primalrc}
produces the factor $2^{|D|}2^{c(D)}2^{-h(D)}$, apart from a constant.
The first factor changes the local odds according to
\begin{equation}\label{c:dualodds}
 \frac{p_e^*}{1-p_e^*}
 =\frac{2(1-\theta_p)}{\theta_p},\qquad e=p^*.
\end{equation}
After normalization, the remaining product is the ordinary FK weight
$2^{c(D)}\prod_{e\in D}p_e^*\prod_{e\notin D}(1-p_e^*)$ multiplied by
$2^{-h(D)}$.  This proves Eq.~\eqref{c:duallaw}.  Both the tilt and its
normalizing sum belong to $[1/8,1]$, proving Eq.~\eqref{c:factor8}.  The calculation
initially assumes finite positive couplings; zero and infinite couplings
follow by continuity on the finite torus.
\end{proof}

The factor $2^{-h(D)}$ accounts for the winding cycles of the dual graph.
Its bounded size allows estimates from periodic FK to be transferred to the
gauge model with a uniform constant.

\begin{lemma}[Uniform subcritical cluster tail]\label{lem:c4}
Fix a background coupling satisfying
\begin{equation}\label{c:subcritical}
 p^*(\beta)=\frac{2}{e^{2\beta}+1}
 <p_c^{\mathrm{FK}}(2;\mathbb Z^3).
\end{equation}
Here $p_c^{\mathrm{FK}}(2;\mathbb Z^3)$ is the critical bond
parameter of the homogeneous random-cluster model on the infinite cubic
lattice with cluster weight $q=2$.
For every inhomogeneous profile $0\leq p_e^*\leq p^*(\beta)$, there are
positive constants $C_\beta,\xi_\beta,c_\beta$, independent of $L$ and
of the profile, such that for every dual vertex $v$ and
$1\leq R\leq L/4$,
\begin{align}
 \nu^*\{v\longleftrightarrow\partial B_R(v)\}
 &\leq C_\beta e^{-R/\xi_\beta},\label{c:onearm}\\
 \nu^*\{\text{some cluster has periodic diameter at least }L/4\}
 &\leq C_\beta L^3e^{-c_\beta L}.\label{c:macrocluster}
\end{align}
Here $B_R(v)$ is a ball in the graph distance of the dual cubic lattice;
for the stated range of $R$ it lifts to a ball in $\mathbb Z^3$. The
connection event means an open path from $v$ to the boundary of this
ball. Periodic diameter refers to the
ambient periodic lattice distance, not the internal path length of a
cluster.
\end{lemma}

\begin{proof}
First work with the ordinary periodic FK measure.  Conditioning on all
edges outside the lifted ball $B_R(v)$ induces a partition of its
boundary vertices according to their connections through the exterior.
For $q=2$, the fully wired boundary condition, which identifies all
boundary vertices, dominates every such partition. These comparisons
and monotonicity in the edge parameters are standard random-cluster
properties~\cite[Chs.~3--4]{SI-Grimmett2006}. They bound the
conditional connection probability by that of the homogeneous wired
ball with parameter $p^*(\beta)$.

The finite wired-volume estimate of Theorem~1.2 in
Ref.~\cite{SI-Duminil2019} is also used as Theorem~36 in the
perimeter-law proof of Ref.~\cite{SI-DuncanSharp2025}. To apply it, take unit
nearest-neighbor couplings $J_e=1$.  Its inverse-temperature parameter,
denoted $b_{\mathrm{FK}}$ here, enters the FK edge odds as
$e^{b_{\mathrm{FK}}}-1$.  Hence the conversion is
\[
 b_{\mathrm{FK}}=-\log[1-p^*(\beta)]
                =-\log\tanh\beta.
\]
Condition~\eqref{c:subcritical} places $b_{\mathrm{FK}}$ below the
critical value $-\log[1-p_c^{\mathrm{FK}}(2;\mathbb Z^3)]$.
The cited theorem gives exponential decay of the center-to-boundary
connection probability in each finite wired graph-distance ball.
The comparison above makes the bound uniform in the exterior
configuration and in the enhanced edge parameters. This is the part of
the cited theorem used here; the conversion to the inhomogeneous torus
and the Wilson estimate require the additional arguments in this note.
Averaging over the exterior and applying the factor-eight comparison in
Eq.~\eqref{c:factor8} proves Eq.~\eqref{c:onearm} for the actual dual
gauge law $\nu^*$.

A cluster of periodic diameter at least $L/4$ contains a vertex with an
open path reaching periodic distance at least $\lfloor L/8\rfloor$.
A union bound over the $L^3$ vertices and Eq.~\eqref{c:onearm} prove
Eq.~\eqref{c:macrocluster}.  Bounded small values of $L$ are absorbed in
the constants.
\end{proof}

Condition~\eqref{c:subcritical} is the strict perimeter-law side of the
transition identified in Ref.~\cite{SI-DuncanSharp2025}, after converting
the action convention.  We require fixed background coupling on this
side; the constants need not remain bounded as the transition is
approached.  The finite wired-volume estimate in
Lemma~\ref{lem:c4} supplies the input needed for distant winding
representatives.  Infinite-volume rectangular-loop asymptotics,
including the low-temperature expansion of Ref.~\cite{SI-Forsstrom2025},
do not by themselves give this estimate on a finite torus.

\subsection{Tube geometry and boundary restoration}

The cluster tail becomes useful once a geometric obstruction is identified.
A small dual cluster can prevent a filling only when it approaches the
Wilson boundary.  Hardening a tube around that boundary therefore leaves
only large, exponentially unlikely obstructions. The linking criterion
in Ref.~\cite[Sec.~7]{SI-DuncanSharp2025} is formulated in the lattice
and box setting. For the two winding representatives considered here,
we construct the obstruction directly on the torus and apply the
intersection argument after lifting each small cluster to a local box.

Take straight $x$-directed representatives at $(y,z)=(0,0)$ and
$(a,0)$, where $\alpha L\leq a\leq(1-\alpha)L$ for fixed
$0<\alpha\leq1/2$.  Let $C$ be their union, $|C|=2L$, and let $S$ be
the intervening flat strip, so that $\partial S=C$.  Choose a fixed
local periodic cell-incidence distance $\dist_L$ between primal and
dual cells and define
\begin{equation}\label{c:tube}
 T_r^*(C)=\{e^*: \dist_L(e^*,C)<r\},
 \qquad
 \mathcal P_r=\{p:p^*\in T_r^*(C)\}.
\end{equation}
Changing to an equivalent local lattice distance changes only geometric
constants. The cell-incidence distance used here and the qubit check-graph
distance of Supplementary Note~3 are comparable by constants independent
of $L$. Consequently a full tube of radius $r$ contains the protected
plaquette ribbons of Lemma~\ref{lem:b7} at a width proportional to $r$.
We assume $1\leq r\leq c_\alpha L$, where
$0<c_\alpha<\min\{\alpha/8,1/16\}$.  In particular,
\begin{equation}\label{c:tubesize}
 |\mathcal P_r|\leq C|C|(r+1)^2.
\end{equation}
The two-dimensional analogue has $|\mathcal P_r|=O(|C|(r+1))$.

\begin{lemma}[A local obstruction must approach the boundary]\label{lem:c5}
Let a connected dual cluster $\Gamma$ have periodic diameter
$s<L/4$.  Suppose it contains an Eulerian chain $\gamma$ with
$\gamma\cdot S=1$, where the dot denotes mod-two intersection.  Then
$\gamma$ is null-homologous and
\begin{equation}\label{c:localobstruction}
 \dist_L(\Gamma,C)\leq c_f(s+1)
\end{equation}
for a geometric constant $c_f$.  In particular, if $\Gamma$ avoids
$T_r^*(C)$, then $s\geq c_g r$ for a fixed $c_g>0$.  An obstruction
at diameter scale $s$ can be anchored among at most
$C|C|(s+1)^2$ dual edges.
\end{lemma}

\begin{proof}
Choose a vertex of $\Gamma$ as origin and lift each cluster vertex to
its nearest coordinate representative.  Because $s<L/4$, these
representatives lie in a cubical box of side $2s+O(1)$.  They preserve
cluster edges: a spurious jump by a full period would require a
coordinate separation at least $L-1$, larger than the span of this
box.  Thus the whole cluster lifts to a contractible cubical box.
Bounded small sizes can be incorporated by changing geometric constants.

The lifted Eulerian chain is a cycle and bounds a dual two-chain
$A_\gamma$ inside the box.  Each filling cell lies within
$c_f(s+1)$ of $\Gamma$.  The boundary--intersection identity on the
torus gives
\begin{equation}\label{c:intersection}
 1=\gamma\cdot S
   =(\partial A_\gamma)\cdot S
   =A_\gamma\cdot\partial S
   =A_\gamma\cdot C.
\end{equation}
Hence the filling meets $C$, proving Eq.~\eqref{c:localobstruction}.
If $\Gamma$ avoids the tube, then $r\leq c_f(s+1)$.  Decreasing a
positive geometric constant $c_g$ covers the bounded small-$r$ cases
and gives $s\geq c_g r$ for an obstruction.

A neighborhood of radius $O(s+1)$ around the two length-$L$ lines
contains at most $C|C|(s+1)^2$ dual edges.  Equation~\eqref{c:localobstruction}
guarantees that $\Gamma$ has an edge in this neighborhood.  Choose the
first such edge under a fixed ordering as its anchor.  This proves the
anchor count without counting the volume of the entire filling $S$.
\end{proof}

\begin{lemma}[Hard boundary enhancement restores the Wilson overlap]\label{lem:c6}
Assume Eq.~\eqref{c:subcritical}, put $\beta_p=\infty$ on $\mathcal P_r$, and
put $\beta_p=\beta$ elsewhere.  There are positive constants, uniform
in $L$ and $r$, such that
\begin{equation}\label{c:hardrestoration}
 1-\langle W(C)\rangle_{\mathrm{hard}}
 \leq C_\beta|C|(r+1)^2e^{-c_g r/\xi_\beta}
       +C_\beta L^3e^{-c_\beta L}.
\end{equation}
The geometric and decay constants may be adjusted from those in the
preceding lemmas.
\end{lemma}

\begin{proof}
Hard protection sets every dual parameter in $T_r^*(C)$ to zero, so
all dual clusters avoid this tube.  Exclude first the event of a
cluster with periodic diameter at least $L/4$.  Its probability is
bounded by the second term in Eq.~\eqref{c:hardrestoration}.

Suppose that $C$ fails to bound a chain of open primal plaquettes.
Linear duality supplies an edge cochain $a$ such that
$\delta a|_A=0$ and $a(C)=1$.  Its dual chain
$z=*\delta a$ is supported on $D$, satisfies $\partial z=0$ and
$[z]=0$, and obeys
\begin{equation}\label{c:oddwitness}
 z\cdot S=(\delta a)(S)=a(\partial S)=1.
\end{equation}
Split $z$ over the connected components of $D$.  Each part is Eulerian,
since different components share no vertices.  On the event under
consideration every component has diameter below $L/4$, so each part
lifts to a local box and is null-homologous.  Equation~\eqref{c:oddwitness}
implies that at least one part has odd intersection with $S$.  This
step includes disconnected witnesses; possible nonlocal components
have already been included in the excluded macroscopic event.

Let $\Gamma$ contain an odd part.  Lemma~\ref{lem:c5} implies that
its diameter is at least $c_g r$.  At a dyadic diameter scale
$[s,2s)$ it has an anchor among $C|C|(s+1)^2$ edges.  From an endpoint
of an anchor, a cluster of diameter at least $s$ contains an open
connection to distance at least $s/2$, up to a fixed lattice constant.
Use Eq.~\eqref{c:onearm} and sum over anchors and dyadic scales.  The
summation is controlled by the elementary bound
\begin{equation}\label{c:dyadicsum}
 \sum_{j\geq0}(2^js_0+1)^2e^{-a2^js_0}
 \leq C_a(s_0+1)^2e^{-as_0/2},
 \qquad s_0\geq1,\quad a>0.
\end{equation}
Indeed, one half of each exponential is bounded by
$e^{-as_0/2}$; the remaining half absorbs the growth of $4^j$ in
the polynomial prefactor and yields a summable series uniformly for
$s_0\geq1$.  Taking $s_0$ of order $\max\{1,c_g r\}$ and adjusting
constants gives the first term in Eq.~\eqref{c:hardrestoration}.
Finally, Eq.~\eqref{c:filling} identifies the filling-failure
probability with the Wilson deficit.
\end{proof}

The anchor count is the reason the leading prefactor depends on the
boundary length and tube cross-section.  The proof does not require
strengthening the plaquettes throughout the strip $S$.  The second
term separately controls obstructions that explore the topology of
the finite torus.

\begin{lemma}[Finite-strength boundary enhancement]\label{lem:c7}
Assume Eq.~\eqref{c:subcritical}.  If
$\beta_p\geq\beta_R\geq\beta$ on $\mathcal P_r$ and $\beta_p=\beta$
elsewhere, then
\begin{equation}\label{c:softrestoration}
 \begin{split}
 1-\langle W(C)\rangle_{\mathrm{soft}}
 \leq{}& C_\beta|C|(r+1)^2e^{-c_g r/\xi_\beta}
       +C_\beta L^3e^{-c_\beta L}\\
       &+16|\mathcal P_r|e^{-2\beta_R}.
 \end{split}
\end{equation}
Consequently, for fixed $b>1$ and $\delta>0$, the choices
\begin{equation}\label{c:softscales}
 r=b\frac{\xi_\beta}{c_g}\log L,
 \qquad
 \beta_R=\left(\frac12+\delta\right)\log L
\end{equation}
(rounding $r$ upward to an integer) suffice for
$\langle W(C)\rangle_{\mathrm{soft}}\to1$ as $L\to\infty$. The
condition $\beta_R\geq\beta$ holds for all sufficiently large $L$.
\end{lemma}

\begin{proof}
Let $E_r$ be the event that at least one protected dual edge is open.
For ordinary FK with cluster weight two, the conditional open
probability of an edge is at most its parameter.  On the tube,
$p_e^*\leq2e^{-2\beta_R}$.  A union bound and
Eq.~\eqref{c:factor8} therefore give
\begin{equation}\label{c:tubeviolation}
 \nu^*(E_r)\leq16|\mathcal P_r|e^{-2\beta_R}.
\end{equation}
Conditional on $E_r^c$, all protected dual edges are closed.  Their
local factors are constant, and the remaining weight in
Eq.~\eqref{c:duallaw} is exactly the hard-tube weight: $c(D)$ and
$h(D)$ depend on the edge configuration, not on the values of its
parameters.  Thus the conditional filling-failure probability is the
hard-tube deficit.  Splitting according to $E_r$ bounds the soft deficit
by that hard deficit plus $\nu^*(E_r)$, proving
Eq.~\eqref{c:softrestoration}.

Using $|C|=2L$ and Eq.~\eqref{c:tubesize}, the three contributions
under Eq.~\eqref{c:softscales} are respectively
$O(L^{1-b}\log^2 L)$, $O(L^3e^{-c_\beta L})$, and
$O(L^{-2\delta}\log^2 L)$.  They vanish, and the required geometric
condition $r\leq c_\alpha L$ holds for all sufficiently large $L$.
\end{proof}

Lemma~\ref{lem:c7} proves the three-dimensional Wilson-restoration
statement in the main text. The same full tubes also restore AQEC for each
chosen encoding under the fixed low-noise condition
$\lambda\leq\lambda_0<1$ of Supplementary Note~3. Indeed, the ribbons
contained in these tubes impose a minimum silent distinguishing size
proportional to $r^2$, by Lemma~\ref{lem:b7}. With $r$ proportional to
$\log L$, the hard-projection activity in Eq.~\eqref{b:toric-hard-activity}
is $O(L e^{-c(\log L)^2})$ for some fixed $c>0$. The soft--hard correction
in Eq.~\eqref{c:aqeccomparison} is
$O(\sqrt{L(r+1)^2}e^{-\beta_R})=O(L^{-\delta}\log L)$. Both vanish,
so the recovery and entropy bounds of Lemma~\ref{lem:a8} apply to each
fixed logical qubit. The preparation uses $O(L\log^2L)$ enhanced
plaquettes out of $3L^3$, a fraction $O(\log^2L/L^2)$.

The radius $O(\log L)$ is sufficient for this joint conclusion, whereas
AQEC alone has the sufficient radius $O(\sqrt{\log L})$ established in
Supplementary Note~3. Comparing these constructions does not prove that
Wilson restoration requires a larger radius. Their assumptions also play
different roles: low physical noise controls recovery, and subcriticality
of the dual bulk controls the Wilson overlap.

\subsection{Encoding distance and recovery with a common channel}
\label{app:operational}

The representative overlap determines both the distinguishability of two
encodings and the cost of recovering without a record of which one was used.
Let $Q$ denote the entire $n$-qubit physical system, and let
$\mathcal H_L$ be the two-dimensional logical input space. Both representatives are enhanced in the same
preparation; $|\psi_\beta\rangle$ therefore includes the full,
inhomogeneous measurement-strength profile. For $i=1,2$, define
\begin{equation}\label{d:encodings}
 V_i:\mathcal H_L\longrightarrow\mathcal H_Q,\qquad
 V_i\bigl(a|0\rangle_L+b|1\rangle_L\bigr)
   =a|\psi_\beta\rangle+bZ^{\ell_i}|\psi_\beta\rangle.
\end{equation}
The index $i$ labels the representative defining the image of
$|1\rangle_L$. A complementary physical logical $X$ operator stabilizes
$|\psi_\beta\rangle$ and anticommutes with both $Z^{\ell_i}$. The two
codewords within each encoding therefore belong to opposite exact
symmetry sectors. They are normalized and orthogonal, so
$V_i^\dagger V_i=I_L$.

Write
$w=\langle\psi_\beta|Z^{\ell_1}Z^{\ell_2}|\psi_\beta\rangle$.
This number belongs to $[0,1]$ by Lemmas~\ref{lem:c2} and~\ref{lem:c3},
including the inhomogeneous and projective limits. The shared zero codeword and
the vanishing cross-sector matrix elements give
\begin{equation}\label{d:crossgram}
 V_1^\dagger V_2=\operatorname{diag}(1,w)
\end{equation}
in the ordered logical basis $(|0\rangle_L,|1\rangle_L)$.
The channel associated with $V_i$ is
$\mathcal V_i(\rho_L)=V_i\rho_L V_i^\dagger$.
All recovery channels below act on the entire physical system and output one
logical qubit.  They are not assumed to be spatially local or efficiently
implementable.

\begin{lemma}[Distance between the encoding channels]
\label{lem:d1}
For these encodings,
\begin{equation}
 \frac12\lVert\mathcal V_1-\mathcal V_2\rVert_\diamond
 =\sqrt{1-w^2}
 \leq\lVert V_1-V_2\rVert_\infty
 =\sqrt{2(1-w)}.
 \label{eq:encoding-diamond-distance}
\end{equation}
In particular, the first expression bounds the trace distance between the two
encoded outputs for every logical input, including inputs entangled with an
arbitrary reference system.
\end{lemma}

\begin{proof}
For discrimination between two channels, the diamond norm can be optimized over pure input states with a reference system~\cite[Theorem~3.51]{SI-Watrous2018}.  Let $\lvert\Omega\rangle_{LR}$ be any such input and let
$\rho_L$ be its reduced state on the logical qubit.  The two encoded outputs are
pure and have overlap
\[
 {}_{LR}\langle\Omega\rvert
 (V_1^\dagger V_2\otimes I_R)\lvert\Omega\rangle_{LR}
 =\operatorname{Tr}\!\left[\rho_L\,\operatorname{diag}(1,w)\right]
 =1-(1-w){}_L\langle1\rvert\rho_L\lvert1\rangle_L.
\]
This real number lies in $[w,1]$. The trace distance between pure states with overlap $z$ is $\sqrt{1-|z|^2}$~\cite{SI-Fuchs1999}, so it is at most $\sqrt{1-w^2}$.
The input $\lvert1\rangle_L$ attains this bound, without a reference system.
The operator-norm equality follows from
$(V_1-V_2)^\dagger(V_1-V_2)=\operatorname{diag}(0,2(1-w))$.
\end{proof}

\begin{lemma}[Transfer of a fixed recovery channel]
\label{lem:d2}
Let $\mathcal N$ be the same subsequent physical noise channel for both
encodings, and let $\mathcal R$ be a fixed recovery channel.  Then
\begin{equation}
 \frac12\lVert\mathcal R\mathcal N\mathcal V_1
              -\mathcal R\mathcal N\mathcal V_2\rVert_\diamond
 \leq\sqrt{1-w^2}.
 \label{eq:recovery-transfer-distance}
\end{equation}
Using the same maximally entangled input as in the main text, now with
$K=2$, write
\[
 F_e(\mathcal T)=\langle\Phi_2\rvert(\mathrm{id}_R\otimes\mathcal T)
       (\lvert\Phi_2\rangle\langle\Phi_2\rvert)\lvert\Phi_2\rangle.
\]
The entanglement fidelities of the two recovered channels satisfy
\begin{equation}
 \left|F_e(\mathcal R\mathcal N\mathcal V_1)
       -F_e(\mathcal R\mathcal N\mathcal V_2)\right|
 \leq\sqrt{1-w^2}.
 \label{eq:recovery-transfer-fidelity}
\end{equation}
Their errors relative to the identity also obey
\[
 \frac12\lVert\mathcal R\mathcal N\mathcal V_2-\mathrm{id}_L\rVert_\diamond
 \leq \frac12\lVert\mathcal R\mathcal N\mathcal V_1-\mathrm{id}_L\rVert_\diamond
       +\sqrt{1-w^2}.
\]
\end{lemma}

\begin{proof}
Postcomposition by a channel contracts the diamond norm~\cite[Proposition~3.48]{SI-Watrous2018}, giving
\eqref{eq:recovery-transfer-distance}.  Apply this bound to the maximally
entangled input.  Measuring the projector
$\lvert\Phi_2\rangle\langle\Phi_2\rvert$ changes its outcome probability by at
most the trace distance, proving \eqref{eq:recovery-transfer-fidelity}.
The final statement follows from the triangle inequality for diamond norm.
No conversion from entanglement fidelity to worst-case fidelity is used.
\end{proof}

The preceding lemma is sufficient for the three-dimensional application:
when the chosen encoding is recoverable and $w\to1$, a recovery selected for
that encoding also works for the second encoding.  This implication uses a
common physical noise channel and does not give the decoder a record of the
representative choice.

Nonvanishing distance between two general encoding maps would not, by itself,
exclude a common recovery: two orthogonal code subspaces, for example, can
carry distinguishable encoding flags.  The present family has the additional
structure of a shared physical codeword $\lvert\psi_\beta\rangle$ for
logical input $\lvert0\rangle_L$. The common-recovery optimum can then
be evaluated exactly.

\begin{lemma}[Common recovery without a representative record]
\label{lem:d3}
In the absence of subsequent noise, suppose that either $\mathcal V_1$ or
$\mathcal V_2$ is used, and that its identity is not supplied to the recovery.
The optimal worst-choice entanglement fidelity of a common recovery is
\begin{equation}
 \max_{\mathcal R}\min_{i=1,2}F_e(\mathcal R\mathcal V_i)
 =\frac{1+\sqrt{(1+w)/2}}{2}.
 \label{eq:optimal-common-recovery}
\end{equation}
Consequently, after any common subsequent noise channel $\mathcal N$,
\begin{equation}
 \inf_{\mathcal R}\max_{i=1,2}
       [1-F_e(\mathcal R\mathcal N\mathcal V_i)]
 \geq\frac{1-\sqrt{(1+w)/2}}{2}.
 \label{eq:noisy-common-recovery-obstruction}
\end{equation}
\end{lemma}

\begin{proof}
Within this proof, write $\lvert a\rangle=V_1\lvert1\rangle_L$ and
$\lvert b\rangle=V_2\lvert1\rangle_L$.  Both are orthogonal to
$\lvert\psi_\beta\rangle$, and $\langle a\vert b\rangle=w$.
First let $w<1$, and define the orthonormal physical states
\[
 \lvert s_+\rangle=\frac{\lvert a\rangle+\lvert b\rangle}{\sqrt{2(1+w)}},
 \qquad
 \lvert s_-\rangle=\frac{\lvert a\rangle-\lvert b\rangle}{\sqrt{2(1-w)}},
 \qquad c=\sqrt{(1+w)/2}.
\]
They are orthogonal to $\lvert\psi_\beta\rangle$, and
$\lvert a\rangle=c\lvert s_+\rangle+\sqrt{1-c^2}\lvert s_-\rangle$,
$\lvert b\rangle=c\lvert s_+\rangle-\sqrt{1-c^2}\lvert s_-\rangle$.
Consider the equally weighted mixture
$\mathcal M=(\mathcal V_1+\mathcal V_2)/2$.  Its action is
\begin{align*}
 \mathcal M(\lvert0\rangle_L\langle0\rvert_L)
   &=\lvert\psi_\beta\rangle\langle\psi_\beta\rvert,\\
 \mathcal M(\lvert1\rangle_L\langle1\rvert_L)
   &=c^2\lvert s_+\rangle\langle s_+\rvert
      +(1-c^2)\lvert s_-\rangle\langle s_-\rvert,\\
 \mathcal M(\lvert0\rangle_L\langle1\rvert_L)
   &=c\lvert\psi_\beta\rangle\langle s_+\rvert.
\end{align*}
For an arbitrary recovery $\mathcal R$ with Kraus operators $R_\mu$,
entanglement fidelity is one quarter of the sum of two correct-output
population terms and two conjugate coherence terms.  Each population term is
at most one.  The relevant coherence obeys
\begin{align*}
 \left|{}_L\langle0\rvert\mathcal R
       (\lvert\psi_\beta\rangle\langle s_+\rvert)\lvert1\rangle_L\right|
 &=\left|\sum_\mu
   {}_L\langle0\rvert R_\mu\lvert\psi_\beta\rangle
   \langle s_+\rvert R_\mu^\dagger\lvert1\rangle_L\right|\\
 &\leq\left(\sum_\mu|{}_L\langle0\rvert R_\mu\lvert\psi_\beta\rangle|^2\right)^{1/2}
       \left(\sum_\mu|{}_L\langle1\rvert R_\mu\lvert s_+\rangle|^2\right)^{1/2}
 \leq1.
\end{align*}
Hence $F_e(\mathcal R\mathcal M)\leq(2+2c)/4$.
By linearity of entanglement fidelity in the channel,
\[
 \min_i F_e(\mathcal R\mathcal V_i)
 \leq\tfrac12\sum_i F_e(\mathcal R\mathcal V_i)
 =F_e(\mathcal R\mathcal M)\leq\frac{1+c}{2}.
\]

To attain the bound, take recovery Kraus operators
\[
 R_+=\lvert0\rangle_L\langle\psi_\beta\rvert
           +\lvert1\rangle_L\langle s_+\rvert,
 \qquad
 R_-=\lvert1\rangle_L\langle s_-\rvert.
\]
Complete these to a trace-preserving channel on the remaining physical
subspace, which the encoded inputs never occupy.  For both $i$, the recovered channel
preserves the logical populations and multiplies the off-diagonal matrix
elements by $c$.  It therefore has entanglement fidelity $(1+c)/2$ for
each encoder, proving \eqref{eq:optimal-common-recovery}.
At $w=1$ the encoders coincide and their isometric inverse recovers perfectly.
Finally, every $\mathcal R\mathcal N$ is itself a physical-to-logical channel;
optimizing over this restricted class cannot improve the noiseless optimum.
This proves \eqref{eq:noisy-common-recovery-obstruction}.
\end{proof}

For the two-dimensional boundary-enhancement regime in which $w\to0$, the
last bound tends to
$(1-1/\sqrt2)/2\simeq0.1464$.
This is a lower bound on the worst-choice entanglement infidelity when the
representative choice is unknown to the recovery.  It is compatible with
vanishing recovery error for each separately specified encoding.  In three
dimensions, $w\to1$ removes this particular obstruction, while
Lemma~\ref{lem:d2} combines with the AQEC upper bound to establish a common
recovery.  Neither statement asserts that two disjoint physical regions each
contain an independently recoverable copy of the logical qubit.

\subsection{Changing the boundary used for a controlled logical operation}

Changing the physical boundary used for a logical operation provides another
way to test representative equivalence.  Here the target encoding is held
fixed, and only the controlled string is changed.  Use $V=V_1$, with $V|0\rangle_L=|\psi_\beta\rangle$ and
$V|1\rangle_L=Z^{\ell_1}|\psi_\beta\rangle$.  These are the same representatives as above:
$Z^{\ell_1}$ and $Z^{\ell_2}$ commute, square to the identity, and anticommute with an
exact $X$ logical operator that stabilizes $|\psi_\beta\rangle$.  We keep the
logical basis fixed, so the physical string $Z^{\ell_1}$ implements logical
$X_L=|0\rangle_L\langle1|_L+|1\rangle_L\langle0|_L$.

\begin{lemma}[Controlled-operation error from the Wilson overlap]
\label{lem:cnot}
Let $w=\langle\psi_\beta|Z^{\ell_1}Z^{\ell_2}|\psi_\beta\rangle\in[0,1]$.  For an ideal control
qubit $c$, define operators on $\mathcal H_c\otimes\mathcal H_Q$, with $I_Q$ the physical-target identity,
\begin{equation}\label{d:controlledstrings}
 G_j=|0\rangle_c\langle0|_c\otimes I_Q
       +|1\rangle_c\langle1|_c\otimes Z^{\ell_j},\qquad j=1,2.
\end{equation}
The ideal gate $\operatorname{CNOT}_{c\to L}$ acts on
$\mathcal H_c\otimes\mathcal H_L$. With $I_c$ and $I_L$ the identities
on the control and logical input, respectively,
\begin{align}
 G_1(I_c\otimes V)&=(I_c\otimes V)\operatorname{CNOT}_{c\to L},
       \label{d:cnotexact}\\
 V^\dagger Z^{\ell_2}V&=wX_L,\qquad
 V^\dagger Z^{\ell_1}Z^{\ell_2}V=wI_L,\label{d:remotecompression}\\
 \left\|G_2(I_c\otimes V)
       -(I_c\otimes V)\operatorname{CNOT}_{c\to L}\right\|_\infty
       &=\sqrt{2(1-w)}.\label{d:cnotnorm}
\end{align}
For control input $|1\rangle_c$, the probability that $G_2$ takes the target
outside the fixed code space is $1-w^2$ for every target input.  These
statements remain valid for logical inputs entangled with an arbitrary
reference system.

Writing $\mathcal G_j$ for the channel from the control and logical target
to the control and physical target induced by $G_j(I_c\otimes V)$, one
also has
\begin{equation}\label{d:cnotdiamond}
 \frac12\|\mathcal G_1-\mathcal G_2\|_\diamond=\sqrt{1-w^2}.
\end{equation}
\end{lemma}

\begin{proof}
The identities $Z^{\ell_1}|\psi_\beta\rangle=V|1\rangle_L$ and
$(Z^{\ell_1})^2|\psi_\beta\rangle=V|0\rangle_L$ give $Z^{\ell_1}V=VX_L$, proving
Eq.~\eqref{d:cnotexact}.  The exact $X$ logical symmetry makes the
expectations of $Z^{\ell_1}$ and $Z^{\ell_2}$ vanish.  Since $Z^{\ell_1}$ and $Z^{\ell_2}$ commute,
the diagonal entries of $V^\dagger Z^{\ell_2}V$ vanish and its off-diagonal
entries both equal $w$.  Similarly, $V^\dagger Z^{\ell_1}Z^{\ell_2}V$ has diagonal
entries $w$ and vanishing off-diagonal entries.  This proves
Eq.~\eqref{d:remotecompression}.

Subtracting the first controlled implementation from the second leaves
only the control-$1$ block.  Its squared norm is determined by
\[
 \bigl[(Z^{\ell_2}-Z^{\ell_1})V\bigr]^\dagger\bigl[(Z^{\ell_2}-Z^{\ell_1})V\bigr]
 =V^\dagger(2I_Q-2Z^{\ell_1}Z^{\ell_2})V=2(1-w)I_L.
\]
The largest singular value is therefore $\sqrt{2(1-w)}$, giving
Eq.~\eqref{d:cnotnorm}.  Projection onto the fixed code is given by
$VV^\dagger$.  Equation~\eqref{d:remotecompression} gives
\[
 V^\dagger Z^{\ell_2}VV^\dagger Z^{\ell_2}V
 =(V^\dagger Z^{\ell_2}V)^\dagger(V^\dagger Z^{\ell_2}V)=w^2I_L.
\]
For any logical state $\rho_L$, the probability to remain in the code
after the remote string is consequently
$\operatorname{Tr}(\rho_Lw^2I_L)=w^2$.  Tensoring these operator
identities with a reference identity proves the corresponding statements
for entangled inputs.

Finally, the overlap of the two control--target isometries is
\[
 (I_c\otimes V^\dagger)G_1^\dagger G_2(I_c\otimes V)
 =|0\rangle_c\langle0|_c\otimes I_L
    +w|1\rangle_c\langle1|_c\otimes I_L.
\]
For an arbitrary pure input including a reference, the output overlap
is $1-(1-w)\Pr(c=1)$ and belongs to $[w,1]$.  The pure-state trace
distance is at most $\sqrt{1-w^2}$ and attains that value for control
$|1\rangle_c$.  Convexity handles mixed inputs, proving
Eq.~\eqref{d:cnotdiamond}.
\end{proof}

Combining this lemma with the main-text Wilson-restoration theorem proves the
encoded-CNOT corollary.  In the enhanced three-dimensional regime,
the two distant boundaries implement the same controlled logical
operation with vanishing error.  In the two-dimensional regime, using
the remote representative in the fixed target encoding instead gives a
leakage probability tending to one when the control is $|1\rangle_c$.
Each representative still acts exactly as a logical $X$ for its own
encoding.  The distinction is whether it can replace the chosen
representative while the target encoding is held fixed.

The control and controlled strings in this comparison are ideal.  A
controlled string can be realized algebraically as a product of
controlled-$Z$ gates on its support; geometric locality, depth, and
fault-tolerant implementation of those gates require a separate circuit
analysis.  The result is an operational consequence of representative
equivalence, and places no general prohibition on other logical-CNOT
constructions in two dimensions.

\makeatletter
\let\auto@bib@innerbib\@empty
\let\SI@label\label
\def\label#1{%
  \def\SI@labelname{#1}\def\SI@lastbib{LastBibItem}%
  \ifx\SI@labelname\SI@lastbib
    \SI@label{LastSIBibItem}%
  \else
    \SI@label{#1}%
  \fi}
\makeatother

\endgroup

\end{document}